\pdfoutput=1

\documentclass[review]{elsarticle}
\usepackage[dvipsnames]{xcolor}

\makeatletter
\def\ps@pprintTitle{%
 \let\@oddhead\@empty
 \let\@evenhead\@empty
 \def\@oddfoot{}%
 \let\@evenfoot\@oddfoot}
\makeatother

\usepackage{amssymb}
\usepackage{mathtools}
\usepackage{subcaption}
\usepackage[section]{placeins}

\usepackage[numbers]{natbib}
\usepackage{listings}

\usepackage{graphicx}
\usepackage{hyperref}
\usepackage{floatrow}
\usepackage{algorithm}
\usepackage{algpseudocode}
\usepackage{amsmath}
\usepackage[size=footnotesize]{subcaption}
\usepackage[textsize=tiny]{todonotes}
\usepackage[capitalise]{cleveref}
\crefname{appendix}{}{}
\usepackage{soul}
\usepackage{geometry}
\usepackage{algorithmicx}
\usepackage{booktabs}
\usepackage{longtable}
\usepackage{siunitx}
\usepackage{stmaryrd}
\usepackage{algpseudocode}
\usepackage{adjustbox}
\usepackage{color, colortbl}
\usepackage{multirow}
\usepackage{xcolor}
\definecolor{mygreen}{RGB}{0,180,0}
\definecolor{myred}{RGB}{220,0,0}
\definecolor{Reviewer2}{rgb}{0.0, 0.0, 0.0}
\definecolor{Reviewer1}{rgb}{0.0, 0.0, 0.0}
\colorlet{Reviewer21}{Cerulean}
\usepackage{ulem}
\usepackage{svg-extract}
\usepackage{svg}

\newcommand{\rhoLENT}{$\rho$LENT }
\newcommand{\Cell}[1]{\Omega_{#1}}

\newcommand{\tn}{t^n}
\newcommand{\tnn}{t^{n+1}}
\newcommand{\Sf}{\mathbf{S}_f}

\renewcommand{\v}{\mathbf{v}}
\newcommand{\x}{\mathbf{x}}
\newcommand{\n}{\mathbf{n}}

\colorlet{Santiago}{black}
\colorlet{Davide}{black}
\colorlet{Tobi}{black}

\colorlet{Reviewer22}{black}
\allowdisplaybreaks

\renewcommand{\emph}[1]{\textit{#1}}

\usepackage{lineno}
\graphicspath{{figures/}{svg-inkscape/}}
\begin{document}

\frenchspacing

\begin{frontmatter}


\author[add1]{Jun Liu}
\ead{liu@mma.tu-darmstadt.de}
\author[add2]{Tobias Tolle}
\ead{tolle@mma.tu-darmstadt.de}
\author[add3]{Davide Zuzio}
\ead{Davide.Zuzio@onera.fr}
\author[add3]{Jean-Luc Estival\`{e}zes}
\ead{Jean-Luc.Estivalezes@onera.fr}
\author[add4]{Santiago Marquez Damian}
\ead{santiagomarquezd@gmail.com}
\author[add1]{Tomislav Mari\'{c}\corref{corr}}
\cortext[corr]{Corresponding author}
\ead{maric@mma.tu-darmstadt.de}

\address[add1]{Mathematical Modeling and Analysis, Technische Universit\"{a}t Darmstadt, Germany}
\address[add2]{Bosch Research, Robert Bosch GmbH; research conducted at MMA, Technische Universit\"{a}t Darmstadt}
\address[add3]{ONERA/DMPE, Universit\'{e} de Toulouse, France}
\address[add4]{Centro de Investigaciones en Mecánica Computacional (CIMEC), UNL/CONICET, Argentina}



\title{Inconsistencies in Unstructured Geometric Volume-of-Fluid Methods for Two-Phase Flows with High Density Ratios}


\begin{abstract}
\textbf{This is the preprint version of the published manuscript \url{https://doi.org/10.1016/j.compfluid.2024.106375} : please cite the published manuscript when refering to the contents
of this document.}

Geometric flux-based Volume-of-Fluid (VOF) methods \citep{Maric2020} are widely considered consistent in handling two-phase flows with high density ratios.
However, although the conservation of mass and momentum is consistent for two-phase incompressible single-field Navier-Stokes equations without phase-change \citep{Liu2023}, discretization may easily introduce  inconsistencies 
that result in very large errors or catastrophic failure.
We apply the consistency conditions derived for the  $\rho$LENT unstructured Level Set / Front Tracking method \citep{Liu2023} to flux-based geometric VOF methods \citep{Maric2020}, and implement our discretization into the plicRDF-isoAdvector geometrical VOF method \citep{Roenby2016}. We find that computing the mass flux by scaling the geometrically computed fluxed phase-specific volume \textcolor{Reviewer1}{can ensure} equivalence between \textcolor{Reviewer1}{ the mass conservation equation and the phase indicator (volume conservation) if consistent discretization schemes are chosen for the temporal and convective term.}
\textcolor{Reviewer1}{Based} on the analysis of discretization errors, we suggest a consistent combination of the temporal discretization scheme and the interpolation scheme for the momentum convection term. \textcolor{Reviewer1}{We confirm the consistency by solving an auxiliary mass conservation equation with a geometrical calculation of the face-centered density \citep{Liu2023}}. We prove the equivalence between these two approaches mathematically and verify and validate their numerical stability for density ratios within [$1$,$10^6$] and viscosity ratios within [$10^2$,$10^5$]. 
\end{abstract}



\begin{keyword}



volume-of-fluid \sep unstructured \sep finite volume \sep high density ratios
\end{keyword}

\end{frontmatter}


\section{Introduction}
\label{sec:intro}

Numerical simulation methods still face challenges when dealing with incompressible two-phase flows that involve fluid phases of significantly different densities (high density ratios).
We have provided a detailed review of two-phase flow simulation methods that address the challenges in handling high density ratios recently in \citep{Liu2023}, and here we only list more current literature contributions.

\citet{Huang2019} proposed the mixed Upwind/Central WENO scheme on a staggered structured grid, an extension of the conventional WENO scheme \citep{Jiang1996} on collocated grids, to spatially discretize the nonlinear convective term in conservative form. The Upwind WENO scheme is used to evaluate velocity at cell faces, while 3 different forms of Central WENO scheme are applied to evaluate mass flux in x-/y-directions as well as density at cell corners. In addition to the mentioned spatial discretization scheme, \citet{Huang2019} proposed also a semi-implict projection scheme to decouple pressure and velocity in momentum transport equation. A backward difference scheme combined with a special treatment for the viscous term is introduced to discretize the momentum equation without the pressure gradient. The \textcolor{Santiago}{viscous} term containing the intermediate velocity is split into two parts, one of which 
comprises the constant arithmetic mean of two phases' viscosity and the intermediate velocity, another one comprises the updated viscosity and the explicit velocity (referring to \citep[Apendix A]{Dodd2014}).  Numerous 2D cases with density ratio $[1,1000]$ are tested, whose results are in good agreement with analytical and experimental results. The interface capturing method in \citep{Huang2019} is selected to phase-field. The authors indicated that the mixed Upwind/Central WENO scheme is able to be coupled with any interface capturing/tracking method.

\citet{Xie2020} introduced a consistent and balanced-force model with level-set and volume of fluid function (CBLSVOF) on polyhedral unstructured grids. A hybrid algebraic/geometric VOF method based on THINC/QQ \citep{Xie2017} is developed to capture interface, while the LS function constructed from volume fraction is exploited to improve the accuracy of interface curvature evaluation.  To achieve the consistency,  the convective term of both the volume-fraction equation and momentum equation in the conservative form are discretized in the same manner, wherein the velocity and volume fraction are reconstructed by using the identical high-order reconstruction scheme based on a quadratic polynomial function. The balanced-force formulation proposed firstly by \citet{Francois2006} is adapted in this work to eliminate the parasitic current at the interface. The interface curvature estimated by the  continuous signed-distance function instead of the abruptly \textcolor{Tobi}{changing} volume fraction function offers computational simplicity and numerical stability. A 3D single bubble, and two bubbles coalescence cases with the density ratio of magnitude $10^3$ are tested. The results agree well with previous experimental results.         

\citet{Desmons2021} proposed a generalized approach known as the HOMP (High-Order Momentum Preserving) method. A high-order temporal and spatial scheme, i.e., RK2 and WENO5,3 \citep{Wang2007}, are used to discretize an auxiliary advection equation of characteristic function $\chi$ that is compatible with the mass equation and independent of the underlying interface-representation function. The momentum equation is then discretized using the same high-order schemes as before, and the density is deduced from the characteristic function $\chi$, which is computed from the auxiliary advection equation rather than from the interface transport step. During the discretization, both the auxiliary advection equation and the momentum equation remain in the conservative form. The authors combined several interface representation methods, i.e., Volume Of Fluid, Level Set Method, and Moment Of Fluid, with HOMP and then tested them. The results from various selected validation cases, including water and air, demonstrate good agreement with the literature results. It is worth noting that no theoretical explanation is offered as to why the auxiliary advection equation is required, rather than just maintaining the consistency between the interface transport equation and the momentum equation. Alternatively, the higher-order WENO scheme requires very large stencils of variable width when used with the finite-volume method, making its parallel implementation inefficient using the domain-decomposition / message-passing parallel programming model.  

\textcolor{Reviewer1}{\citet{Ouafa2021monolithic} devised a fully coupled solver for simulating incompressible two-phase flows characterized by large density and viscosity ratios on  staggered structured meshes. The interface is captured by a Piecewise Linear Interface Calculation (PLIC)-Volume of Fluid (VOF) method. In this solver, the linearized momentum and continuity equations arising from the implicit solution of the fluid velocities and pressure are solved simultaneously to avoid the errors arising from operator splitting. Cases featuring density ratios up to $10^6$ and viscosity ratios up to $10^{10}$ are tested, whose results demonstrate a high level of stability and accuracy. %
} 

\textcolor{Reviewer1}{\citet{yang2021robust} introduced a robust methodology that integrates the consistent mass-momentum convection approach from \citet{nangia2019} with the CLSVOF interface capturing method by \citet{sussman2000} to tackle the challenge posed by high-density ratio problems encountered in high-Reynolds-number flows. At each time step, the interface evolution is initiated using the CLSVOF method to provide an initial value for the mass equation. Subsequently, the conservative form of the momentum equation and mass equation are simultaneously solved employing consistent temporal (second-order Runge-Kutta (RK2)) and spatial (third-order cubic upwind interpolation (CUI)) discretization schemes at all cell faces of the staggered structured mesh. The authors emphasized the importance of employing identical densities in the discretized mass and momentum equations at the two substeps of the RK2 method to ensure robust simulation of high-Reynolds-number two-fluid flows with high density ratios. The proposed method yields accurate predictions for both two-dimensional and three-dimensional wave breaking cases with a density ratio of $10^3$ and Reynolds number of $10^8$.
} 

\textcolor{Reviewer1}{\citet{li2020numerical} proposed a straightforward and robust method to simulate high density ratio interfacial flows. Similar to the method in \citep{yang2021robust}, the auxiliary mass equation is solved together with the momentum equation. However, instead of using the consistent discretization schemes in space and time, this method applies the identical cell-center velocity updated by solving the mass equation and the mass flux to the discretized momentum equation to maintain consistency between them. Three different schemes of VOF, i.e, PLIC-VOF, the spatial filtering VOF, and the THINC-VOF, are implemented in the work to transport the interface. The improvements of stability and accuracy for all three schemes in canonical cases like heavy droplet advection and falling droplet show the generality of the consistent method to deal with high density ratios problem.}

\textcolor{Reviewer1}{\citet{zeng2023consistent} transferred the consistent treatment of conservative mass and momentum equations on staggered Cartesian grids introduced by \citet{nangia2019} to multilevel collocated grids. The adapted level set method with a multilevel reinitialization technique is applied to capture the interface. The consistent scheme achieves a numerically stable and reasonably accurate solution to realistic multiphase ﬂows, such as breaking waves with a high Reynolds number.} 

In this manuscript, we apply the consistency requirements imposed by the single-field Navier-Stokes equations derived in \citep{Liu2023} to flux-based geometric Volume-of-Fluid methods \citep{Maric2020}, using the isoAdvector method to verify and validate our findings \citep{Roenby2016,Scheufler2019,Scheufler2023twophaseflow}. The consistency requirements from \citep{Liu2023} indicate that geometrical flux-based VOF methods are inherently consistent in handling high density ratios. On the modeling level, an exact solution of the volume fraction equation is equivalent to solving the mass conservation equation. On the discrete level, flux-based VOF methods consistently compute the mass flux needed in the implicitly discretized single-field momentum convection term, by scaling the fluxed phase-specific volume. However, we show that inconsistencies easily arise through the choice of an inconsistent combination of the temporal integration scheme and the interpolation scheme for the two-phase momentum conservation term, leading to significant errors for small density ratios and catastrophic failures for large density ratios.

In the following sections, we briefly review the single-field formulation of incompressible two-phase Navier-Stokes equations and analyze their discretization using the collocated unstructured Finite Volume method \citep{OFprimer}. In \Cref{sec:Connection_mass_vof_transport}, we show that a consistent combination of the temporal discretization scheme and the interpolation scheme for the momentum convection term is necessary for ensuring stable solutions with high density ratios. In \Cref{sec:introduction_solver,sec:submerged_areaFraction}, we introduce an auxiliary mass conservation equation with a geometrical calculation of the face-centered density to the geometrical VOF method isoAdvector\textcolor{Tobi}{\citep{Roenby2016,Scheufler2019,Scheufler2023twophaseflow}}. We prove the equivalence between these two approaches and verify and validate their numerical stability for density ratios of [$1$,$10^6$] and viscosity ratios of [$10^2$,$10^5$] in the results \Cref{sec:validations-results}.
\section{Mathemathical model}
\begin{figure}[!htb]
    \centering
    \def\svgwidth{0.6\textwidth}
\begingroup%
  \makeatletter%
  \providecommand\color[2][]{%
    \errmessage{(Inkscape) Color is used for the text in Inkscape, but the package 'color.sty' is not loaded}%
    \renewcommand\color[2][]{}%
  }%
  \providecommand\transparent[1]{%
    \errmessage{(Inkscape) Transparency is used (non-zero) for the text in Inkscape, but the package 'transparent.sty' is not loaded}%
    \renewcommand\transparent[1]{}%
  }%
  \providecommand\rotatebox[2]{#2}%
  \newcommand*\fsize{\dimexpr\f@size pt\relax}%
  \newcommand*\lineheight[1]{\fontsize{\fsize}{#1\fsize}\selectfont}%
  \ifx\svgwidth\undefined%
    \setlength{\unitlength}{475.33460242bp}%
    \ifx\svgscale\undefined%
      \relax%
    \else%
      \setlength{\unitlength}{\unitlength * \real{\svgscale}}%
    \fi%
  \else%
    \setlength{\unitlength}{\svgwidth}%
  \fi%
  \global\let\svgwidth\undefined%
  \global\let\svgscale\undefined%
  \makeatother%
  \begin{picture}(1,0.57363443)%
    \lineheight{1}%
    \setlength\tabcolsep{0pt}%
    \put(0,0){\includegraphics[width=\unitlength,page=1]{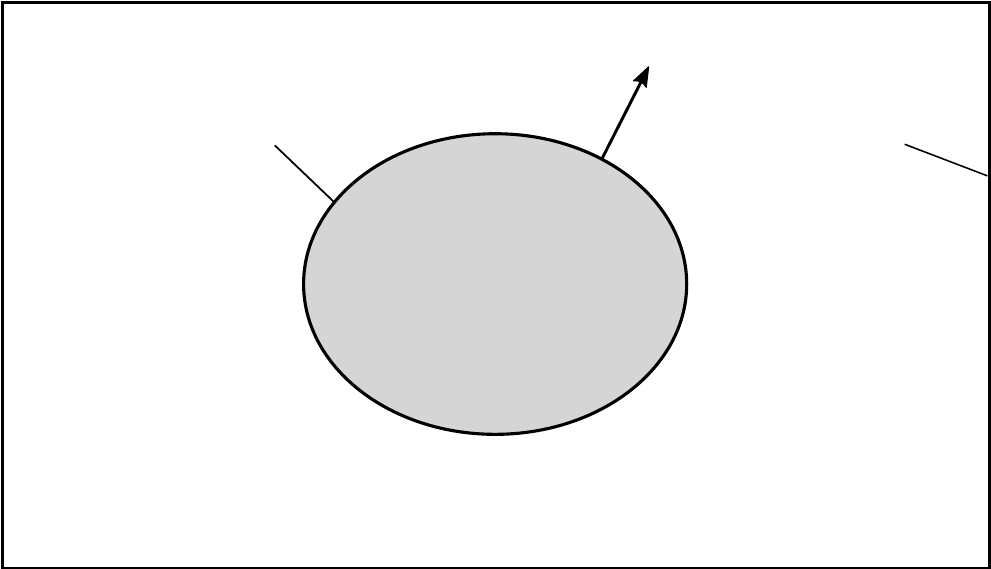}}%
    \put(0.41135156,0.33339009){\color[rgb]{0,0,0}\makebox(0,0)[lt]{\smash{\begin{tabular}[t]{l}$\Omega^-(t)$\end{tabular}}}}%
    \put(0.41067232,0.28531297){\color[rgb]{0,0,0}\makebox(0,0)[lt]{\smash{\begin{tabular}[t]{l}$\chi(\mathbf{x}, t) = 1$\end{tabular}}}}%
    \put(0.06215877,0.07651644){\color[rgb]{0,0,0}\makebox(0,0)[lt]{\smash{\begin{tabular}[t]{l}$\chi(\mathbf{x}, t) = 0$\end{tabular}}}}%
    \put(0.8689053,0.44140486){\color[rgb]{0,0,0}\makebox(0,0)[lt]{\smash{\begin{tabular}[t]{l}$\partial \Omega$\end{tabular}}}}%
    \put(0.64858374,0.51490576){\color[rgb]{0,0,0}\makebox(0,0)[lt]{\smash{\begin{tabular}[t]{l}$\boldsymbol{n}_{\Sigma}$\end{tabular}}}}%
    \put(0.06311562,0.12014973){\color[rgb]{0,0,0}\makebox(0,0)[lt]{\smash{\begin{tabular}[t]{l}$\Omega^+(t)$\end{tabular}}}}%
    \put(0.24701592,0.4468515){\color[rgb]{0,0,0}\makebox(0,0)[lt]{\smash{\begin{tabular}[t]{l}$\Sigma(t)$\end{tabular}}}}%
  \end{picture}%
\endgroup%

    \caption{The domain $\Omega$, split by the fluid interface $\Sigma(t)$ into two sub-domains $\Omega^\pm$.}
    \label{fig:drop-gas}
\end{figure}
\Cref{fig:drop-gas} illustrates an incompressible two-phase flow system without phase change. The flow domain $\Omega\subset \mathbb{R}^3$ with a boundary $\partial \Omega$ is partitioned into two subdomains $\Omega^+(t)$ and $\Omega^-(t)$, occupied by two different fluid phases. The boundary between $\Omega^\pm(t)$, referred to as the fluid interface, is denoted by $\Sigma(t)$ and has the normal $\mathbf{n}_{\Sigma}$ pointing, say, outwards of the domain $\Omega^-(t)$. 

A phase indicator function 
\begin{equation}
  \chi(\x,t) := 
    \begin{cases}
      1, & \x \in \Omega^-(t), \\ 
      0, & \x \in \Omega^+(t), 
    \end{cases}
  \label{eq:indicator}
\end{equation}
indicates $\Omega^\pm(t)$ and formulates single-field density and dynamic viscosity as
\begin{equation}
    \rho(\x,t) = (\rho^-  - \rho^+) \chi(\x,t) + \rho^+, \label{eq:rhoindicator} 
\end{equation}
\begin{equation}
    \mu(\x,t) = (\mu^- - \mu^+) \chi(\x,t) + \mu^+, \label{eq:nuindicator} 
\end{equation} 
where $\rho^\pm$ and $\mu^\pm$ are the constant densities and dynamic viscosities in $\Omega^\pm(t)$. The one-fluid density and dyamic viscosity are used in the Navier-Stokes equations without phase change, namely
\begin{align}
    \nabla\cdot\v &= 0 \label{eq:volume-transport}, \\
     \partial_t(\rho \v)+\nabla\cdot(\rho \v \otimes \v)& = -\nabla p - (\mathbf{g}\cdot \x) \nabla \rho + \nabla \cdot (\mu\left(\nabla\v + (\nabla\v)^T\right)) + \mathbf{f}_\Sigma. \label{eq:momentum-transport}
\end{align}
where $p$ is the \textcolor{Santiago}{modified} pressure defined as the subtraction of hydrostatic pressure from the total pressure, i.e., $p=P-\rho \mathbf{g}\cdot\x$.
The surface tension force $\mathbf{f}_\Sigma$ is exerted only on the interface, as
\begin{equation}
    \mathbf{f}_\Sigma = \sigma \kappa \mathbf{n}_\Sigma\delta_\Sigma,
    \label{eq:csf-model}
\end{equation}
where $\sigma$ is the constant surface tension coefficient, $\kappa$ is twice the local mean curvature of the interface $\Sigma(t)$, and $\delta_\Sigma$ is the interface Dirac distribution.

This singe-field formulation of incompressible two-phase Navier-Stokes Equations (NSE) without phase change is especially relevant for engineering applications because it provides a solid modeling basis for two-phase flows with fluid interfaces that can arbitrarily deform, breakup and merge, provided that the method responsible for advecting the phase indicator does not impose its own restrictions regarding fluid interface deformation and topological changes. Single-field NSE also embed strict consistency requirements \citep{Liu2023}: an equivalence between the conservation of mass and the phase-indicator (volume) conservation. These requirements translate on the discrete level into requirements for the computation of the mass flux in the discretized two-phase momentum convection term in \cref{eq:momentum-transport}, analyzed in the following section.

\section{Numerical methodology}
\label{sec:numerical-method-and-implementation}

\subsection{\textcolor{Reviewer22}{Consistent mass, volume, and momentum conservation}}
\label{sec:Connection_mass_vof_transport}

The \textcolor{Reviewer22}{unstructured Finite Volume} discretization of the single-field two-phase momentum convection term from \cref{eq:momentum-transport} \textcolor{Reviewer22}{results in}
\begin{equation}
    \int_{\Omega_c} \nabla \cdot( \rho \v \otimes \v) \, dV
    = \int_{\partial \Omega_c} ( \rho \v \otimes \v) \cdot \n \, ds
    = \sum_{f \in F_c} \rho_f F_f \v_f + \mathbf{e}_{\rho\v}(h^2),
    \label{eq:momterm}
\end{equation}
for the \textcolor{Reviewer2}{solution domain} $\Omega$, discretized by $|C|$ non-overlapping control volumes $\Omega:=\cup_{c\in C}\Omega_c$. Each finite volume $\Omega_c$ is bounded by a number ($|F_c|$) of non-planar surfaces \textcolor{Reviewer22}{(faces)} $S_f$, i.e. $\partial \Omega_c=\cup_{f \in F_c}S_f$, oriented outwards of $\Omega_c$. Linearizing the convective term for the velocity $\mathbf{v}$ and using the second-order accurate collocated Unstructured Finite Volume Method (UFVM) \citep{Jasak1996, OFprimer, moukalled2016finite} introduces the face-centered volumetric flux
\begin{equation}
    F_f := \v_f \cdot \Sf,
    \label{eq:faceVolumetricFlux}
\end{equation}
where $\v_f$ is the second-order accurate face-centered velocity average and \textcolor{Reviewer22}{$\mathbf{e}_{\rho\v}(h^2)$ is the discretization error of the momentum convection term}. The collocated UFVM denotes with $f$ second-order accurate face-centered area-averaged quantities, and with $c$ second-order accurate cell-centered volume-averaged quantities.

\textcolor{Reviewer22}{The mass flux $\rho_f F_f$ is constricted in an incompressible two-phase flow by volume conservation, i.e. the transport of the volume fraction, defined as a volumetric average of the phase-indicator function (\ref{eq:indicator}), i.e.}
\begin{equation}
  \alpha_c(t) := \frac{1}{|\Cell{c}|}\int_{\Cell{c}} \chi(\x, t) \, dV.
  \label{eq:volfracdef}
\end{equation}

The volume fraction advection equation \cite{Maric2020}

\begin{equation}
    \frac{d}{dt}\int_{\Omega_c} \chi \, dV = |\Omega_c| \frac{d}{dt}\alpha_c(t) 
    = - \int_{\partial \Omega_c} \chi \v \cdot \n \, dS,
    \label{eq:alphaeqn1}
\end{equation}
 is equivalent to the advection equation for the phase $\Omega^+(t)$ indicated by $1 - \chi(\x,t)$, namely

\begin{equation}
    \frac{d}{dt}\int_{\Omega_c} (1 - \chi) \, dV = - |\Omega_c| \frac{d}{dt}\alpha_c(t) 
    =  -\int_{\partial \Omega_c} (1 - \chi) \v \cdot \n \, dS,
    \label{eq:alphaeqn2}
\end{equation}
since $\frac{d}{dt} \int_{\Omega_c} \, dV= 0$, as $\Omega_c\ne\Omega_c(t)$, and $\int_{\partial \Omega_c} \v\cdot \n \, dS = \int_{\Omega_c} \nabla \cdot \v \, dV = 0$ for incompressible two-phase flows without phase change. 

The mass conservation with the single-field density \cref{eq:rhoindicator} in a fixed control volume $\Omega_c$ states

\begin{equation}
    \frac{d}{dt} \int_{\Omega_c} \rho \, dV = - \int_{\partial \Omega_c} \rho \v \cdot \n \, dS. \label{eq:massconserv}
\end{equation}

Inserting \cref{eq:rhoindicator} into the r.h.s. of \cref{eq:massconserv} and integrating over the time step $[\tn, \tnn]$ results in 

\begin{equation}
    \begin{aligned}
   \rho_c^{n+1} & = \rho_c^n + \frac{1}{|\Omega_c|}
    \left[
     -\rho^-  \int_{\tn}^{\tnn}\sum_{f \in F_c}\int_{S_f} \chi \v \cdot \n \, dS \, dt
     - \rho^+  \int_{\tn}^{\tnn}\sum_{f \in F_c}\int_{S_f} (1 - \chi) \v \cdot \n \, dS \, dt,
    \right]
    \end{aligned}
    \label{eq:massintegral}
\end{equation}

Inserting the \cref{eq:alphaeqn1,eq:alphaeqn2} in the \cref{eq:massintegral} to replace the sums of surface integrals results in 

\begin{equation}
    \rho_c^{n+1} = \rho_c^n + \frac{\rho^-}{|\Omega_c|}|\Omega_c|\int_{\tn}^{\tnn}\frac{d}{dt}\alpha_c(t) \, dt
    + \frac{\rho^+}{|\Omega_c|}|\Omega_c|\int_{\tn}^{\tnn}-\frac{d}{dt}\alpha_c(t) \, dt, 
    \label{eq:rho-alpha-intermediate}
\end{equation}
leading finally to

\begin{equation}
    \rho_c^{n+1} = \rho_c^{n} + (\rho^- - \rho^+) (\alpha_c^{n+1} - \alpha_c^n).
    \label{eq:rho-alpha}
\end{equation}

\Cref{eq:rho-alpha} shows that solving \textcolor{Reviewer22}{the mass conservation equation} (\ref{eq:massconserv}) for $\rho_c^{n+1}$ is equivalent to solving \textcolor{Reviewer22}{the volume fraction equation (\ref{eq:alphaeqn1})} scaled with $(\rho^- - \rho^+)$. 

Alternatively, integrating the single-field density model (\ref{eq:rhoindicator}) over the control volume $\Omega_c$ gives

\begin{equation}
    \int_{\Omega_c} \rho \, dV = \int_{\Omega_c} \rho^-\chi + \rho^+(1 - \chi) \, dV.
    \label{eq:rhoindicatorintegral}
\end{equation}

Dividing \cref{eq:rhoindicatorintegral} by $|\Omega_c|$ \textcolor{Santiago}{using \cref{eq:volfracdef}} results in the discrete single-field density model,

\begin{equation}
    \rho_c(t) = \rho^-\alpha_c(t) + \rho^+(1 - \alpha_c(t)),
    \label{eq:rhoalpha}
\end{equation}
which, evaluated at $\tnn$ and $\tn$ and subsequently subtracted, results in 
\begin{equation*}
    \begin{aligned}
   \rho_c^{n+1} - \rho_c^n &= \rho^-\alpha_c^{n+1} + \rho^+(1 - \alpha_c^{n+1}) - 
   \rho^-\alpha_c^n - \rho^+(1 - \alpha_c^n) \\
   & = (\rho^- - \rho^+)(\alpha_c^{n+1} - \alpha_c^n),
   \end{aligned}
   \end{equation*}
which is \cref{eq:rho-alpha}. \textcolor{Reviewer22}{Therefore, solving the mass conservation equation (\ref{eq:massconserv}), solving the volume fraction advection equation (\ref{eq:alphaeqn1}) scaled by $(\rho^- - \rho^+)$, and using the single-field density model to compute the cell centered density from volume fractions by \cref{eq:rhoindicator} are equivalent}.

Note that all equations \cref{eq:volfracdef} - \cref{eq:rhoalpha} are exact, as $\partial \Omega_c := \bigcup_{f \in F_c} S_f$, where $S_f$ are non-linear surfaces that bound the control volume $\Omega_c$, and reformulating the integration in time is exact by the fundamental theorem of calculus. 

\citet{Ghods2013} introduce the auxiliary mass conservation equation as a means for ensuring the consistency of the two-phase momentum convection, and in \cite{Liu2023} we provide the theoretical reasoning for the auxiliary mass conservation equation. Contrary to \citet{Ghods2013}, \cref{eq:rho-alpha} and  \cref{eq:rhoindicatorintegral} both demonstrate, that the solution of the volume fraction equation in the context of the VOF method \citep{Maric2020} is exactly equivalent to the solution of the mass conservation equation, rendering an auxiliary mass conservation equation unnecessary. 

However, the following question arises: if flux-based algebraic/geometric VOF methods inherently ensure numerical stability for the two-phase momentum convection with high density ratios, where do the numerical inconsistencies reported throughout the literature \citep{fuster2018,Zuzio2020,Pal2021,Arrufat2021,Liu2022,Jin2022} come from? 

Although \cref{eq:rho-alpha,eq:rhoindicatorintegral} show the inherent consistency \textcolor{Reviewer22}{between the volume and mass conservation in} VOF methods in the mathematical model, the discrete computation of $\alpha_c^{n+1}$, the approximation of the mass flux $\rho_f F_f$, and the choice of the temporal scheme and the flux limiting scheme, \textcolor{Tobi}{can potentially cause inconsistencies.}

We \textcolor{Reviewer22}{focus first on the consistency between the mass and volume conservation equations on the discrete level}. For a second-order accurate flux-based VOF method, the approximations applied to the temporal and surface integrals for the fluxed phase-specific volumes $V_f^\alpha$ when solving \cref{eq:alphaeqn1} lead to
\begin{equation}
    \alpha_c^{n+1} = \alpha_c^n - \frac{1}{|\Omega_c|} \sum_{f \in F_c} \int_{\tn}^{\tnn}\int_{S_f} \chi \v \cdot \n \, dS \, dt
    = \alpha_c^n - \frac{1}{|\Omega_c|}\sum_{f \in F_c} |V_f^\alpha|_s
    + e_{\alpha_t}(\Delta t^p) + e_{\alpha_h}(h^2),
    \label{eq:alpha-discrete}
\end{equation}
with $e_{\alpha_t}(\Delta t^p)$ and $e_{\alpha_h}(h^2)$ as temporal and spatial volume fraction discretization errors. Note that 
\begin{equation}
    |V_f^\alpha|_s := sgn(F_f)|V_f^\alpha|
\end{equation}
is a signed magnitude of a \emph{phase-specific volume} $V_f^\alpha$ \citep{Maric2020}, whose sign is determined by the volumetric flux $\rho_f F_f$.

The phase-specific volume $|V_f^\alpha|_s$ can be used to approximate the mass flux $\rho_f F_f$ using \cref{eq:rho-alpha} and \cref{eq:alpha-discrete}.
Reordering \cref{eq:alpha-discrete}, results in 
\begin{equation}
     \alpha_c^{n+1} - \alpha_c^n = \frac{1}{|\Omega_c|}\sum_{f \in F_c}  |V_f^\alpha|_s
      + e_{\alpha_t}(\Delta t^p) + e_{\alpha_h}(h^2)
    \label{eq:alpha-diff}
\end{equation}
Inserting \cref{eq:alpha-diff} into \cref{eq:rho-alpha} results in 
\begin{equation}
    \rho_c^{n+1} - \rho_c^{n} = \frac{\rho^- - \rho^+}{|\Omega_c|} \sum_{f \in F_c} |V_f^\alpha|_s + (\rho^- - \rho^+)[e_{\alpha_t}(\Delta t^p) + e_{\alpha_h}(h^2)].
    \label{eq:rhodiffvfalpha}
\end{equation}
Equivalently, integrating \cref{eq:massconserv} over $[t^n, t^{n+1}]$ results in
\begin{equation}
\begin{aligned}
    \rho_c^{n+1} - \rho_c^{n} & = \frac{1}{|\Omega_c|} \sum_{f \in F_c} \int_{t^n}^{t^{n+1}} \rho_f F_f \, dt  \\ 
    & = \frac{1}{|\Omega_c|} \sum_{f \in F_c} |M_f|_s + e_{\rho_t}(\Delta t^s) + e_{\rho_h}(h^2),
    \label{eq:rhodiffmf}
\end{aligned}
\end{equation}
with $e_{\rho_t}(t^s), e_{\rho_h}(h^2)$ as the temporal and spatial discretization errors of the mass conservation equation. The right-hand sides of \cref{eq:rhodiffvfalpha,eq:rhodiffmf} express the mass fluxed through $\partial \Omega_c$ over $[t^n, t^{n+1}]$, i.e. $|M_f|_s$, as the phase-specific volume $|V_f^\alpha|_s$ scaled by the density difference, thus connecting the fluxed mass with the fluxed phase specific volume.

Consistency of mass conservation \textcolor{Reviewer22}{and volume conservation} on the discrete level requires the equivalence of \cref{eq:rhodiffvfalpha,eq:rhodiffmf}. \Cref{eq:rhodiffvfalpha,eq:rhodiffmf} will be exactly the same, only if their errors on the r.h.s. cancel out. Error cancellation is impossible if \textcolor{Reviewer22}{an auxiliary density equation is actually solved, and}  \cref{eq:rhodiffvfalpha,eq:rhodiffmf} are using different numerical schemes \textcolor{Reviewer22}{when integrating} $|V_f^\alpha|_s$ and $|M_f|_s$. In other words, if we use a specific VOF method for $|V_f^\alpha|_s$, we should compute the fluxed mass  from $|V_f^\alpha|_s$. This is hypothetical, of course, since there is no need to actually solve two equations that are equivalent, \textcolor{Reviewer22}{and computing the new cell-centered density $\rho_c^{n+1}$ using  \cref{eq:rhoalpha} from $\alpha_c^{n+1}$ suffices for VOF methods}.

Consistency of the mass \textcolor{Reviewer22}{and volume conservation} can be shown \textcolor{Reviewer22}{exemplary for a simplified first-order geometrical VOF method}, which uses the "Euler"  temporal integration (rectangle quadrature) of $|V_f^\alpha|_s$, i.e.
\begin{equation}
\begin{aligned}
    |V_f^\alpha|_s^{Euler}  & = \int_{\tn}^{\tnn} \int_{S_f} \chi \v \cdot \n \, dS \, dt 
    = \frac{F_f^n}{|S_f|} \int_{\tn}^{\tnn} \int_{S_f} \chi \, dS \, dt  + e_{\alpha_t}(\Delta t^2) + e_{\alpha_h}(h^2) \\
    & = F_f^n \int_{\tn}^{\tnn} \alpha_f(t) \, dt + e_{\alpha_t}(\Delta t^2) + e_{\alpha_h}(h^2) \\
    & = F_f^n \alpha_f^n \Delta t + e_{\alpha_t}(\Delta t^2) + e_{\alpha_h}(h^2),
\end{aligned}
  \label{eq:vfalphaeuler}
\end{equation}
\textcolor{Santiago}{where we define the fraction of the wetted face area $A_f(t)$ as} $\alpha_f := A_f(t) / S_f$, and $N_t e_{\alpha_t}(\Delta t^2)=\frac{T}{\Delta t}e_{\alpha_t}(\Delta t^2) \propto \Delta t$ results in a first-order temporal quadrature error over the simulated physical time $T$. We emphasize that \cref{eq:vfalphaeuler} is a simplified scheme used here only to discuss the consistency \textcolor{Reviewer22}{of mass and volume conservation} on the discrete level, it is not a practically usable scheme for advecting $\alpha_c$, because it is significantly less accurate than modern geometrical schemes \citep{Maric2020}. Multiplying  $|V_f^\alpha|_s^{Euler}$ from \cref{eq:vfalphaeuler} with $(\rho^- - \rho^+)$ to obtain $|M_f|_s$ ensures the consistency of \cref{eq:rhodiffvfalpha,eq:rhodiffmf}, if \cref{eq:rhodiffmf} is integrated using the Euler scheme, namely
\textcolor{Reviewer22}{
\begin{equation}
\begin{aligned}
    \frac{(\rho^- - \rho^+)|V_f^\alpha|_s^{Euler}}{\Delta t} & =  (\rho^- - \rho^+)F_f^n \alpha_f^n + (\rho^- - \rho^+)(e_{\alpha_t}(\Delta t) + \frac{1}{\Delta t}e_{\alpha_h}(h^2)).
\label{eq:massfluxeulerscale}
\end{aligned}
\end{equation}}
The temporal accuracy lost by dividing by $\Delta t$ is recovered when the mass flux is integrated over (multiplied with) $\Delta t$, from the temporal term in the momentum conservation equation.

\textcolor{Reviewer22}{We move on to the approximation of the mass flux in the discretized convective term of the momentum equation.}

\textcolor{Reviewer22}{The unstructured Finite Volume Method linearizes the volumetric flux when discretizing the convective term in \cref{eq:momentum-transport}, introducing outer iterations $"o"$ to the equation solution algorithm. We further use an Euler implicit temporal discretization, for example, which leads to} 
\begin{equation}
     \frac{\Delta t}{|\Omega_c|} 
     \left(
     \int_{\Omega_c} \nabla \cdot (\rho \mathbf{v} \otimes \mathbf{v}) \, dV 
    \right)_{t^{n+1}}
    \approx \frac{\Delta t}{|\Omega_c|}\sum_{f \in F_c} \rho_f^o F_f^o \mathbf{v_f}^{o+1},
    \label{eq:momconveuler}
\end{equation}
with the factor $\frac{\Delta t}{|\Omega_c|}$ resulting from the finite-difference approximation of the cell-centered average of the temporal derivative term, i.e. $(\partial_t \rho \mathbf{v})_c$, recovering the first-order temporal accuracy in \cref{eq:massfluxeulerscale} for the implicit Euler temporal discretization scheme.
Note that \cref{eq:momentum-transport} is solved (discretized) iteratively in a segregated solution algorithm (e.g., \citep{Tolle2020,Liu2023}), so the linearized mass flux is denoted with the outer iteration index $1 \le o \le N_o$.
  Once the solution algorithm converges, $
 \rho_f^o F_f^o = \rho_f^{n+1} F_f^{n+1}$.

\textcolor{Reviewer22}{The mass flux $\rho_f^o F_f^o$ is uniquely determined by $\chi(\x,t)$ and $\mathbf{v}(\x,t)$ \citep{Liu2023}. Namely, at any time $t$, omitted here for brevity, from \cref{eq:rhoindicator}, we get
\begin{equation}
    \int_{S_f} \rho \mathbf{v} \cdot \mathbf{n} \, dS =: (\rho_f F_f)^{\rho} 
    \approx \frac{F_f}{|S_f|} \int_{S_f} [(\rho^- - \rho^+)\chi(\mathbf{x}) + \rho^+] \, dS 
    = (\rho^- - \rho^+)\alpha_fF_f + \rho^+ F_f
    \label{eq:massfluxrho}
\end{equation}
with superscript $\rho$ in $(\rho_f F_f)^{\rho}$ denoting the mass flux estimated directly from the model for $\rho$ (\ref{eq:rhoindicator}) and the volumetric flux given by \cref{eq:faceVolumetricFlux}.}

\textcolor{Reviewer22}{The contribution $\rho^+F_f$ to \cref{eq:massfluxrho} and its role in the approximation of the mass flux in the momentum equation must be carefully addressed.}

\textcolor{Reviewer22}{The term $\rho^+F_f$ is a zero-sum contribution to mass conservation (\ref{eq:massconserv}), and therefore, volume conservation (\ref{eq:rho-alpha}). Namely, inserting 
\begin{equation}
    \rho \mathbf{v} = (\rho^- - \rho^+) \chi \mathbf{v} + \rho^+ \mathbf{v}
\end{equation}
from \cref{eq:rhoindicator} multiplied by $\mathbf{v}$, into \cref{eq:massconserv}, and integrating over $[\tn, \tnn]$, and applying divergence-free velocity condition (\ref{eq:volume-transport}) results in
\begin{equation}
\begin{aligned}
 \rho_c^{n+1} & = \rho_c^n - \frac{1}{|\Omega_c|}\int_{t^n}^{t^{n+1}} \left[\int_{\partial \Omega_c} (\rho^- - \rho^+) \chi \mathbf{v} \cdot \mathbf{n} \, dS \, dt +  \rho^+ \int_{\partial \Omega_c} \mathbf{v} \cdot \mathbf{n} \, dS \right] \, dt \\
 & = \rho_c^n - \frac{1}{|\Omega_c|}\int_{t^n}^{t^{n+1}} \left[\int_{\partial \Omega_c} (\rho^- - \rho^+) \chi \mathbf{v} \cdot \mathbf{n} \, dS +  \rho^+ \int_{\Omega_c} (\nabla \cdot \mathbf{v}) dV \right] \, dt \\
 & = \rho_c^n - \frac{1}{|\Omega_c|}\int_{t^n}^{t^{n+1}} \left[\int_{\partial \Omega_c} (\rho^- - \rho^+) \chi \mathbf{v} \cdot \mathbf{n} \, dS \right] \, dt \\
 & = \rho_c^n - 
 \frac{(\rho^- - \rho^+)}{|\Omega_c|}\sum_{f\in F_c} \int_{t^n}^{t^{n+1}}  \chi \mathbf{v} \cdot \mathbf{n} \, dS \\
 & = \rho_c^n - 
 \frac{(\rho^- - \rho^+)}{|\Omega_c|}\sum_{f\in F_c} |V_f^\alpha|_s.
 \end{aligned} 
 \label{eq:rhofluxconsistent}
\end{equation}
The main point of \cref{eq:rhofluxconsistent} (and equivalently \cref{eq:rhodiffvfalpha,eq:rhodiffmf}) is that the additional term $\rho^+\mathbf{v}$, discrete $\rho^+F_f$ in \cref{eq:massfluxrho}, does not impact consistency of mass and volume conservation given by  \cref{eq:rhodiffmf,eq:rhodiffvfalpha}. More importantly, \cref{eq:rhofluxconsistent} shows this is true irrespective of the VOF method used to approximate $|V_f^\alpha|_s$, in our case, the plicRDF-isoAdvector method \citep{Scheufler2019}, or, in the example from \cref{eq:vfalphaeuler}, $|V_f^\alpha|_s^{Euler}$ .}

\textcolor{Reviewer22}{However, while $\rho^+F_f$ from \cref{eq:massfluxrho} is a zero contribution in mass and volume conservation \cref{eq:rhodiffmf,eq:rhodiffvfalpha,eq:rhofluxconsistent}, it is a non-zero contribution to the discretized convective term (\ref{eq:momconveuler}), i.e.  
\begin{equation}
     \frac{\Delta t}{|\Omega_c|}\sum_{f \in F_c} \rho_f^o F_f^o \mathbf{v_f}^{n+1} = 
    \frac{\Delta t}{|\Omega_c|}\left(\sum_{f \in F_c} (\rho^- - \rho^+) \alpha_f^o F_f^o \mathbf{v}_f^{n+1}  + \rho^+\sum_{f \in F_c} F_f^o \mathbf{v}_f^{n+1} \right),
    \label{eq:momentummassflux}
\end{equation}
because it models the single-phase momentum convection in the bulk of each phase $\Omega^\pm(t)$.}

\textcolor{Reviewer22}{The final choice of approximating $(\rho^- - \rho^+)\alpha_f^oF_f^o$ is constricted by accuracy and the temporal scheme chosen for Navier-Stokes equations. Concretely, if the Euler implicit scheme is used to discretize Navier-Stokes equations, scaling the fluxed phase-specific volume $|V_f^\alpha|_s$ in \cref{eq:massfluxeulerscale} to approximate $(\rho^- - \rho^+)\alpha_f^o F_f^o$ will be consistent, with an Euler-implicit discretization of an equivalent mass conservation equation.}
The consistency of the mass conservation and volume fraction advection is crucial in the momentum conservation equation, as a deviation from volume conservation increases the the source-term $\sum_{f \in F_c} \left(\frac{1}{a_c}\right)_f\left[\mathbf{H}(F^o, \rho^o, \mathbf{v}^{i-1})\right]_f\cdot \mathbf{S}_f$ in the pressure Poisson equation, with $i$ denoting the inner iterations of the pressure equation (cf. \citep{Tolle2020,Liu2023} for details). In other words, errors in the mass flux artificially \textcolor{Tobi}{accelerate or decelerate} the fluid as the pressure equation tries to ensure volume (mass) conservation. Examining the errors in  \cref{eq:massfluxeulerscale}, we see that an error in the mass flux scaled from $|V_f^\alpha|_s$ will be multiplied by $(\rho^- - \rho^+)$: small errors in the volumetric flux (fluxed phase-specific volume) are scaled with the density difference, and this leads to large errors in the velocity field and catastrophic failures for large density differences. \textcolor{Reviewer22}{Therefore, a more accurate $|V_f^\alpha|_s$, e.g., a second-order geometrical $|V_f^\alpha|^{isoAdvector}_s$, results in a more accurately ensured consistency between mass and volume conservation (\ref{eq:rhofluxconsistent}).} 

\textcolor{Reviewer22}{However, there is a downside in approximating the mass flux $\rho_f^oF_f^o$ by scaling a  highly accurate geometrical phase-specific volume, i.e., $|V_f^\alpha|^{isoAdvector}_s$.} 
\textcolor{Reviewer22}{Consider} an alternative temporal integration scheme for \cref{eq:momentum-transport}, say, Crank-Nicolson scheme, resulting in contributions from the convective term in the form of 
\begin{equation}
    0.5\frac{\Delta t}{|\Omega_c|}\left(\sum_{f \in F_c} \rho_f^n F_f^n \mathbf{v_f}^{n} + 
    \sum_{f \in F_c} \rho_f^o F_f^o \mathbf{v_f}^{n+1}\right)
    \label{eq:cranknicolsonmomentum}
\end{equation}
\textcolor{Reviewer22}{In this case, it is impossible to ensure consistency by scaling a single value of $|V_f^\alpha|_s$ to obtain two values $\rho_f^nF_f^n$ and $\rho_f^oF_f^o$. The phase-specific volume is integrated in time at $t^n$ over the time step $\Delta t$, making it a constant value over the time step, and we obtain $(\rho_f F_f)^{Euler}$ from \cref{eq:massfluxeulerscale} as a single average quantity over $\Delta t$, so we do not have two mass fluxes for \cref{eq:cranknicolsonmomentum}, making \cref{eq:massfluxeulerscale} inconsistent with the Crank-Nicolson scheme \cref{eq:cranknicolsonmomentum} already in the first time step.}

The main takeway point is that computing the mass flux by scaling the fluxed phase-specific volume over $\Delta t$ limits the temporal discretization to schemes that utilize a single mass flux term within $\Delta t$ - e.g., Euler explicit or implicit, or 2nd-order backward implicit schemes. Any other temporal discretization scheme (e.g., Runge-Kutta) that utilises a linear combination of different mass fluxes within $\Delta t$ are inconsistent with the mass flux scaling given by \cref{eq:massfluxeulerscale}.

Additionally, any modification of the scaled mass flux causes inconsistencies. Concretely, \emph{limiting the mass flux} in the discretized \cref{eq:momentum-transport} causes a hidden inconsistency between mass and volume fraction conservation. The inconsistency is hidden because \textcolor{Reviewer22}{limiting the mass flux in \cref{eq:momconveuler} creates a mass flux that does not correspond to the mass and volume conservative flux used to update $\rho_c^o$ from $\alpha_c^o$ in \cref{eq:rho-alpha}, i.e., the consistent approximation of $(\rho^- - \rho^+) \alpha_f^o F_f^o$ as a part of the total mass flux in \cref{eq:rhodiffvfalpha}}. 

\begin{figure}[htb]
    \centering
     \includegraphics{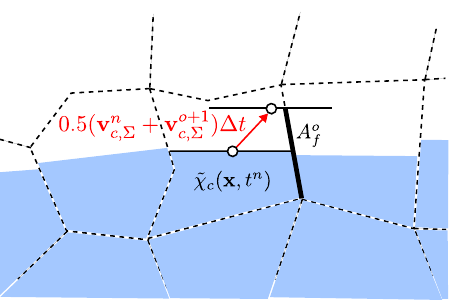}
         \caption{Geometric upwinding for $|V_f^\alpha|^{isoAdvector}_s$ in plicRDF-isoAdvector \citep{Scheufler2019}.}
         \label{fig:isoadvector}
\end{figure}

We now turn our attention from the simplified VOF scheme \cref{eq:vfalphaeuler} to a more complex, geometric isoAdvector scheme \citep{Roenby2016}, that computes
\begin{equation}
\begin{aligned}
    |V_f^\alpha|_s^{isoAdvector}  
    & = \frac{0.5(F_f^n + F_f^o)}{|\mathbf{S}_f|} \int_{\tn}^{t^o} \int_{S_f} \chi(\x,t) \, dS \, dt + e_{\alpha_t}(\Delta t^3) + e_{\alpha_h}(h^2) \\
    & = \frac{0.5(F_f^n + F_f^o)}{|\mathbf{S}_f|} \int_{\tn}^{t^o} A_f(t) \, dt + e_{\alpha_t}(\Delta t^3) + e_{\alpha_h}(h^2) \\
    & = 0.5(F_f^n + F_f^o) \int_{\tn}^{t^o} \alpha_f(t) \, dt + e_{\alpha_t}(\Delta t^3) + e_{\alpha_h}(h^2)
\end{aligned}
\label{eq:isoadveuler}
\end{equation}
with $\int_{\tn}^{t^o} \alpha_f(t) dt$ computed geometrically by \textcolor{Reviewer22}{kinematically tracking} the piecewise-linear interface from the upwind cell, using the \textcolor{Reviewer22}{interpolated interface velocity $\v_{\Sigma,c}$ associated with the PLIC centroid and the second-order accurate displacement approximation $0.5(\v_{\Sigma,c}^n + \v_{\Sigma,c}^{o}) \Delta t$, schematically shown in \cref{fig:isoadvector}}. \textcolor{Reviewer22}{If the implicit Euler temporal scheme is used for Navier-Stokes equations,} the Euler temporal integration of the displacement $\v_c \Delta t$ for the evaluation of  $\int_{\tn}^{t^o} \alpha_f(t) dt$ on the  r.h.s. of \cref{eq:isoadveuler} makes it possible to consistently compute \textcolor{Reviewer22}{the $(\rho^- - \rho^+) \alpha_f F_f$ contribution to the total mass flux $\rho_f F_f$} by scaling the phase-specific fluxed volume 
\begin{equation}
    \textcolor{Reviewer22}{((\rho^- - \rho^+) F_f^o \alpha_f^o)^{isoAdvector}} = \dfrac{(\rho^- - \rho^+)|V_f^\alpha|^{isoAdvector}_s}{\Delta t},
    \label{eq:massfluxisoadvector}
\end{equation}
equivalently to \cref{eq:massfluxeulerscale}. Even though $\int_{\tn}^{t^o} \alpha_f(t) dt$ is evaluated using exact geometric integration (cf. \citep{Scheufler2019} for details), the $A_f^{o}$, used as the end-point of the geometric integration is approximated by first-order \textcolor{Reviewer22}{accurate displacement when implicit Euler scheme is used for the Navier-Stokes equations, making} the average displacement velocity of the fluid interface constant over $\Delta t$, and, consequently, \textcolor{Reviewer22}{the mass flux contribution from \cref{eq:massfluxisoadvector} consistent w.r.t. \cref{eq:rho-alpha} and \cref{eq:rhodiffvfalpha}}. 

The same consistency would be ensured \textcolor{Reviewer22}{for the implicit Euler method}, if $|V_f^\alpha|_s$ would be constructed using a flux-based VOF method that maps $S_f$ using the reverse flow-map (e.g., \citep{Maric2018}), \emph{if the approximation of the flow-map is first-order accurate}. In other words, if $|V_f^\alpha|_s$ is constructed geometrically using displacements given by $\v(\x,t^n)\Delta_t$, regardless of the geometrical approximation, $|V_f^\alpha|_s$ is constructed using displacements constant over $\Delta t$, so dividing the volume $V_f^\alpha$ with $\Delta t$ results in a volumetric flux \textcolor{Reviewer22}{consistent with the Euler method, and} constant over $\Delta t$. \textcolor{Reviewer22}{In following sections, we describe in detail the consistency and equivalency of the scaled $|V_f^\alpha|_s$ and the solution of an auxiliary density equation, in the context of the solution algorithm for one-field Navier-Stokes equations, for the implicit Euler temporal discretization, and confirm our findings with extensive verification and validation.}


\subsection{Collocated segregated solution algorithm with the auxiliary density equation} \label{sec:introduction_solver}

We base our consistent solution algorithms for the unstructured Volume-of-Fluid methods on the plicRDF-isoAdvector method \citep{Scheufler2019} and implement the auxiliary density equation solution into \textcolor{Davide}{ the segregated algorithm (i.e. solver) "interIsoRhoFoam", which is summarized by \cref{alg:interIsoRhoFoam}}. 

In \citep{Liu2023}, we show on the level of the mathematical model why solving an auxiliary mass conservation equation plays a key role in reducing numerical inconsistency caused by high density ratios.

We solve the mass conservation equation in the outer loop of the segregated algorithm \cref{alg:interIsoRhoFoam} in the following discrete form
\begin{equation}
    \rho^{o+1}_c = \rho^o_c + \frac{\Delta t}{|V_{\Cell{c}}|}\sum_f \rho^{o}_f F^o_f.
    \label{eq:rho_update_Euler}
\end{equation}
\textcolor{Davide}{\Cref{eq:rho_update_Euler} is the auxiliary mass conservation (density equation). It is solved after updating the volume fraction in the outer loop to $\alpha_c^{o}$ by utilizing \cref{eq:alpha-discrete} with any flux-based VOF method.} Interface reconstruction computes $\tilde{\chi}_c(\mathbf{x},t^{o})$ in every finite volume $\Omega_c$ intersected by the fluid interface. The piecewise-linear interface approximation $\tilde{\chi}_c(\mathbf{x},t^{o})$, together with \cref{eq:rhoindicator}, provides $\rho_f^{o}$ for \cref{eq:rho_update_Euler}. Since $F_f^o$ is also available, \cref{eq:rho_update_Euler} can be explicitly evaluated. To compute the mass flux, we utilize the consistency relationship between the phase indicator and the face-centered density $\rho_f$, derived in \citep{Liu2023} for the Level Set / Front Tracking method. It can be applied on any two-phase flow simulation method that is using a phase indicator $\chi$. The surface integral of mass flux at time step $o$ is expressed as
\begin{equation}
    \int_{\partial \Omega_c} \rho^{o} \v^o \cdot \n dS = 
        \int_{\partial \Omega_c} [\rho^- \chi^{o} + \rho^+(1 - \chi^{o})] \v^o \cdot \n \, dS,
\end{equation}
and discretized as
\begin{equation}
    \begin{aligned}
        \sum_{f \in F_c} \rho_f^{o} F_f^o
        & = 
        \sum_{f \in F_c} 
            \left[
                \rho^- \alpha_f^{o} 
                + \rho^+ (1 - \alpha_f^{o})
            \right] F_f^o, 
    \end{aligned}
\label{eq:rhodiscreteflux}
\end{equation}
where 
\begin{equation}
    \alpha_f^{o} = \frac{1}{|S_f|}\int_{S_f} \chi(\x, t^o) \, dS 
    = \frac{|A_f(t^o)|}{|S_f|}.
\label{eq:update_alpha_f}
\end{equation} 
is used to define the face-centered density
\begin{equation}
    \rho_f^{o} = \rho^-\alpha_f^{o} + \rho^+(1-\alpha_f^{o})
\label{eq:update_rho_f}
\end{equation}
in the mass flux. The momentum equation is discretized as
\begin{equation}
    \rho^{o}_c \v^{o} - \rho^n_c \v^n + \frac{\Delta t}{|\Cell{c}|}\sum_f \rho^{o}_f F^o_f \v_f^{o} = \frac{\Delta t}{|\Cell{c}|}  \mathbf{M}.
\label{eq:rhov_update_Euler}
\end{equation}
The source term $\mathbf{M}$ is a shorthand term that contains all the remaining contributions from the discretizaiton, used here for brevity. \textcolor{Davide}{The volume fractions $\alpha_c^o$ are already solved for using \cref{eq:alpha-discrete} at the beginning of the "$o$" outer iteration}, and used to approximate the phase indicator $\tilde{\chi}_c(\x, t^o)$ for the mass flux $\rho_f^oF_f^o$ using \cref{eq:rhodiscreteflux}, used in the same way in the auxiliary density equation \cref{eq:rho_update_Euler}, solved for $\rho_c^o$, as $\rho_f^oF_f^o$ in \cref{eq:rhov_update_Euler}, \emph{without using flux limiters}. \textcolor{Davide}{The values computed from the last outer loop are regarded as the new values at the time step. At last, we need to restore the density field with respect to the volume fraction to maintain consistency between them.} \textcolor{Reviewer1}{\Cref{alg:interIsoRhoFoam} uses a combination of SIMPLE and PISO algorithms in OpenFOAM with a residual-based control to terminate outer iterations, which tests if a maximal number of iterations has been reached, the final domain-maximal residuals of the pressure equation $r_p$ are below absolute tolerance, or the ratio of the final and initial domain-maximal residuals $\frac{r_p}{r_p^i}$ is smaller than a user-prescribed relative tolerance.}
\begin{center}
\begin{algorithm}[H]
    \centering
    \caption{The solution algorithm interIsoRhoFoam.}
    \label{alg:interIsoRhoFoam}
    {\small
    \centering
    \begin{algorithmic}[1]
        \While{ $t \le t_{end}$ }
            \State $t^{n+1} = t^{n} + \Delta t$ 
            \State $o = 0$
            \While{$o < N_{outer}$ or 
                   $max(r_p) > t_p $ or 
                   $\frac{r_p}{r_p^i} > t^r_p$}
                \State $o = o + 1$
                \State Solve volume fraction for $\alpha_c^o$ \Comment \Cref{eq:alpha-discrete} 
                \State Reconstruct the phase indicator $\tilde{\chi}_c(\x, t^o)$ 
                \State Compute \textcolor{Santiago}{$\rho_f^o$ from $\tilde{\chi}_f(\x, t^o)$} \Comment \Cref{eq:update_rho_f}
                \State Compute the mass flux \textcolor{Santiago}{$\rho_f^o F_f^o:=\rho_f^o(\v_f^{o-1}\cdot\Sf)$} 
                \State Solve the density for $\rho_c^{o}$ with $\rho_f^o F_f^o$ \Comment \Cref{eq:rho_update_Euler}
                \State Discretize momentum equation (\Cref{eq:momentum-transport}) with $\rho_c^{o}$ and $\rho_f^o F_f^o$.
                \For{$i = 0$; $i < N_{inner}$; ++$i$}
                    \State Solve the pressure equation for $p_c^i$. \Comment Cf. \citep{Tolle2020,Liu2023}.
                    \State Compute $F_f^i$ and $\v_c^i$ from $p_c^i$. \Comment Cf. \citep{Tolle2020,Liu2023}.
                \EndFor
            \EndWhile
            \State Restore $\rho_c^{n+1}$ consistent with $\alpha_c^{n+1}$, i.e.\ $\rho_c^{n+1}=(\rho^- - \rho^+) \alpha_c^{n+1} + \rho^+$ \Comment \Cref{eq:rhoindicator}.
        \EndWhile
    \end{algorithmic}
    }
\end{algorithm}
\end{center}

\subsection{Phase-specific face area calculation}
\label{sec:submerged_areaFraction}

\begin{figure}[!htb]
    \begin{subfigure}[t]{.49\textwidth}
    \centering
    \def\svgwidth{.96\linewidth}
\begingroup%
  \makeatletter%
  \providecommand\color[2][]{%
    \errmessage{(Inkscape) Color is used for the text in Inkscape, but the package 'color.sty' is not loaded}%
    \renewcommand\color[2][]{}%
  }%
  \providecommand\transparent[1]{%
    \errmessage{(Inkscape) Transparency is used (non-zero) for the text in Inkscape, but the package 'transparent.sty' is not loaded}%
    \renewcommand\transparent[1]{}%
  }%
  \providecommand\rotatebox[2]{#2}%
  \newcommand*\fsize{\dimexpr\f@size pt\relax}%
  \newcommand*\lineheight[1]{\fontsize{\fsize}{#1\fsize}\selectfont}%
  \ifx\svgwidth\undefined%
    \setlength{\unitlength}{259.56770304bp}%
    \ifx\svgscale\undefined%
      \relax%
    \else%
      \setlength{\unitlength}{\unitlength * \real{\svgscale}}%
    \fi%
  \else%
    \setlength{\unitlength}{\svgwidth}%
  \fi%
  \global\let\svgwidth\undefined%
  \global\let\svgscale\undefined%
  \makeatother%
  \begin{picture}(1,0.84521113)%
    \lineheight{1}%
    \setlength\tabcolsep{0pt}%
    \put(0,0){\includegraphics[width=\unitlength,page=1]{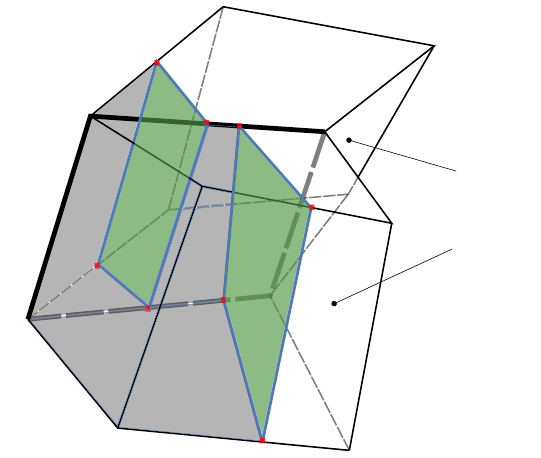}}%
    \put(0.85589321,0.37466019){\color[rgb]{0,0,0}\makebox(0,0)[lt]{\lineheight{1.25}\smash{\begin{tabular}[t]{l}$\tilde{\Omega}_D^+(t^{o})$\end{tabular}}}}%
    \put(0,0){\includegraphics[width=\unitlength,page=2]{cuttedCell.pdf}}%
    \put(0.00236309,0.03815911){\color[rgb]{0,0,0}\makebox(0,0)[lt]{\lineheight{1.25}\smash{\begin{tabular}[t]{l}$\tilde{\Omega}_D^-(t^{o})$\end{tabular}}}}%
    \put(0,0){\includegraphics[width=\unitlength,page=3]{cuttedCell.pdf}}%
    \put(0.50075368,0.73304571){\color[rgb]{0,0,0}\makebox(0,0)[lt]{\lineheight{1.25}\smash{\begin{tabular}[t]{l}$\tilde{\Sigma}^{n+1}(t^{o})$\end{tabular}}}}%
    \put(-0.05322902,0.60588047){\color[rgb]{0,0,0}\makebox(0,0)[lt]{\lineheight{1.25}\smash{\begin{tabular}[t]{l}$\tilde{\Omega}_U^-(t^{o})$\end{tabular}}}}%
    \put(0.86128521,0.51936801){\color[rgb]{0,0,0}\makebox(0,0)[lt]{\lineheight{1.25}\smash{\begin{tabular}[t]{l}$\tilde{\Omega}_U^+(t^{o})$\end{tabular}}}}%
  \end{picture}%
\endgroup%

    \caption{Upwind and downwind cells intersected with piecewise-linear VOF interfaces.}
    \label{fig:cuttedCell}
   \end{subfigure} \hfill
    \begin{subfigure}[t]{.49\textwidth}
    \centering
    \def\svgwidth{.96\linewidth}
\begingroup%
  \makeatletter%
  \providecommand\color[2][]{%
    \errmessage{(Inkscape) Color is used for the text in Inkscape, but the package 'color.sty' is not loaded}%
    \renewcommand\color[2][]{}%
  }%
  \providecommand\transparent[1]{%
    \errmessage{(Inkscape) Transparency is used (non-zero) for the text in Inkscape, but the package 'transparent.sty' is not loaded}%
    \renewcommand\transparent[1]{}%
  }%
  \providecommand\rotatebox[2]{#2}%
  \newcommand*\fsize{\dimexpr\f@size pt\relax}%
  \newcommand*\lineheight[1]{\fontsize{\fsize}{#1\fsize}\selectfont}%
  \ifx\svgwidth\undefined%
    \setlength{\unitlength}{220.96899093bp}%
    \ifx\svgscale\undefined%
      \relax%
    \else%
      \setlength{\unitlength}{\unitlength * \real{\svgscale}}%
    \fi%
  \else%
    \setlength{\unitlength}{\svgwidth}%
  \fi%
  \global\let\svgwidth\undefined%
  \global\let\svgscale\undefined%
  \makeatother%
  \begin{picture}(1,0.69308983)%
    \lineheight{1}%
    \setlength\tabcolsep{0pt}%
    \put(0,0){\includegraphics[width=\unitlength,page=1]{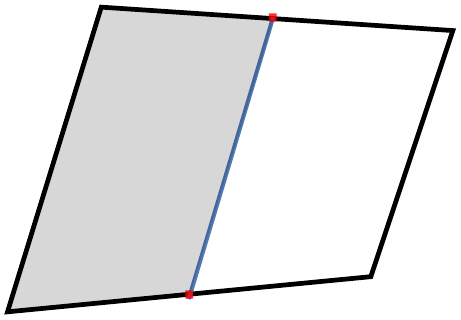}}%
    \put(0.24851666,0.36053205){\color[rgb]{0,0,0}\makebox(0,0)[lt]{\lineheight{1.25}\smash{\begin{tabular}[t]{l}$(\alpha_f)_U(t^{o}) $\end{tabular}}}}%
    \put(0,0){\includegraphics[width=\unitlength,page=2]{cuttedArea.pdf}}%
    \put(0.56607871,0.36017893){\color[rgb]{0,0,0}\makebox(0,0)[lt]{\lineheight{1.25}\smash{\begin{tabular}[t]{l}$(\alpha_f)_D(t^{o}) $\end{tabular}}}}%
  \end{picture}%
\endgroup%

    \caption{A face $S_f$ intersected with downwind (D) and upwind (U) piecewise-linear VOF interfaces.}
    \label{fig:cuttedArea}
   \end{subfigure} 
   \caption{Interface reconstructed as $\tilde{\Sigma}^o_{U,D}$ in upwind ($U$) and downwind ($D$) cells. Green polygons are interface polygons $\tilde{\Sigma}^o_{U,D} \cap \Omega_{U,D}$. Blue lines are intersection line segments $\tilde{\Sigma}^o_{U,D} \cap S_f$. Red points are intersection points $\tilde{\Sigma}^o_{U,D} \cap \partial S_f$. }
   \label{fig:reconstructedInterface}
\end{figure}

The integral \Cref{eq:update_alpha_f} leaves room for alternative discretizations, and therefore requires attention.
Flux-based geometrical Volume-of-Fluid methods advect volume fractions using geometrical upwind advection schemes, transporting $\tilde{\Omega}^-(t) \approx \Omega^-(t)$ geometrically from the upwind cell, denoted with $U$ in \Cref{fig:cuttedCell}, to the downwind cell, denoted with $D$ in \Cref{fig:cuttedCell}.
Flux-based geometrical VOF methods therefore already provide the intersection points and the intersection line segments denoted as red dots and the blue line segments in \Cref{fig:reconstructedInterface}, are available, simplifying the area fraction calculation in \Cref{eq:update_alpha_f}. Since the geometrical VOF methods approximate the phase indicator as piece-wise continuous, with a jump discontinuity not only across the fluid interface $\Sigma(t)$ but also across the finite volume boundary $\partial \Omega_c$, the line segments forming the intersection of the piece-wise continuous interface and a face $S_f$ it intersects (see \Cref{fig:cuttedCell}), do not overlap. \Cref{fig:cuttedArea} shows schematically the two intersection line segments on a cell face. The intersection line segment of the interface and the upwind cell are selected as the submerged area border to calculate the area fraction. The evaluation of $\alpha_f$ and, from it, $\rho_f, \mu_f$, is summarized by \Cref{alg:calc_alphaf}. 
 
\begin{center}
\begin{algorithm}[H]
    \centering
    \caption{Sub-algorithm for calculating $\alpha_f, \rho_f, \mu_f$.}
\label{alg:calc_alphaf}
    {\small
    \centering
    \begin{algorithmic}[1]
        \State Initialize $\rho_f$, $\alpha_f$ with upwind cell-centered values $\rho_U$, $\alpha_U$.
        \State $\alpha_f = \frac{1}{|S_f|}\int_{S_f}\tilde{\chi}_U(\x) \, dS = \frac{|\tilde{\Omega}^-_U \cap S_f|}{|S_f|}$ \Comment \Cref{eq:update_alpha_f} with upwind $\chi$.
        \State $\rho_f=(\rho^- - \rho^+) \alpha_f + \rho^+$. \Comment \Cref{eq:rhoindicator}.
        \State $\mu_f=(\rho^-\nu^- -  \rho^+\nu^+)\alpha_f + \rho^+\nu^+$. \Comment \Cref{eq:nuindicator}.
    \end{algorithmic}
    }
\end{algorithm}
\end{center}
It is important to note that $\mu_f$ is geometrically evaluated from a geometrical $\alpha_f$, and not interpolated, as we found that interpolation leads to large errors in simulations with large differences in dynamic viscosity. 

\subsection{Consistency of VOF methods for
Two-Phase Flows with High Density Ratios}
\label{subsec:consistency}

Here we summarize our findings that lead to an equation discretization with flux-based geometrical VOF methods that remains consistent for very high density ratios:
\begin{itemize}
    \item Computing the mass flux $\rho_fF_f$ by scaling the fluxed phase-specific volume $|V_f^\alpha|_s$ with $(\rho^- - \rho^+)/\Delta t$ approximates only one constant average mass flux over $\Delta t$. 
    \item The constant mass flux $\rho_f F_f$ scaled from $|V_f^\alpha|_s$ with $\Delta t$ \textcolor{Davide}{, as an average value over $\Delta t$, }can only be consistently used in first-order schemes, i.e. use a single mass flux value over $\Delta t$. 
    \item Computing the mass flux $\rho_f F_f$ from the density model \Cref{eq:rhoindicator} disconnects mass conservation from volume fraction advection, which uses geometrical integration, i.e. a geometrically integrated phase-specific volumetric flux. This requires a solution of an additional (auxiliary) density equation (\citep{Ghods2013,Liu2023}) for cell-centered density, discarded at the end of the time step.
    \item Upwinding geometric VOF methods that use the Euler temporal scheme to approximate point displacements are equivalent to using $\rho_f F_f$ from \Cref{eq:massfluxrho} and solving the density equation for $\rho_c^o$, used further in $p-\v$ coupling to obtain a divergence-free cell-centered velocity. In other words, combining Euler temporal integration scheme with upwind scheme for the momentum equation guarantees numerical consistency for any flux-based VOF method, which uses temporally first-order accurate displacements in its geomterical integration of the fluxed phase-specific volume.
\end{itemize}
These points we verify and validate in the following section.

\section{Verification and validation}
\label{sec:validations-results}

Data archives of the implementation of the interIsoRhoFoam algorithm, input data, post-processing software and secondary data are publicly available \citep{rhoVoFCodes2023,rhoVoFDatas2023}.  The method is actively developed in a publicly available git repository \citep{rhoVoFRepository}.

\subsection{Time step size}
The time step is limited by the CFL condition in the explicit plicRDF-isoAdvector method \citep{Scheufler2019}, 
 \begin{equation}
  \Delta t_{CFL} \textcolor{Reviewer2}{=} \dfrac{CFL\ h}{|\v|},
 \end{equation}
where $h$ is the discretizaiton length, and we use $CFL=0.2$ from \citep{Scheufler2019}. Another restriction for the time step size considers the propagation of capillary waves on fluid interfaces,  
\begin{equation}
   \Delta t_{cw} \textcolor{Reviewer2}{=} \sqrt{\dfrac{(\rho^+ +\rho^-)h^3}{2\pi\sigma}}.
\end{equation}
This time step constraint was introduced first by \citet{Brackbill1992}, and revised by \citet{Denner2015}. We restrict the time step using the relation from \citet{Tolle2020}, namely
\begin{equation}
    \Delta t = \textcolor{Reviewer21}{min}\,(k_1\Delta t_{cw},\ k_2\Delta t_{CFL})
    \label{eq:timeStepSize}
\end{equation}
where $k_1$ and $k_2$ are scaling factors. We set $k_1=1$, $k_2=0.5$ as the default value.

\subsection{Translating droplet in ambient flow}

A canonical test case, originally introduced by \citet{Bussmann2002} in 2D, involves a moving droplet in quiescent ambient flow. \citet{Zuzio2020} extended the 2D case to 3D using a density ratio of $10^6$. Following the setup in \citep{Zuzio2020}, the droplet of radius $R=0.15$ has the initial velocity of $(0,0,10)$. 
\textcolor{Reviewer2}{To smooth the velocity field and avoid the perturbation caused by sudden acceleration of still ambient flow, the initial constant velocity is assigned not only for the cells of the droplet and the interface layer but also the interface cell layer adjecent to interface cells.}
The droplet with initial center location $(0.5\ 0.5\ 0.5)$ translates to a distance of $L=1$ and the simulation time is then  $t_{end}=0.1$. The computational domain has dimensions $L_z=L_x=L_y=1$. The periodic boundary condition is applied to all boundary patches. During testing with the periodic boundary condition, we found that the plicRDF-isoAdvector method \citep{Scheufler2019} implemented in \citep{Scheufler2021twophaseflow} has an inconsistency at the periodic boundary, which we fixed as described in \cref{Appdix:cyclic_BC}. The mesh setup from \citep{Zuzio2020} is also followed: the mesh resolution is in a range of  $N\in(32,\ 48,\ 64)$ per unit side-length of the computational domain, resulting in $\approx(10,\ 15,\ 20)$ mesh cells per droplet diameter. Surface tension force is neglected in this case, so only $\Delta t_{CFL}$ from \cref{eq:timeStepSize} is taken into account.
 The viscosity and gravitational forces are neglected to highlight the numerically consistent behavior of the mass and momentum convection. We adopted three error norms to evaluate the results quantitatively, for mass, momentum, and sphericity:
 \begin{align}
     E_{mass} &= \frac{M(t)-M(0)}{M(0)} =\frac{\sum_k m_k(t)-\sum_k m_k(0)}{\sum_k m_k(0)}= \frac{\sum_k\rho_k(t)V_k}{\sum_k\rho_k(0)V_k}, \label{eq:sumAlpha} \\
     E_{mom} &= \frac{|\sum_k m_k(t) \v_k(t)| - |\sum_k m_k(0) \v_k(0)|}{|\sum_k m_k(0) \v_k(0)|}, \label{eq:sumMOM} \\
      E_{sph} &= \textcolor{Reviewer2}{\left|\sum_{c \in C} S_c(t) - \sum_{c \in C} S_c(0)\right|}, 
     \label{eq:evaSph}
 \end{align}
where the subscript $k$ indicates that the value is extracted from the cell $\Omega_k$ and the $m_k$, $V_k$ denote the mass and volume of $\Omega_k$. In \cref{eq:evaSph}, \textcolor{Reviewer2}{$S_c$ is the area of the PLIC-VOF interface polygon in the cell $\Omega_c$.} 
\Cref{eq:sumAlpha,eq:sumMOM} represent the time evolution of the normalized error of the global sum of the heavy phase mass and momentum. In this case, $E_{mass}$ is expected to be near zero because there is no source and dissipation for both mass, and flux-based geometrical VOF method have a very high degree of local volume (mass) conservation. 
 
In the absence of force terms on both sides of momentum transport equation, i.e, \cref{eq:momentum-transport}, assuming the periodic boundary condition is applied to all boundary patches, momentum conservation dictates that the deviation $E_{mom}$ should also theoretically remain at zero over time.  \citet{Bussmann2002} proposed that a droplet, characterized by a large density ratio ($10^6$), should undergo translation without deformation in an ambient flow, much like a solid sphere moving through a void.  This conclusion has been widely accepted and corroborated by several publications  (e.g., \citep{Desjardins2010,Raessi2012,Ghods2013,Chenadec2013,Vaudor2014}). However, these studies qualitatively assessed droplet deformation based on visual representations of droplet shape. In this work, we employ the sphericity error $E_{sph}$ as a quantitative measure, to characterize the deviation from the initial droplet shape.

%

As discussed in \cref{subsec:consistency}, when the first-order accurate Euler and Gauss upwind scheme are employed to discretize momentum conservation \cref{eq:momentum-transport}, the mass flux $(\rho_f F_f)^{isoAdvector}$ from \cref{eq:massfluxisoadvector} will be consistent. Since the choice of discretization schemes ensures consistency of the discretization, there is no need to modify the implementation of the numerical method. We have verified the analysis from \cref{sec:numerical-method-and-implementation} for the "interIsoFoam" solver and compared it with the "interIsoRhoFoam" solution \cref{alg:interIsoRhoFoam} that implements the auxiliary density equation. The normalized mass error \cref{eq:sumAlpha} is \textcolor{Reviewer1}{nearing machine epsilon, and is therefore much smaller than than the linear solver tolerance (set for this case to $10^{-12}$), for both interIsoFoam and interIsoRhoFoam, showing excellent conservation of mass for both configurations. The consistency of the mass and phase indicator transport, as well as the consistency of the mass flux approximation of the interIsoFoam and interIsoRhoFoam with Euler+upwind schemes is reflected in an equivalent accuracy and stability for the momentum: \cref{eq:sumMOM} remains equally much smaller than the linear solver tolerance, nearing machine epsilon, and remains stable}. We therefore verify the consistency \textcolor{Reviewer1}{and equivalence} of the "interIsoFoam" and "interIsoRhoFoam", \textcolor{Reviewer1}{provided the  Euler+upwind schemes are used}. 

Note that this verification case is extremely challenging, since it is an inviscid case - there is no viscous force available in this case to dampen the errors resulting from inconsistent two-phase mass and momentum transport.

Next, we demonstrate \textcolor{Tobi}{inconsistencies} leading to large errors and often to catastrophic failure when more than one mass flux is used over $\Delta t$ when discretizing \cref{eq:momentum-transport}, or the mass flux is limited in the discretized \cref{eq:momentum-transport}.



\subsubsection{Comparison of different schemes}

\begin{table}[]
\begin{adjustbox}{width=0.95\textwidth}
\small
\renewcommand*{\arraystretch}{1.2}
\begin{tabular}{cccccc}
\hline
Time schemes       & Gaussian convection schemes & Order of convection accuracy & Category of convection & Boundedness of convection & Mass flux consistency\\ \hline
Crank-Nicolson\citep{CrankNicolson1947}         & upwind                      & first-order                  & NVD/TVD                & Bounded & no                   \\ \hline
\multirow{9}{*}{Euler} & upwind                      & first-order                  & NVD/TVD                & Bounded  & yes                  \\
                       & cubic\citep{Rubin1976}      & second-order                 & non-NVD/TVD            & Unbounded  & no                 \\
                       & limitedLinearV              & first-/second-order           & NVD/TVD                & Unbounded  & no                 \\
                       & linear                      & second-order                 & non-NVD/TVD            & Unbounded   & no                \\
                       & LUST                        & second-order                 & non-NVD/TVD            & Unbounded    & no               \\
                       & MUSCL\citep{Vanleer1979101} & second-order                 & NVD/TVD                & Unbounded    & no               \\
                       & QUICK\citep{Leonard1979}    & second-order                 & NVD/TVD                & Unbounded    & no               \\
                       & SuperBee\citep{Roe1986}     & second-order                 & NVD/TVD                & Unbounded    & no               \\
                       & vanLeer\citep{Vanleer1974361}& second-order                 & NVD/TVD                & Unbounded    & no              \\ \hline
\end{tabular}
\end{adjustbox}
\caption{The combinations of different time and convection schemes used to test the effect of numerical consistency.}
\label{table:test_schemes_combi}
\end{table}

Using the Crank-Nicolson scheme to discretize \cref{eq:momentum-transport} reveals that the temporal scheme involving implicit mass flux $\rho_f F_f$ results in \textcolor{Tobi}{a} mismatch between mass convection and volume fraction convection scaled by $(\rho^- - \rho^+)$,  due to the fact that the Navier–Stokes equation using a segregated method is solved iteratively within a time step, and the interface's advection velocity thus cannot be updated simultaneously with $\mathbf{v}^{n+1}$ from the previous $p-v$ coupling iteration. The inconsistency is amplified by \textcolor{Tobi}{the density-ratio}. Combinations of schemes listed in \cref{table:test_schemes_combi}, are tested to verify their effect on the mass flux inconsistency. 

\begin{figure}
    \centering
    \includegraphics[width=0.96\textwidth]{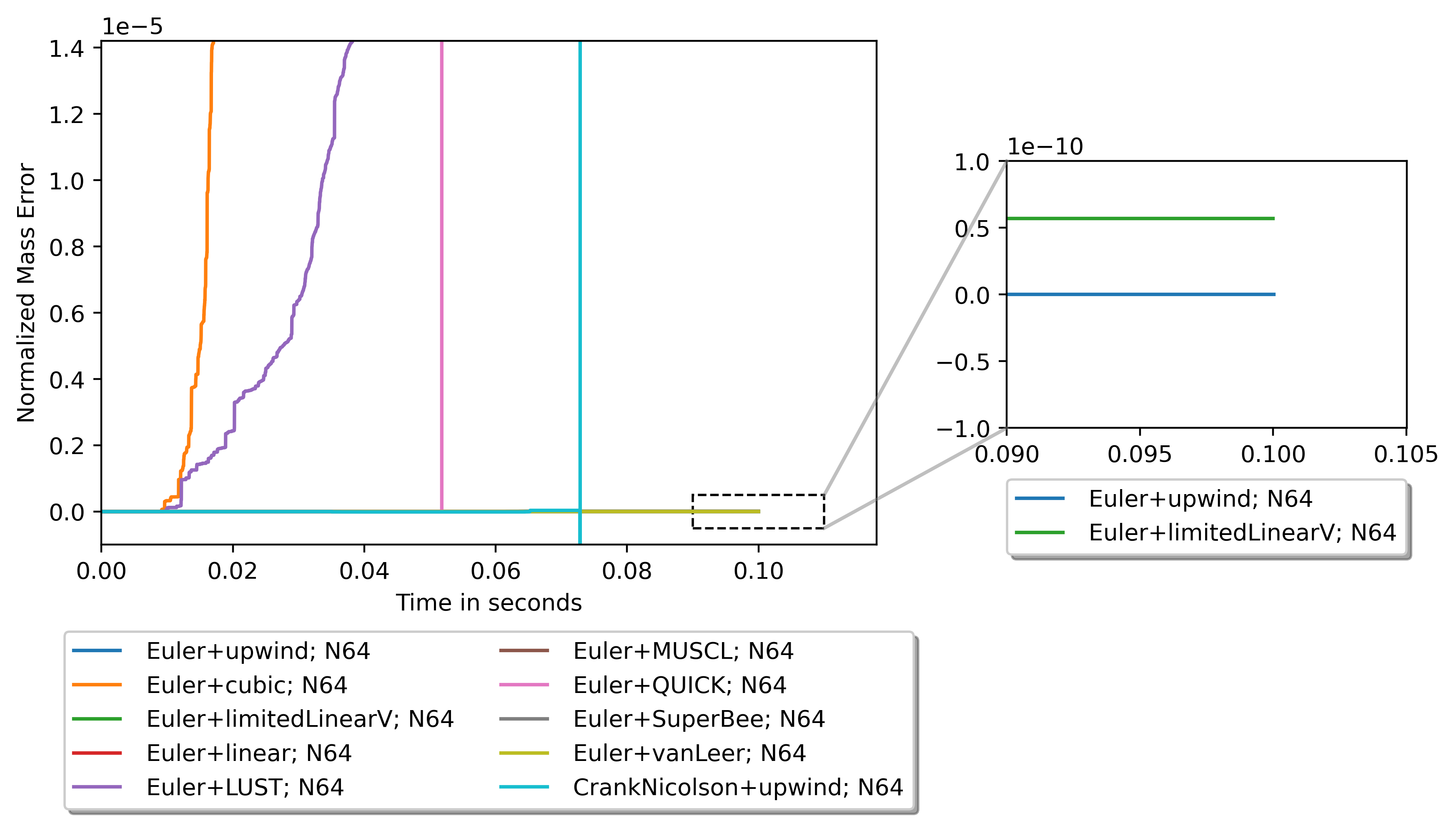}
    \caption{Temporal evolution of normalized mass conservation error with different schemes: interIsoFoam, $N=64$.}
    \label{fig:Schemes_comp_massConservation}
\end{figure}

\begin{figure}
    \centering
    \includegraphics[width=0.96\textwidth]{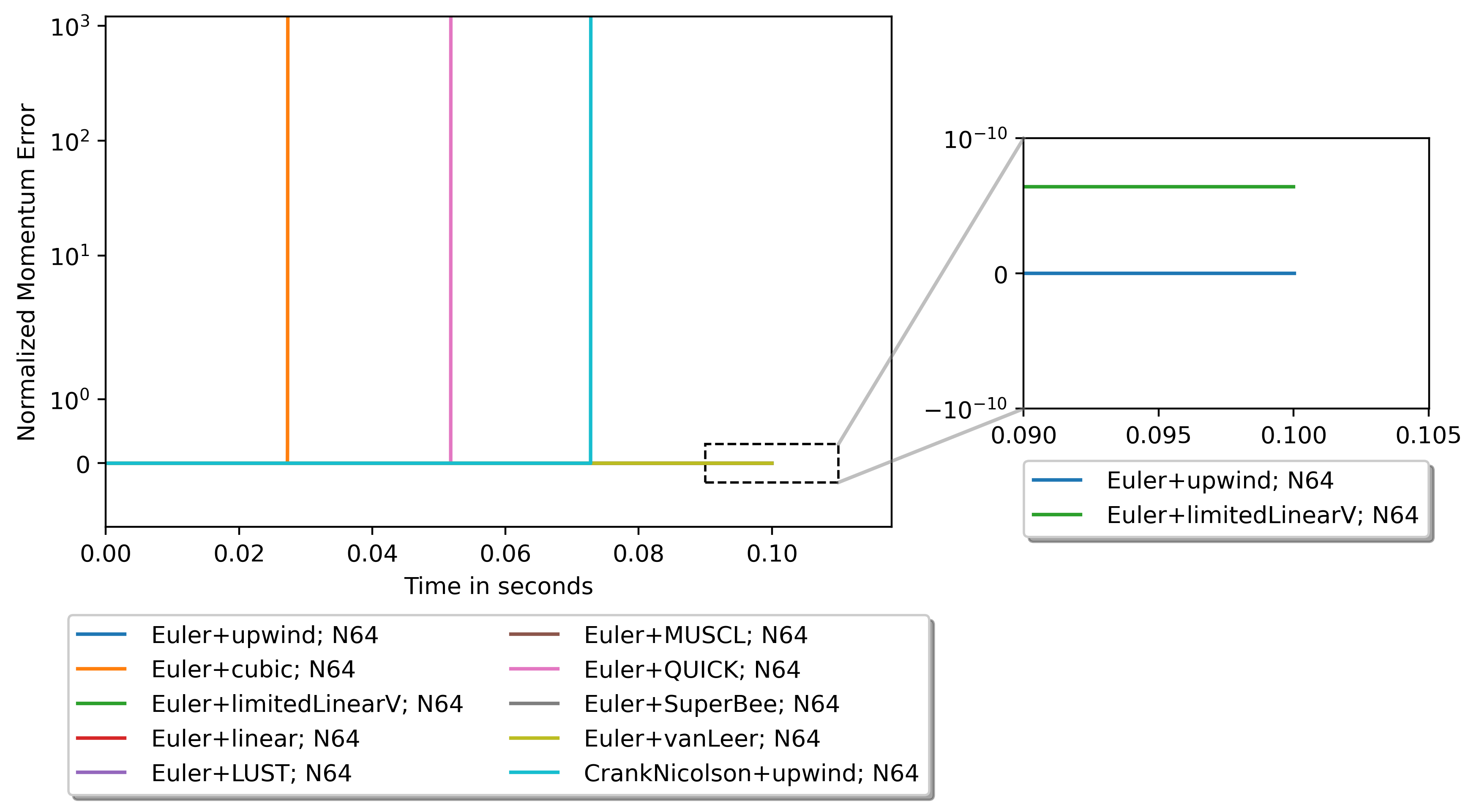}
    \caption{Temporal evolution of normalized momentum conservation error with different schemes: interIsoFoam, $N=64$.}
    \label{fig:Schemes_comp_MOMConservation}
\end{figure}

\Cref*{fig:Schemes_comp_massConservation,fig:Schemes_comp_MOMConservation} represent the temporal evolution of $E_{mass}$, and $E_{mom}$. It is evident from \cref{fig:Schemes_comp_massConservation} that mass conservation is not maintained when utilizing the Crank-Nicolson time \textcolor{Tobi}{discretization} scheme, and divergence schemes cubic,  \textcolor{Reviewer1}{Linear-Upwind Stabilised Transport (LUST)} and Quadratic Upstream Interpolation for Convective Kinematics (QUICK). These simulations terminate at an early stage with catastrophic failure. However, for cases that can run until the final time, the magnitude of mass errors, as shown in the zoomed subfigure of \cref{fig:Schemes_comp_massConservation}, is on the order of $10^{-10}$, indicating mass conservation. This observation aligns with the inherent characteristic of the volume of fluid method, which is known for its mass conservation property. 
The vertical lines from the results of cubic, QUICK and Crank-Nicolson can be observed in \cref{fig:Schemes_comp_MOMConservation}. Some combinations deliver stable momentum errors. A closer examination of the stable cases in the zoomed view of \cref{fig:Schemes_comp_MOMConservation} 
confirms the use of the Euler temporal scheme.    

\begin{figure}
    \centering
    \includegraphics[width=0.96\textwidth]{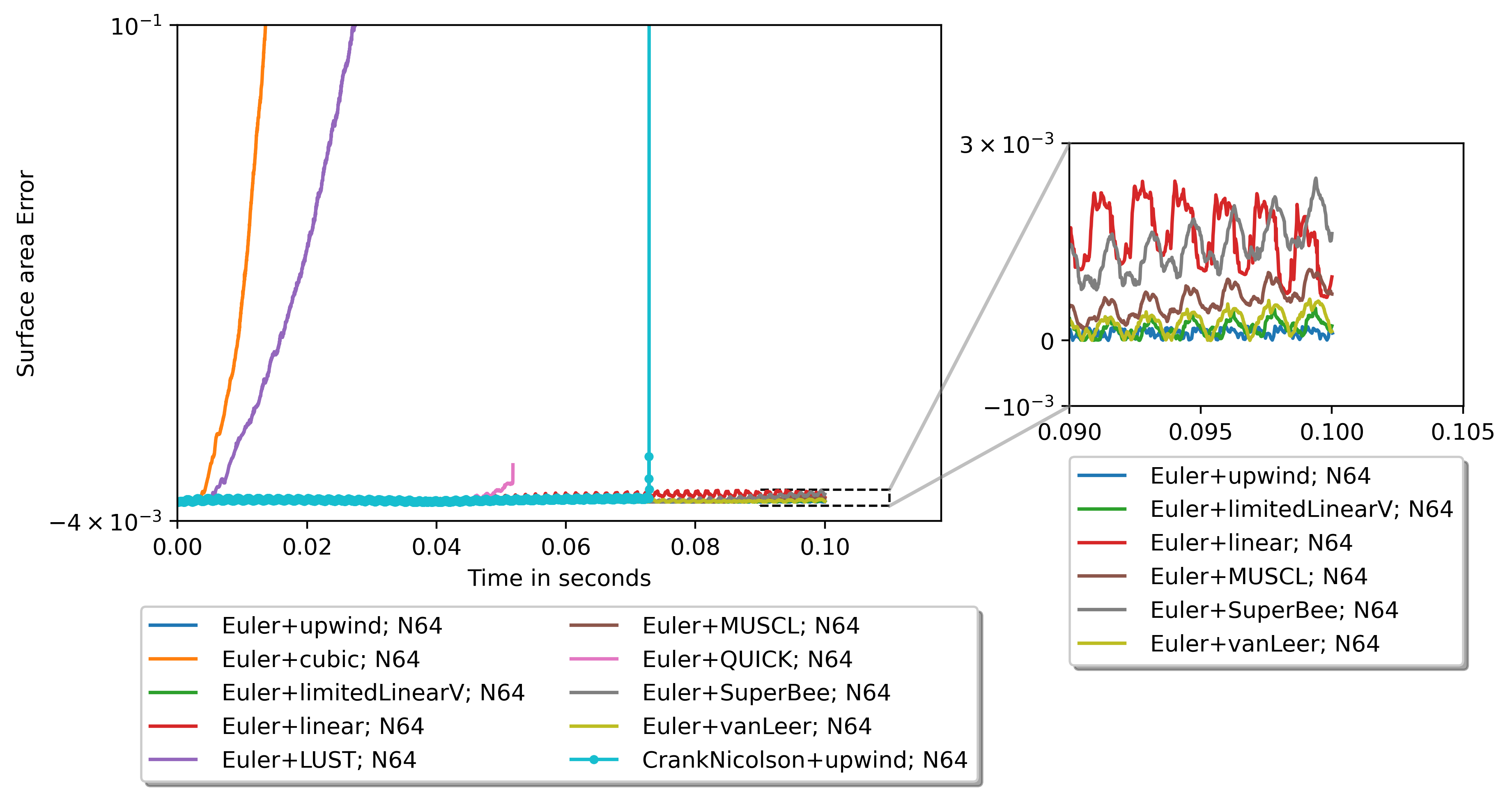}
    \caption{\textcolor{Reviewer2}{Temporal evolution of the sphericity error (\ref{eq:evaSph}) with different schemes: interIsoFoam, $N=64$.}}
    \label{fig:Schemes_comp_sphericity}
\end{figure}

\begin{figure}[!htb]
     \begin{subfigure}{.48\textwidth}
      \centering
      \includegraphics[width=\linewidth]{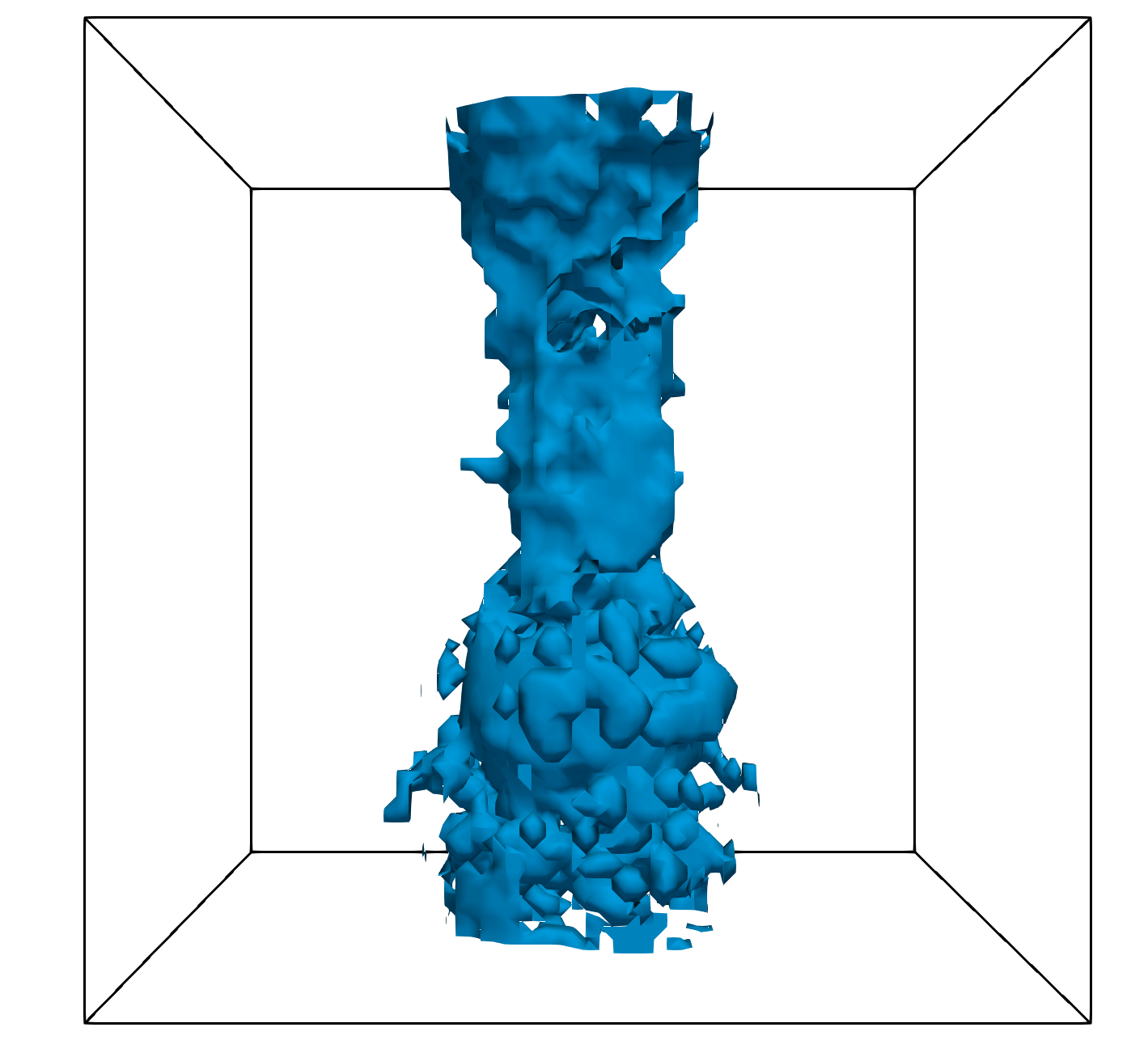}
      \caption{Crank-Nicolson $+$ upwind}
      \label{fig:Zuzio_CN_upwind}
     \end{subfigure}
     \hfill
     \begin{subfigure}{.48\textwidth}
      \centering
      \includegraphics[width=\linewidth]{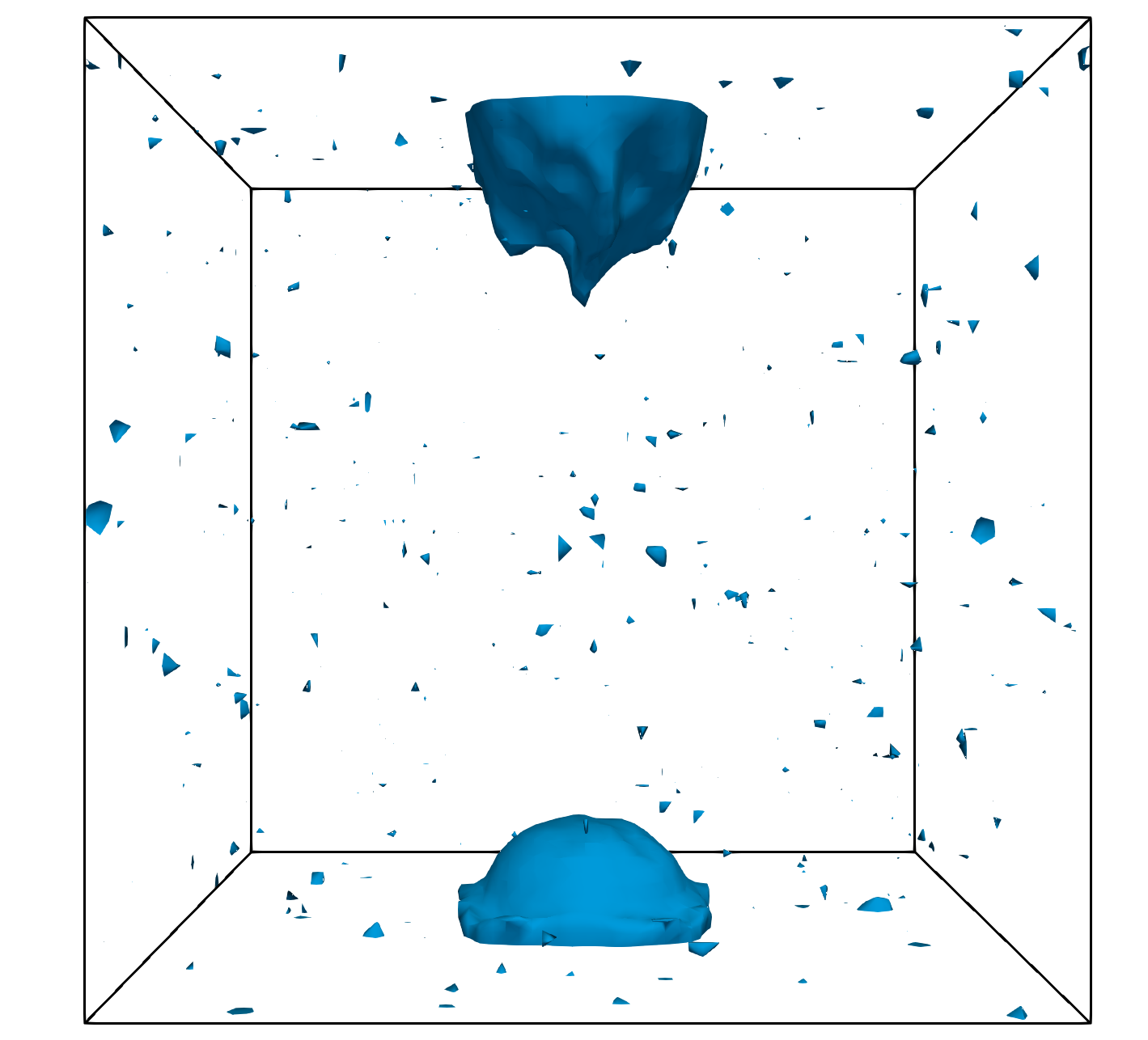}
      \caption{Euler $+$ LUST}
      \label{fig:Zuzio_Euler_LUST}
     \end{subfigure}    
          \begin{subfigure}{.48\textwidth}
      \centering
      \includegraphics[width=\linewidth]{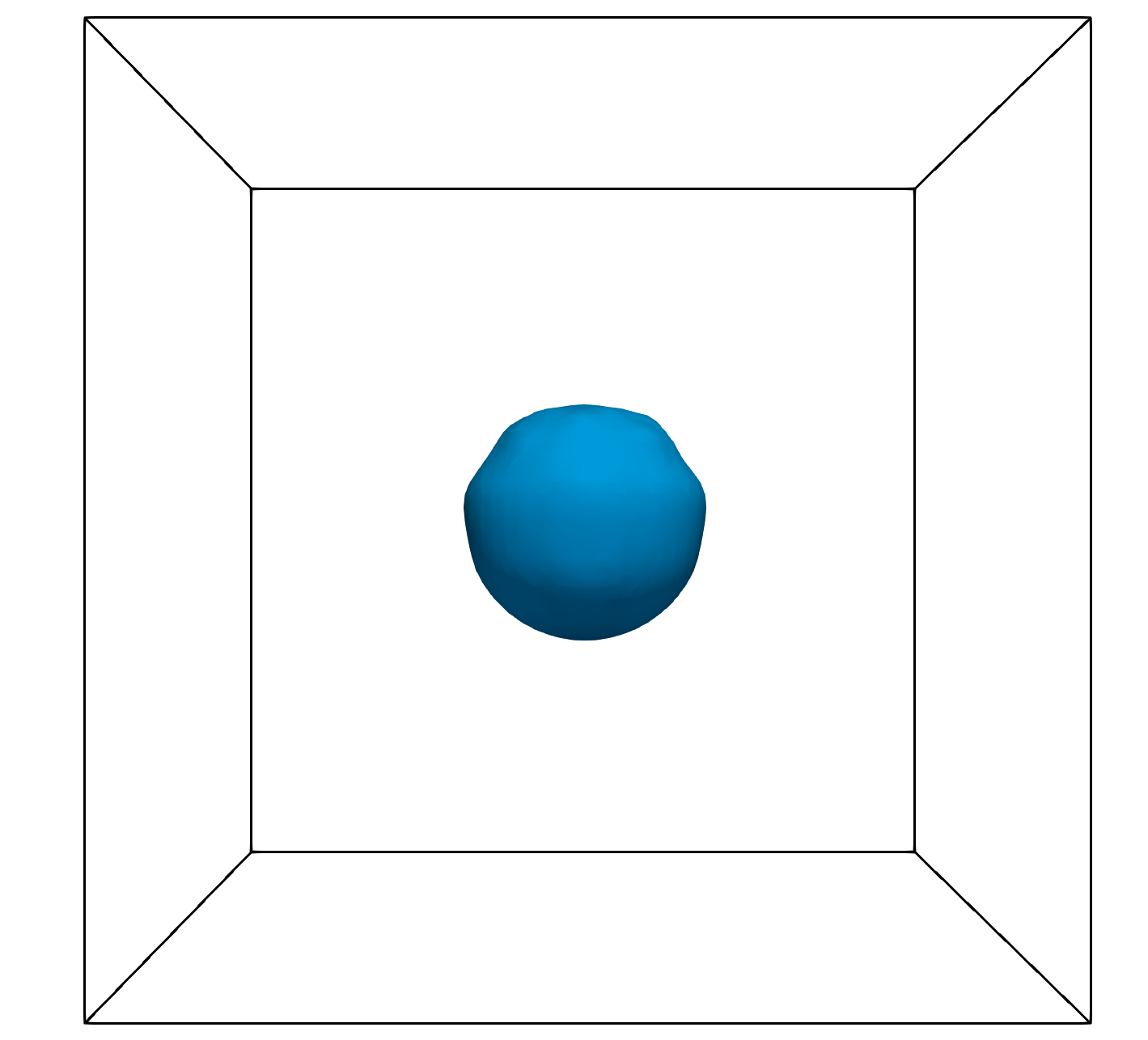}
      \caption{Euler $+$ limitLinearV}
      \label{fig:Zuzio_Euler_limitedLinearV}
     \end{subfigure}
     \hfill
     \begin{subfigure}{.48\textwidth}
      \centering
      \includegraphics[width=\linewidth]{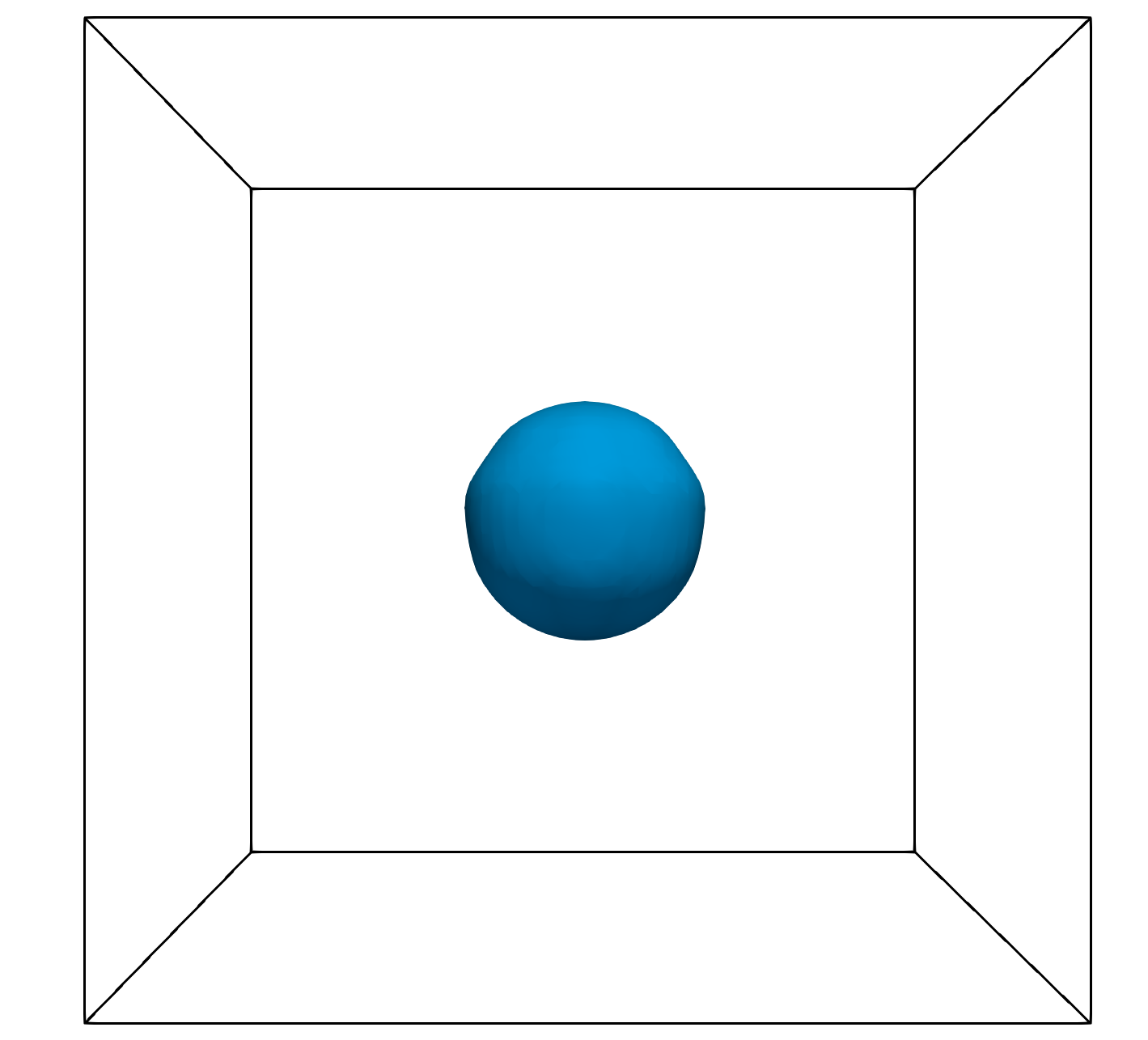}
      \caption{Euler $+$ upwind}
      \label{fig:Zuzio_Euler_upwind}
     \end{subfigure} 
    \caption{Final shape of the droplet calculated by interIsoFoam with different schemes: $N=64$.}
    \label{fig:Zuzio_schemes_comp_droplet_shape}
\end{figure}

The temporal evolution of sphericity error is depicted in \cref{fig:Schemes_comp_sphericity} where the steepness of the curves indicates the extent of droplet deformation from its initial shape. The results obtained from four unstable scheme combinations show significantly larger deviations in sphericitiy. 
Moreover, the zoomed view reveals that, apart from Euler$+$upwind, all other stable scheme combinations display varying degrees of divergence. This suggests that if the simulations were to continue for a longer duration, these cases would likely crash. \Cref{fig:Zuzio_schemes_comp_droplet_shape} presents the final shapes of the droplets simulated using interIsoFoam with two stable scheme combinations and two unstable combinations. For the unstable combinations, a common occurrence is observed: the original droplets disintegrate into small, irregularly scattered pieces. A comparison between the final shapes obtained using the inconsistent but in this verification case, for these parameters, still stable scheme combination Euler $+$ limitedLinearV (\cref{fig:Zuzio_Euler_limitedLinearV}) and the consistent scheme combination Euler $+$ upwind (\cref{fig:Zuzio_Euler_upwind}) corroborates the findings from the sphericity errors in \cref{fig:Schemes_comp_sphericity}: an evident "crown" with irregular bumps forms on the top part of the droplet in \cref{fig:Zuzio_Euler_limitedLinearV}, leading to a reduction in sphericity. On the contrary, in \cref{fig:Zuzio_Euler_upwind}, only a slight shrinkage occurs in the neck region of the droplet, while the top surface remains smooth.

\begin{figure}
    \centering
    \includegraphics[width=0.66\textwidth]{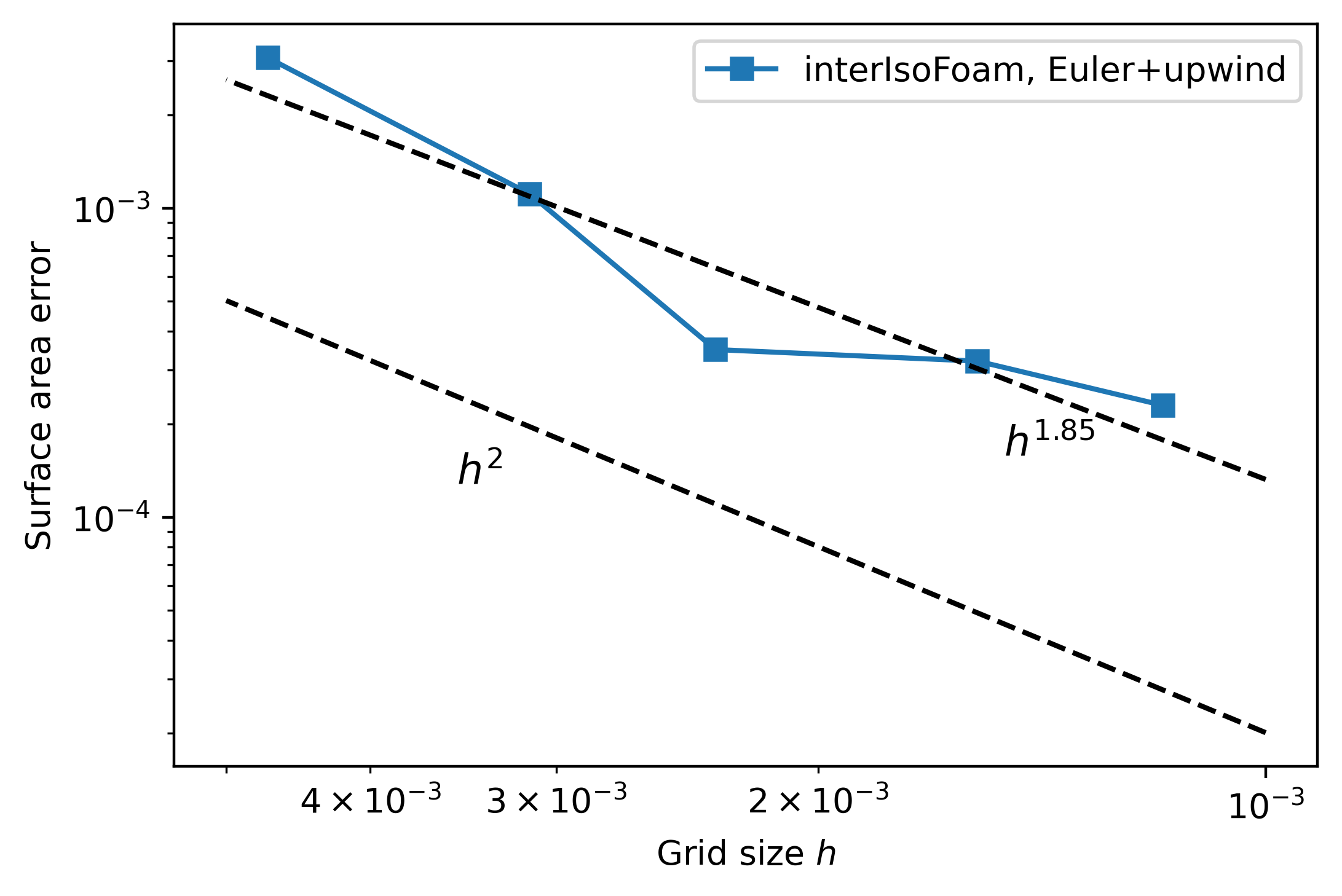}
    \caption{\textcolor{Reviewer2}{Mesh convergence study for sphericity error: interIsoFoam, Euler+upwind, $N\in(36,48,64,96,128)$.}}
    \label{fig:meshConvergenceStudy_sphericity}
\end{figure}

\Cref{table:translatingDropletInQuiscentFlow} summarizes these findings, providing information on early termination of scheme combinations that are inconsistent, as well as mass, momentum and sphericity errors for different mesh resolutions. \textcolor{Reviewer2}{Two finer resolutions, i.e., $N \in (96,128)$, are tested for interIsoFoam with Euler+upwind schemes to verify the mesh convergence of the consistent method, reported in  \cref{fig:meshConvergenceStudy_sphericity}. All scheme combinations are tested with interIsoFoam with density ratio $1$. As shown in \cref{fig:densityRatio1}, all cases with the same densities remain stable.
}

\begin{table}[]
\begin{adjustbox}{width=1\textwidth}
\small
\begingroup
\begin{tabular}{llllllllll}
\toprule
                     &            & \multicolumn{4}{l}{interIsoRhoFoam}                                                   & \multicolumn{4}{l}{interIsoFoam}                          \\
Scheme combination   & Resolution & End time & Mass error & Momentum error        & \multicolumn{1}{l|}{Sphericity Error} & End time & Mass error & Momentum error & Sphericity Error \\ \midrule
Euler+upwind         & $32$         & $0.1$      & $0.0$        & $0.0$                   & \multicolumn{1}{l|}{$3.062\times 10^{-3}$}        & $0.1$      & $0.0$        & $0.0$            & $3.062\times 10^{-3}$        \\
                     & $48$         & $0.1$      & $0.0$        & $0.0$                   & \multicolumn{1}{l|}{$1.111\times 10^{-3}$}        & $0.1$      & $0.0$        & $0.0$            & $1.110\times 10^{-3}$        \\
                     & $64$         & $0.1$      & $0.0$        & $0.0$                   & \multicolumn{1}{l|}{$3.484\times 10^{-4}$}        & $0.1$      & $0.0$        & $0.0$            & $3.484\times 10^{-4}$        \\
                     & \textcolor{Reviewer2}{$96$}         & \textcolor{Reviewer2}{*}     & \textcolor{Reviewer2}{*}        & \textcolor{Reviewer2}{*}                   & \multicolumn{1}{l|}{\textcolor{Reviewer2}{*}}        & \textcolor{Reviewer2}{$0.1$}     & \textcolor{Reviewer2}{$0.0$}        & \textcolor{Reviewer2}{$0.0$}            & \textcolor{Reviewer2}{$3.197\times 10^{-4}$}       \\
                     & \textcolor{Reviewer2}{$128$}          & \textcolor{Reviewer2}{*}     & \textcolor{Reviewer2}{*}        & \textcolor{Reviewer2}{*}                    & \multicolumn{1}{l|}{\textcolor{Reviewer2}{*}}        & \textcolor{Reviewer2}{$0.1$}     & \textcolor{Reviewer2}{0.0}        & \textcolor{Reviewer2}{$0.0$}            & \textcolor{Reviewer2}{$2.298\times 10^{-4}$}        \\ \midrule
Euler+cubic          & $32$         & $0.0206$    & $4.840$   & $1.028\times 10^{18}$              & \multicolumn{1}{l|}{$-0.1299$}          & $0.067$    & $8.807\times 10^{-3}$   & $1.440\times 10^{10}$       & $0.3501$       \\
                     & $48$         & $0.01$    & $-2.925\times 10^{-2}$   & $3.581\times 10^{22}$              & \multicolumn{1}{l|}{$-0.2341$}        & $0.0260$    & $1.6346$        & $2.390\times 10^{6}$       & $0.1024$        \\
                     & $64$         & $0.0065$    & $-2.190\times 10^{-3}$     & $2.233\times 10^{22}$              & \multicolumn{1}{l|}{$-0.2267$}          & $0.0273$      & $0.8301$   & $3.142\times 10^{8}$         & $0.6643$        \\ \midrule
Euler+limitedLinearV & $32$         & $0.1$    & $0.0$        & $5.511\times 10^{-4}$             & \multicolumn{1}{l|}{$6.773\times 10^{-3}$}        & $0.1$    & $0.0$        & $-2.656\times 10^{-9}$            & $5.882\times 10^{-3}$        \\
                     & $48$         & $0.1$    & $0.0$        & $4.793\times 10^{-4}$             & \multicolumn{1}{l|}{$2.852\times 10^{-3}$}        & $0.1$    & $0.0$        & $-4.995\times 10^{-9}$            & $3.671\times 10^{-3}$        \\
                     & $64$         & $0.1$    & $0.0$        & $3.290\times 10^{-4}$                   & \multicolumn{1}{l|}{$2.343\times 10^{-3}$}         & $0.1$      & $5.695\times 10^{-11}$        & $6.407\times 10^{-11}$            & $2.216\times 10^{-4}$         \\ \midrule
Euler+linear         & $32$         & $0.0384$    & $-4.520\times 10^{-4}$        & $3.603\times 10^{21}$             & \multicolumn{1}{l|}{$-0.1932$}        & $0.1$    & $0.0$        & $-2.540\times 10^{-10}$            & $2.234\times 10^{-2}$        \\
                     & $48$         & $0.0306$    & $-0.1954$        & $4.271\times 10^{24}$             & \multicolumn{1}{l|}{$-0.2330$}        & $0.1$    & $0.0$        & $2.210\times 10^{-7}$            & $1.843\times 10^{-2}$        \\
                     & $64$         & $0.0239$    & $-0.0676$        & $2.604\times 10^{25}$                   & \multicolumn{1}{l|}{$-0.2481$}         & $0.1$      & $1.210\times 10^{-9}$        & $3.915\times 10^{-10}$            & $9.681\times 10^{-4}$        \\ \midrule
Euler+LUST           & $32$         & $0.0221$    & $598.0$     & $3.036\times 10^{21}$             & \multicolumn{1}{l|}{$-0.2055$}           & $0.0993$    & $0.1097$   & $2.162\times 10^{19}$      & $0.4242$        \\
                     & $48$         & $0.0166$    & $1.026\times 10^{55}$  & $3.272\times 10^{120}$             & \multicolumn{1}{l|}{$0.2428$}          & $0.0258$      & $1.047\times 10^{13}$        & $1.797\times 10^{39}$       & $-0.1762$        \\
                     & $64$         & $0.002$    & $0.0$        & $9.254\times 10^{-5}$             & \multicolumn{1}{l|}{$5.042\times 10^{-4}$}         & $0.0485$    & $2.006\times 10^{-5}$   & $-2.664\times 10^{-5}$      & $0.2856$          \\ \midrule
Euler+MUSCL          & $32$         &  $0.0046$    & $0.0$        & $4.727\times 10^{8}$             & \multicolumn{1}{l|}{$-0.0932$}        & $0.1$    & $0.0$        & $1.143\times 10^{-5}$      & $7.383\times 10^{-3}$        \\
                     & $48$         & $0.0064$    & $0.0$        & $-0.1949$             & \multicolumn{1}{l|}{$-3.500\times 10^{-3}$}        & $0.0485$    & $2.227\times 10^{12}$        & $2.566\times 10^{36}$      & $-0.1434$        \\
                     & $64$         & $0.0092$    & $0.0$        & $1.059\times 10^{-3}$             & \multicolumn{1}{l|}{$-1.202\times 10^{-3}$}         & $0.1$      & $3.075\times 10^{-9}$        & $1.780\times 10^{-7}$      & $7.070\times 10^{-4}$        \\ \midrule
Euler+QUICK          & $32$         & $0.0203$    & $0.0$        & $5.455\times 10^{6}$               & \multicolumn{1}{l|}{$-0.2096$}          & $0.1$    & $0.0$        & $5.177\times 10^{-4}$      & $9.6670\times 10^{-3}$        \\
                     & $48$         & $0.0066$    & $-3.198\times 10^{-9}$        & $6.806\times 10^{82}$             & \multicolumn{1}{l|}{$-0.2699$}        & $0.0185$    & $2.417\times 10^{-6}$        & $4.164\times 10^{51}$      & $0.0118$        \\
                     & $64$         & $0.0079$    & $-5.097\times 10^{-7}$        & $1.862\times 10^{24}$             & \multicolumn{1}{l|}{$-0.1086$}         & $0.0518$    & $0.1564$        & $5.537\times 10^{21}$        & $7.812\times 10^{-3} $       \\ \midrule
Euler+SuperBee       & $32$         & $0.0237$    & $0.0$        & $5.732\times 10^{15}$             & \multicolumn{1}{l|}{$-0.1570$}        & $0.1$    & $0.0$        & $-2.632\times 10^{-5}$      & $5.094\times 10^{-3}$        \\
                     & $48$         & $0.0022$    & $0.0$        & $0.2876$             & \multicolumn{1}{l|}{$-4.630\times 10^{-4}$}        & $0.0631$    & $1.645\times 10^{-10}$        & $8.588\times 10^{42}$      & $0.0120$        \\
                     & $64$         & $0.0075$    & $0.0$        & $-1.206\times 10^{-4}$                   & \multicolumn{1}{l|}{$-8.018\times 10^{-4}$}         & $0.1$      & $1.477\times 10^{-8}$        & $2.970\times 10^{-7}$      & $1.628\times 10^{-3}$        \\ \midrule
Euler+vanLeer        & $32$         & $0.1$    & $0.0$       & $4.190\times 10^{-4}$            & \multicolumn{1}{l|}{$-4.514\times 10^{-4}$}        & $0.1$    & $0.0$        & $4.916\times 10^{-7}$      & $3.495\times 10^{-3}$        \\
                     & $48$         & $0.1$    & $0.0$        & $3.480\times 10^{-4}$             & \multicolumn{1}{l|}{$1.417\times 10^{-3}$}        & $0.1$    & $0.0$        & $3.535\times 10^{-7}$      & $4.868\times 10^{-3}$        \\
                     & $64$         & $0.0896$    & $0.0$        & $-9.273\times 10^{-5}$             & \multicolumn{1}{l|}{$0.0107$}         & $0.1$      & $3.310\times 10^{-9}$        & $5.478\times 10^{-8}$      & $1.388\times 10^{-4}$        \\ \midrule
CrankNicolson+upwind & $32$         & $0.0096$    & $0.0$        & $4.131\times 10^{-4}$              & \multicolumn{1}{l|}{$-9.018\times 10^{-4}$}         & $0.0043$    & $-0.2219$        & $7.627\times 10^{21}$     & $-0.1632$        \\
                     & $48$         & $0.0049$    & $0.0$        & $3.238\times 10^{-3}$ & \multicolumn{1}{l|}{$-3.139\times 10^{-4}$}         & $0.0037$    & $32.30$        & $2.203\times 10^{24}$            & $-0.1965$        \\
                     & $64$         & $0.0032$    & $6.001\times 10^{-3}$   & $6.280\times 10^{-4}$               & \multicolumn{1}{l|}{$-2.093\times 10^{-3}$}          & $0.0729$    & $9.124\times 10^{25}$   & $2.576\times 10^{60}$        & $0.3248$    \\ \bottomrule     
\end{tabular}
\endgroup
\end{adjustbox}
\caption{\textcolor{Reviewer22}{The terminated time and the final normalized errors of interIsoRhoFoam and interIsoFoam with different scheme combinations and mesh resolutions for translating droplet in ambient flow case.}}
\label{table:translatingDropletInQuiscentFlow}
\end{table}

\subsection{Translating sub-millimeter droplet with realistic physical properties}\label{sec: small-droplet-cases}

\begin{table}[H]
\begin{adjustbox}{width=1\textwidth}
\small
\begingroup
\renewcommand{\arraystretch}{1.1}
\begin{tabular}{cccccl}
\hline
\rowcolor[HTML]{EFEFEF}
materials/properties (\SI{25}{\celsius}) & density (\si[per-mode=symbol]{\kilogram\per\cubic\metre}) & kinematic viscosity (\si[per-mode=symbol]{\square\metre\per\second}) & surface tension (\si[per-mode=symbol]{\newton\per\metre}) & density ratio \\
\hline
air                                                 & $1.1839$                                                    & $---$                                                             & $---$                                            & $---$                                   \citep{Adamson1997}\\
mercury                                             & $13.5336\times10^3$                                                & $---$                                                             & $---$ (in air)                                & $11431.37$(in air)                      \citep{Adamson1997}\\
 \hline
\end{tabular}
\endgroup
\end{adjustbox}
\caption{Realistic densities of the mercury droplet/air ambient pair.}
\label{table:physical-properties} 
\end{table}

The realistic densities of the mercury droplet and air ambient pair are selected for the verification case of the translating sub-millimeter droplet. The density ratio displayed in \cref{table:physical-properties} is around $10^4$. The rest setups are same as in \citep{Liu2023}. A spherical droplet of radius $R=$ \SI{0.25}{\milli\metre} translates a distance of three diameters with velocity \SI[per-mode=symbol]{0.01}{\metre\per\second} in $z$-direction of the rectangular solution domain ($L_x=L_y=5R,L_z=\textcolor{Reviewer2}{15R}$). The ambient flow has the same velocity, that is $\v_a=(0,0,0.01)$. Three resolutions are tested in this case: $N\in(16,\ 32,\ 64,\ \textcolor{Reviewer2}{96},\ \textcolor{Reviewer2}{128})$. The initial centroid position of the droplet is $(2.5R,2.5R,2R)$. Surface tension and viscous forces are not considered in this case. Since the droplet translates with the ambient flow and there is no sink or source for the droplet moving, the velocity field should keep unchanged.

The error norm $L_\infty$ is employed to measure the maximal deviation between the numerical velocity and the analytical one among all cells, i.e.,\  

\begin{equation}
    L_{\infty}(\mathbf{v})= \max_i\left(\frac{\|\mathbf{v}_i-\mathbf{v}_{\infty}\|}{\|\mathbf{v}_{\infty}\|}\right),
    \label{eq:linfty}
\end{equation}
where $\mathbf{v}_i$ denotes the velocity of the cell $i$, and the analytical velocity value is $\mathbf{v}_{\infty}=\v_a=(0,0,0.01)$. The expected exact value of $L_\infty$ is $0$\textcolor{Reviewer1}{; however, in practice, the absolute accuracy is limited by the absolute tolerance of the linear solver used to solve the pressure Poisson equation.}

\begin{figure}[!htb]
     \begin{subfigure}[t]{.49\textwidth}
      \centering
      \includegraphics[width=\linewidth]{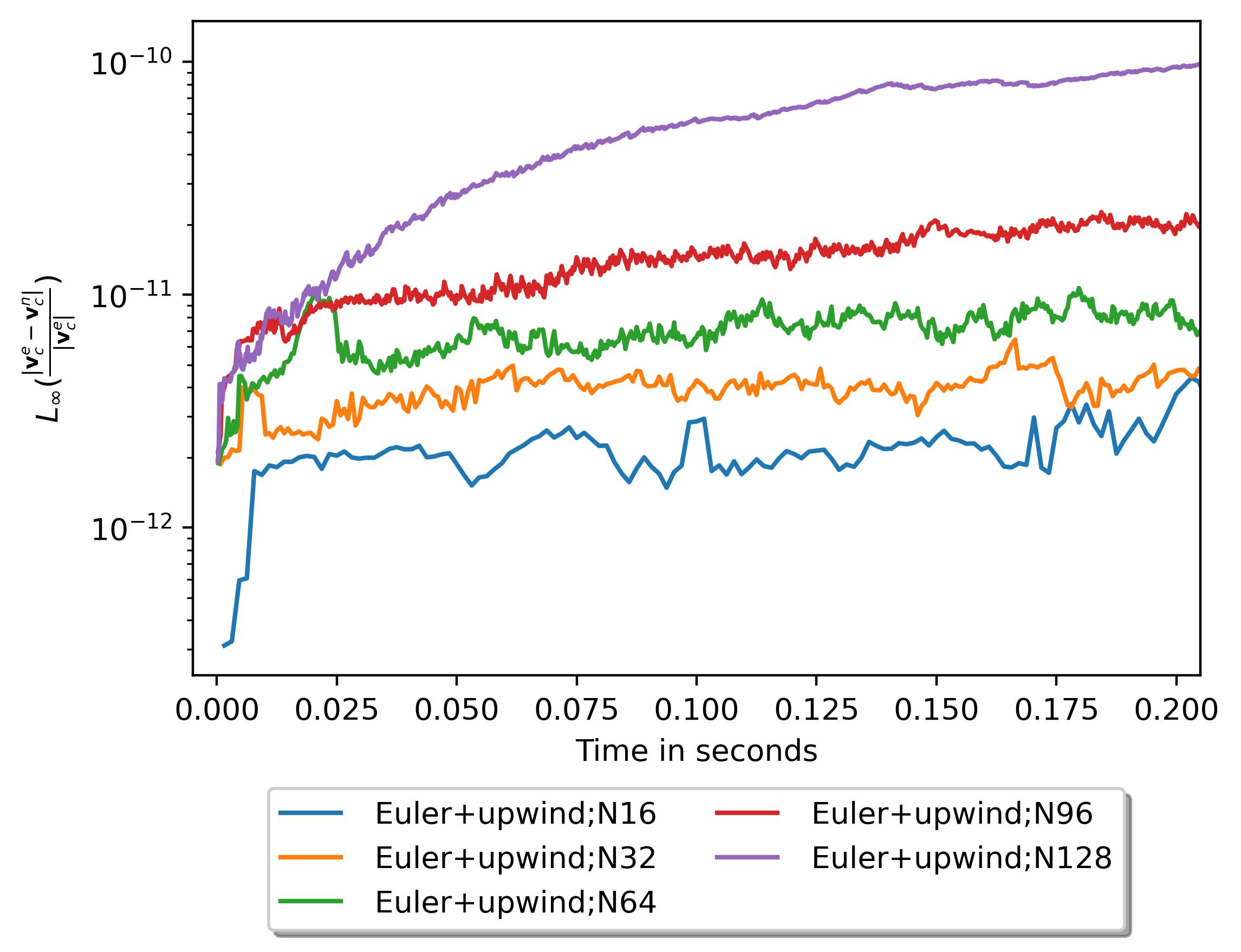}
      \caption{interIsoFoam}
      \label{fig:MercuryAir_interIso}
     \end{subfigure}
     \hfill
     \begin{subfigure}[t]{.49\textwidth}
      \centering
      \includegraphics[width=\linewidth]{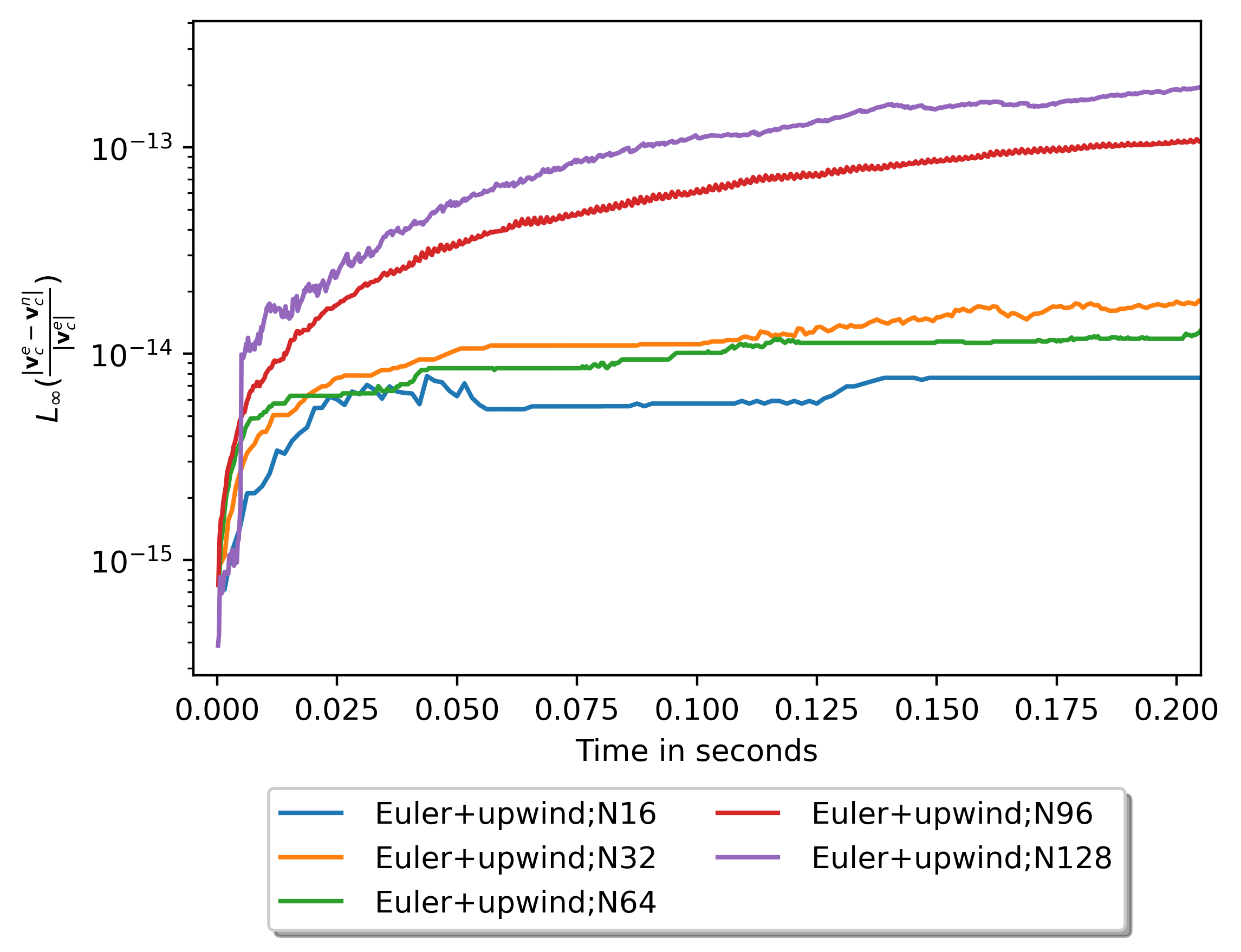}
      \caption{interIsoRhoFoam}
      \label{fig:MercuryAir_interIsoRho}
     \end{subfigure}    
    \caption{Temporal evolution of the velocity error norm $L_{\infty}(\v)$ with pure advection: Euler, Gauss upwind,  density ratio: $10^4$, mesh resolution: $N\in(16,\ 32,\ 64,\ \textcolor{Reviewer2}{96},\ \textcolor{Reviewer2}{128})$.}
    \label{fig:MercuryAir}
\end{figure}

\begin{figure}
    \centering
    \includegraphics[width=.58\textwidth]{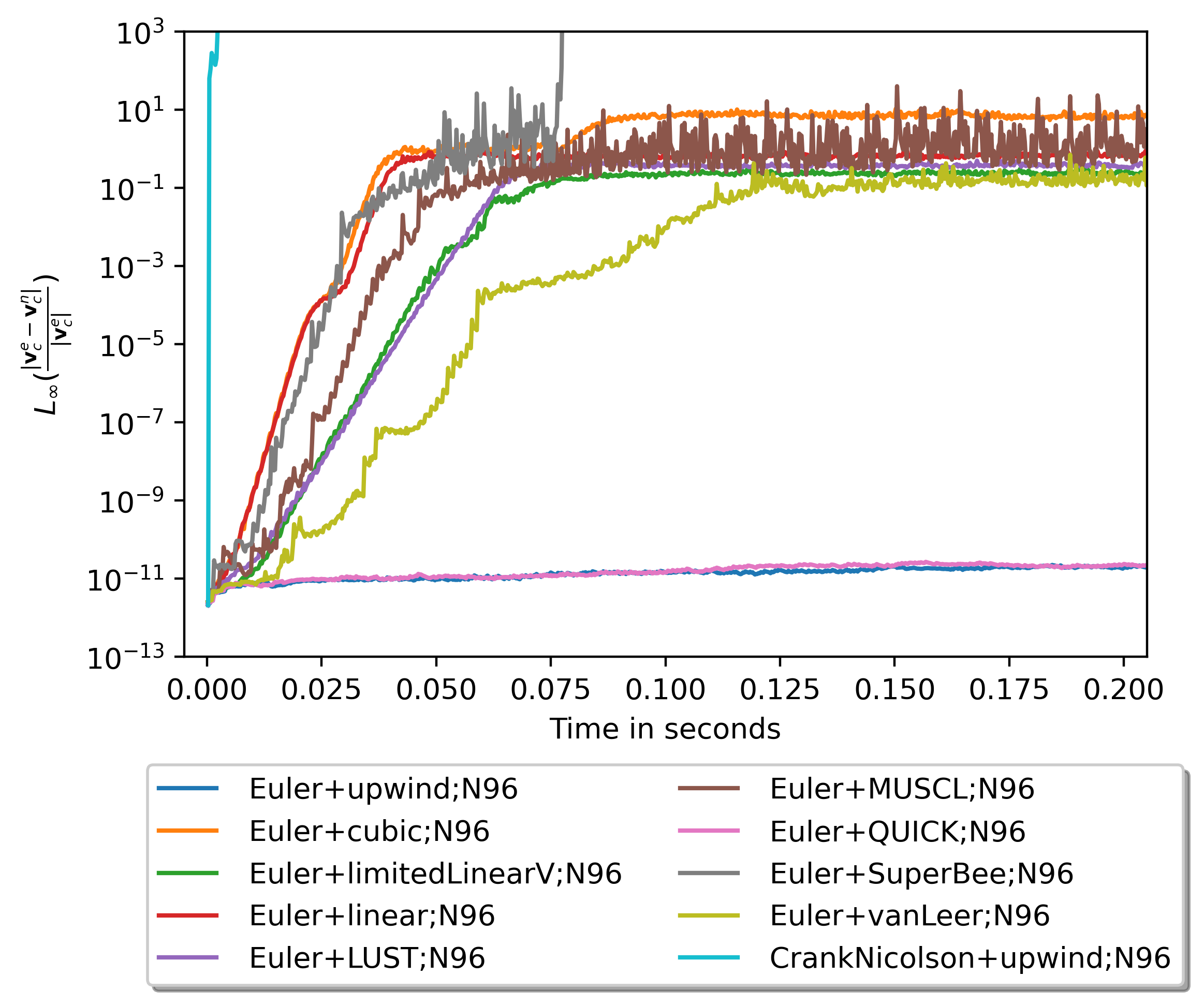}
    \caption{Temporal evolution of the velocity error norm $L_{\infty}(\v)$ with pure advection - combining $10$ schemes,  density ratio is $10^4$, mesh resolution is $N=\textcolor{Reviewer2}{96}$. Only Euler and upwind(ing) schemes remain consistent and stable.}
    \label{fig:TranslatingDroplet3D_schemes_comp}
\end{figure}

\textcolor{Reviewer1}{\Cref{fig:MercuryAir} presents the temporal evolution of $L_{\infty}(\v)$. The same $L_{\infty}(\v)$ calculated for both solvers reveals a very close numerical equivalence between the volume fraction and mass conservation equation using the Euler+upwind combination of schemes. Errors of both solvers remain stable. Absolute errors of interIsoFoam are somewhat larger; however, they remain in the realm of numerical noise, significantly below the linear tolerance for the pressure Poisson equation, ensuring consistency.} A notable outcome from \cref{fig:MercuryAir} is the value of the final converged $L_{\infty}(\v)$, which is at the magnitude of \textcolor{Santiago}{$1\times10^{-11}$} and \textcolor{Reviewer1}{for interIsoRhoFoam almost reaches the machine epsilon, confirming numerical stability and consistency of a very high degree for this challenging verification case}. 

We also conducted tests on all schemes listed in \cref{table:test_schemes_combi} for this particular case setup. The corresponding results are presented in \cref{fig:TranslatingDroplet3D_schemes_comp}. The results from Crank-Nicolson temporal scheme \textcolor{Reviewer2}{and SuperBee convection scheme} show significant numerical instability at the early stage of the simulation. In contrast, the errors obtained from all other schemes remain stable throughout the entire simulation. Notably, there is a substantial variation in accuracy among these schemes. The velocity errors from Euler $+$ QUICK and upwind remain at magnitudes around $10^{-11}$, while the errors from the other schemes initially increase and stabilize at values that are $10^10$ to $10^{11}$ times larger. \textcolor{Reviewer2}{Additionally, the same cases are tested with density ratio $1$ to investigate the effects of the density ratio on the numerical instability. As shown in \cref{fig:TranslatingDroplet3D_schemes_comp_densityRatio1}, all cases keep stable with the low density ratio at the final stage. The velocity errors of all tests are reduced to magnitudes of $10^{-14}$, even for the most critical schemes, i.e. Crank-Nicolson + upwind.}

\subsection{Mixing layer}\label{sec:mixing-layer-2D}

\begin{figure}[!htb]
    \centering
    \def\svgwidth{0.4\textwidth}
    \footnotesize
    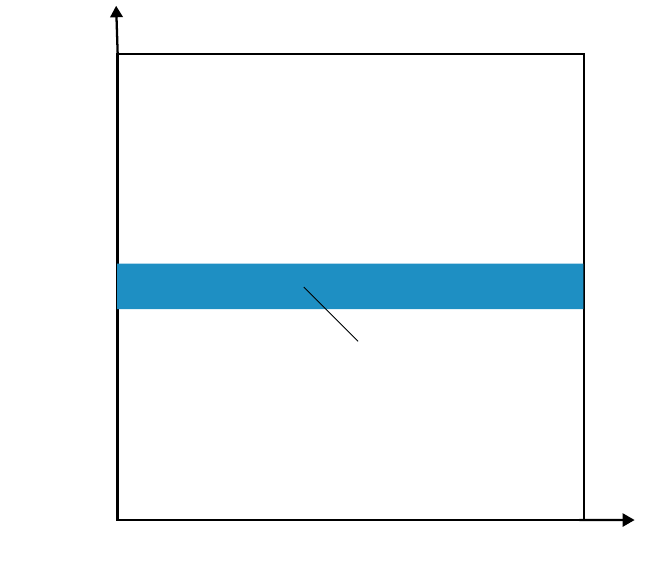%
    \caption{2D mixing layer}
    \label{fig:2D-mixing-layer}
\end{figure}

In this case, a 2D mixing layer case is tested. The 2D computational domain as depicted in \cref{fig:2D-mixing-layer} has the same length $L= \SI{3}{\mm}$ in both $x-$ and $y-$direction. The liquid with high density $\rho^-= \SI{1000}{\kg\per\m^3}$ flows in the middle region $\SI{-0.15}{\mm}<y<\SI{0.15}{\mm}$ of the square computational domain with relatively low initial velocity $v_x^-= \SI[per-mode=symbol]{2}{\m\per\s}$, while the gas with the density $\rho^+= \SI{1}{\kg\per\m^3}$ flows on both sides of the liquid area with very high initial velocity $v_x^+= \SI[per-mode=symbol]{30}{\m\per\s}$. A spatial velocity perturbation is initialized in the internal field and has the distribution 
 \begin{equation*}
  v_y(y)= 0.01v^-_x\sin{2\pi\frac{x}{L}}\exp{-(\frac{2y}{h})^2}, 
 \end{equation*}
where $h$ indicates the thickness of the liquid region, i.e. \SI{0.3}{\mm}. The simulations are tested with a resolution of \textcolor{Reviewer1}{$N_x\times N_y \times N_z = 256\times 256 \times 1$.} The duration of the simulation is set to \SI{0.003}{\s}, allowing for sufficient number of time steps for the inconsistencies to develop. To highlight the dominant impacts of the convection, the surface tension force, gravity, and viscosity are excluded from the simulations. The periodic condition is employed for all boundaries. We compared our results with the results computed by the ONERA DYJEAT codes \citep{couderc2007developpement,ZUZIO2011339,ZUZIO2018285,Xavier2020toward,Zuzio2020}, which upholds consistent mass-momentum transport through solving the temporary density equations together with momentum equations on staggered meshes.  
 
\begin{figure}[!htb]
    \centering
    \includegraphics[width=.8\linewidth]{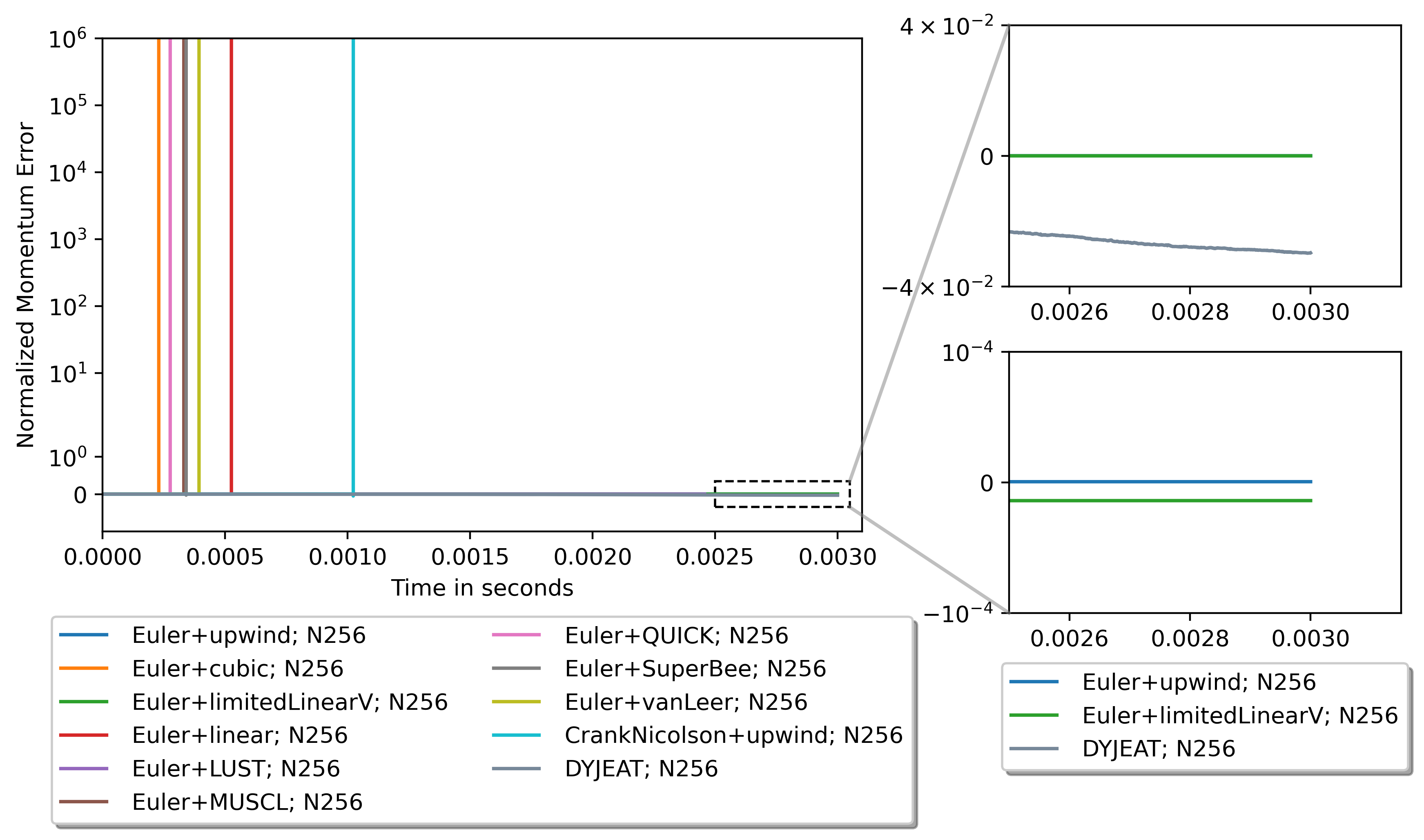}
    \caption{Time evolution of normalized momentum error of mixing layer with different schemes and DYJEAT codes, density ratio: $10^3$, resolution: \textcolor{Reviewer1}{$N=256$}.}
    \label{fig:MixingLayer2D_schemes_comp}
\end{figure}

The \cref{fig:MixingLayer2D_schemes_comp} provides a quantitative comparison among multiple schemes in \cref{table:test_schemes_combi} and with the DYJEAT code, focusing on the temporal evolution of the normalized momentum error evaluated using \cref{eq:sumMOM}. From the plot, it is evident that only \textcolor{Reviewer1}{two} cases using interIsoFoam remain stable, namely, as expected using Euler $+$ upwind, but also \textcolor{Reviewer1}{Euler $+$ limitedLinearV}. 
The zoomed-in view in \cref{fig:MixingLayer2D_schemes_comp} highlights the accuracy of the results for these three stable cases. The errors calculated from DYJEAT are larger than errors from stable cases using interIsoFoam, which are around $4\%$. As shown in the detail in \cref{fig:MixingLayer2D_schemes_comp}, the errors from consistent Euler $+$ upwind are minimal, with respect to another combination of schemes. 


\subsection{Validation of a single rising bubble}
In the present study, we investigate the performance of the proposed method by applying it to the simulation of a single bubble rising in a quiescent viscous liquid. To validate our approach, we adopt the configuration presented by \citet{Anjos2014}, who \textcolor{Tobi}{simplified} the rising bubble experiments originally conducted by \citet{Bhaga_weber_1981}. In their work, \citet{Anjos2014} select three distinct viscosities for comparative analysis. Our focus, in particular, lies on the cases, characterized by larger liquid viscosities. The specific cases correspond to a Morton number \textcolor{Reviewer2}{$Mo=g\nu_l^4/\rho_l\sigma^3=(848, 41.1, 1.31)$}, where $g$ represents the gravitational acceleration, and $\nu_l$, $\rho_l$, and $\sigma$ denote the viscosity, density of the ambient liquid, and surface tension coefficient, respectively. Three resolutions are tested: $N\in(64, 96, 128)$.

For our simulations, we initialize the air bubble with an idealized spherical shape, possessing a diameter of $D=\SI{2.61}{\cm}$. The air properties are defined by a viscosity of $1.78\times\SI[per-mode=symbol]{e-5}{\kilogram\per\metre\per\second}$ and a density of $\SI[per-mode=symbol]{1.225}{\kilogram\per\cubic\metre}$, while the liquid properties encompass viscosities of \textcolor{Reviewer2}{$(2.73, 1.28, 0.54)\,\si[per-mode=symbol]{\kilogram\per\metre\per\second}$} and a density of $\SI[per-mode=symbol]{1350}{\kilogram\per\cubic\metre}$. Furthermore, the surface tension between the air bubble and the liquid is $\textcolor{Reviewer2}{\SI{0.078}{\N\per\meter}}$. The computational domain is defined as $(-4D,-4D,-2D)\times(4D,4D,6D)$, where the positions of the space diagonal vertices of the computational domain are delineated, with the initial position of the bubble set at the origin, $(0,0,0)$. A set of dimensionless normalized variables is introduced as follows:

\begin{equation}
    \mathbf{w}=\frac{\v}{\sqrt{gD}},\ t=\sqrt{\frac{g}{D}}\tau,
\end{equation}
where $\tau$ indicates the \textcolor{Tobi}{physical time in seconds}.

\begin{figure}[!htb]
     \begin{subfigure}{.48\textwidth}
      \centering
      \includegraphics[width=\linewidth]{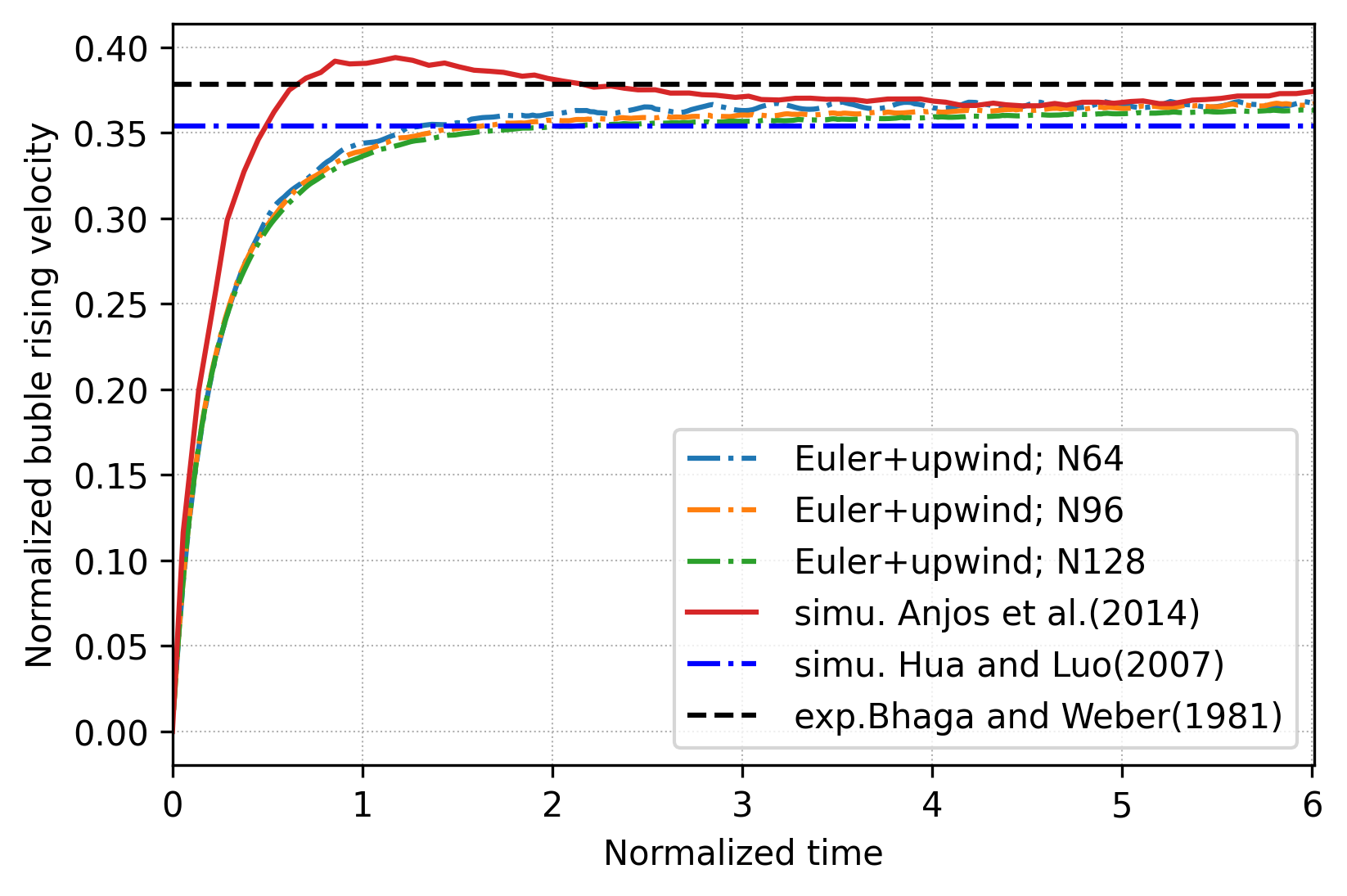}
      \caption{interIsoFoam: Mo$=848$}
      \label{fig:risingBubble_isoAdv_848}
     \end{subfigure}    
     \hfill
     \begin{subfigure}{.48\textwidth}
      \centering
      \includegraphics[width=\linewidth]{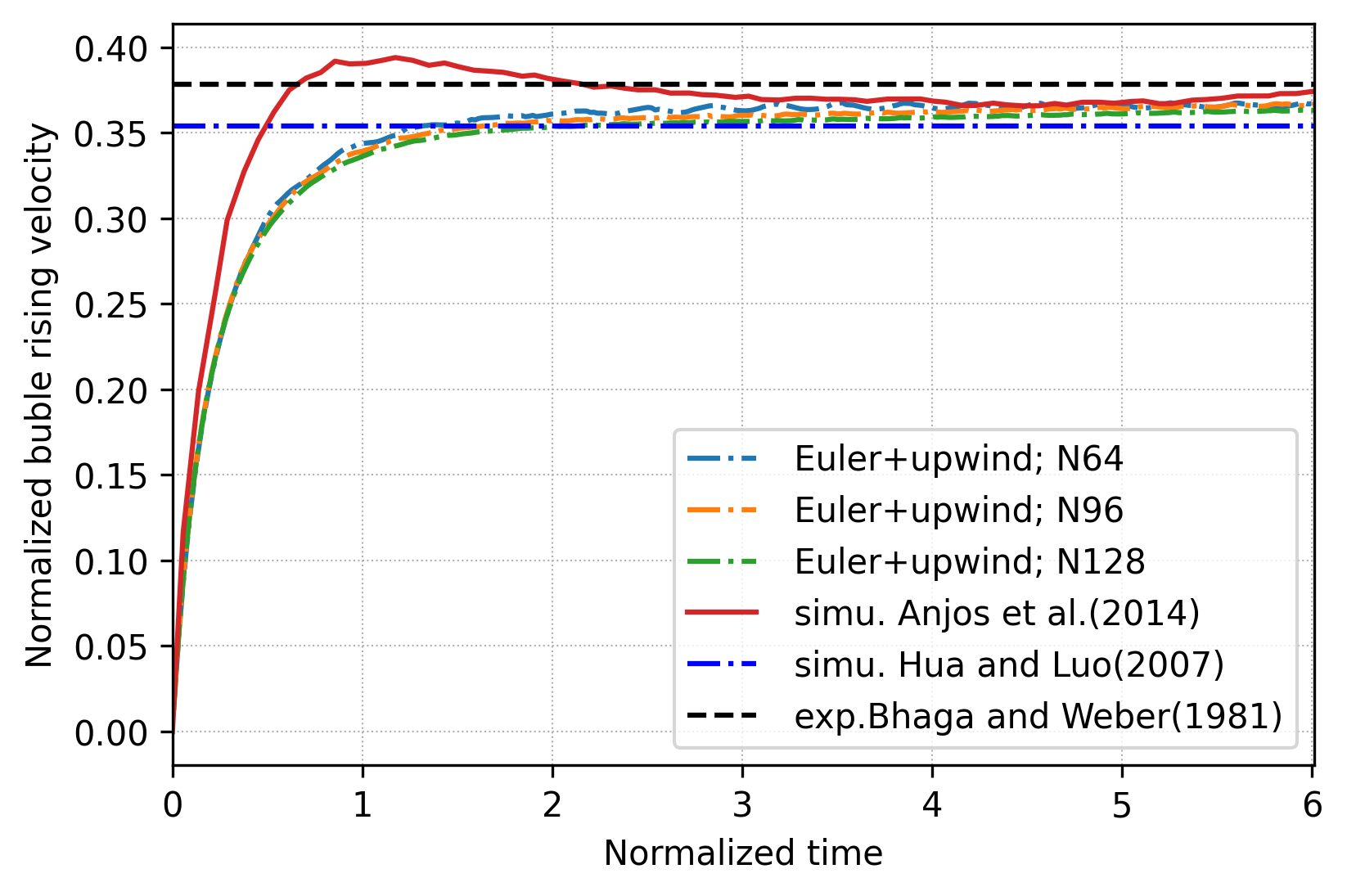}
      \caption{interIsoRhoFoam: Mo$=848$}
      \label{fig:risingBubble_isoRho_848}
     \end{subfigure} 
     \begin{subfigure}{.48\textwidth}
      \centering
      \includegraphics[width=\linewidth]{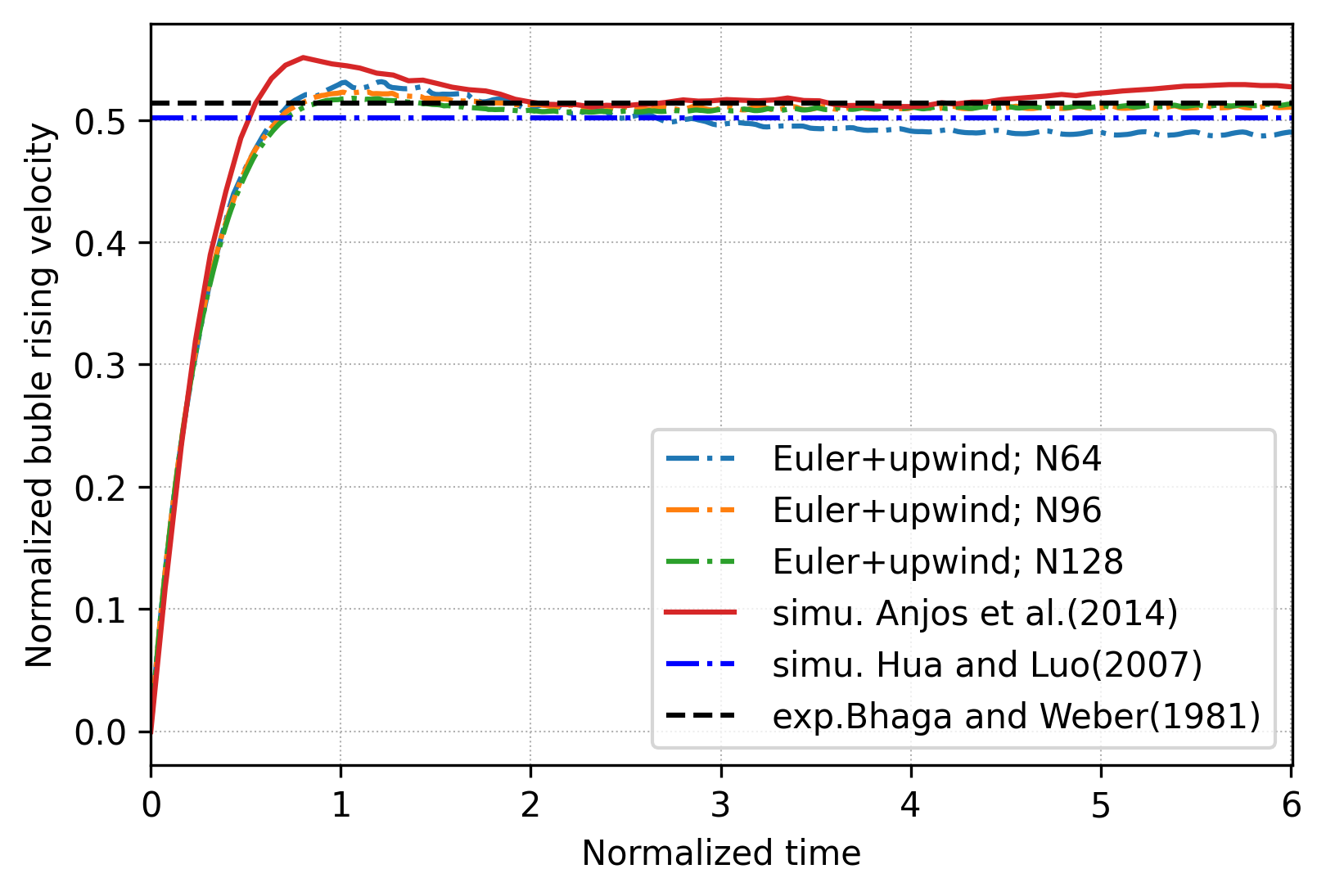}
      \caption{interIsoFoam: Mo$=41.4$}
      \label{fig:risingBubble_isoAdv_41d1}
     \end{subfigure}
     \hfill
     \begin{subfigure}{.48\textwidth}
      \centering
      \includegraphics[width=\linewidth]{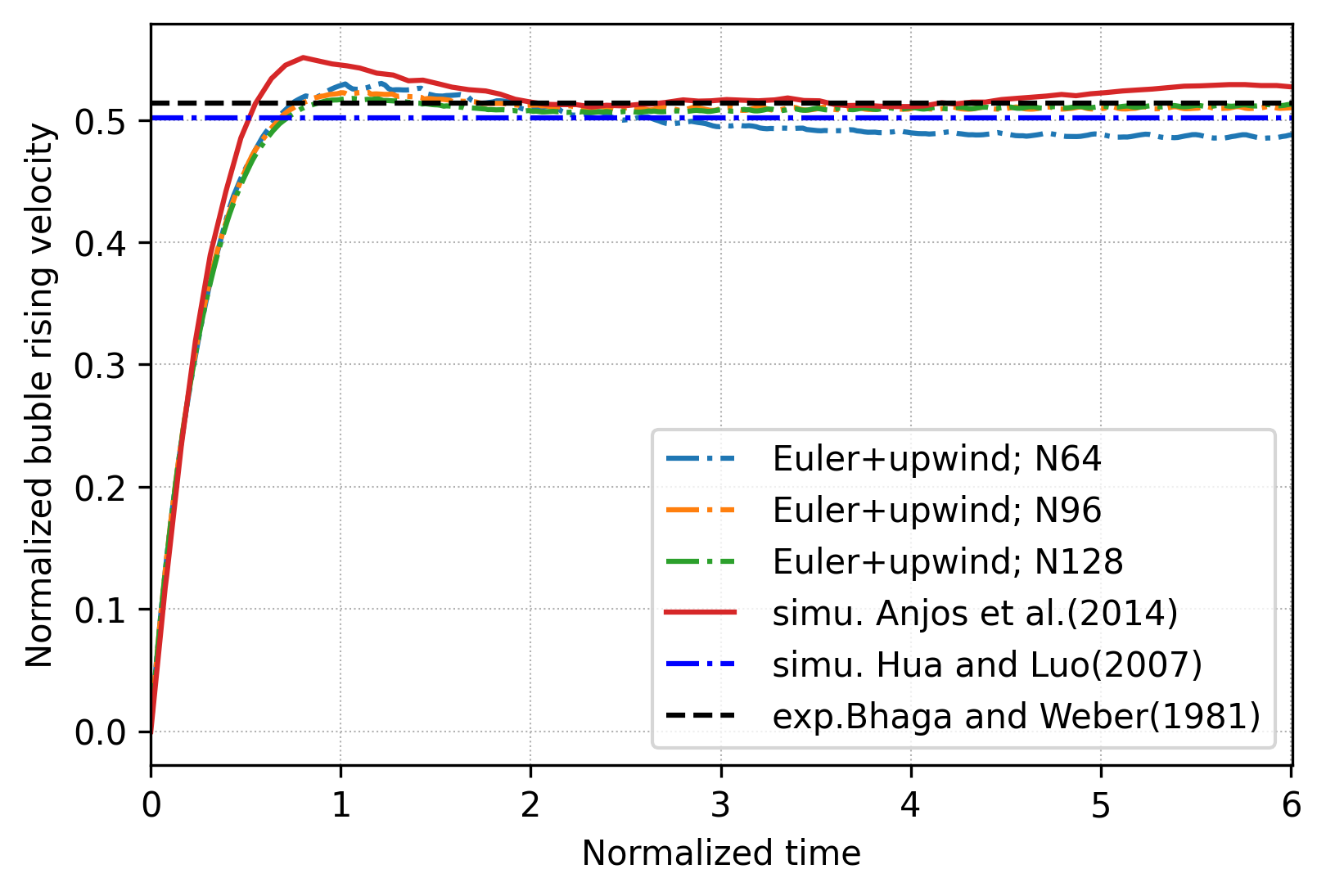}
      \caption{interIsoRhoFoam: Mo$=41.4$}
      \label{fig:risingBubble_isoRho_41d1}
     \end{subfigure}
     \begin{subfigure}{.48\textwidth}
      \centering
      \includegraphics[width=\linewidth]{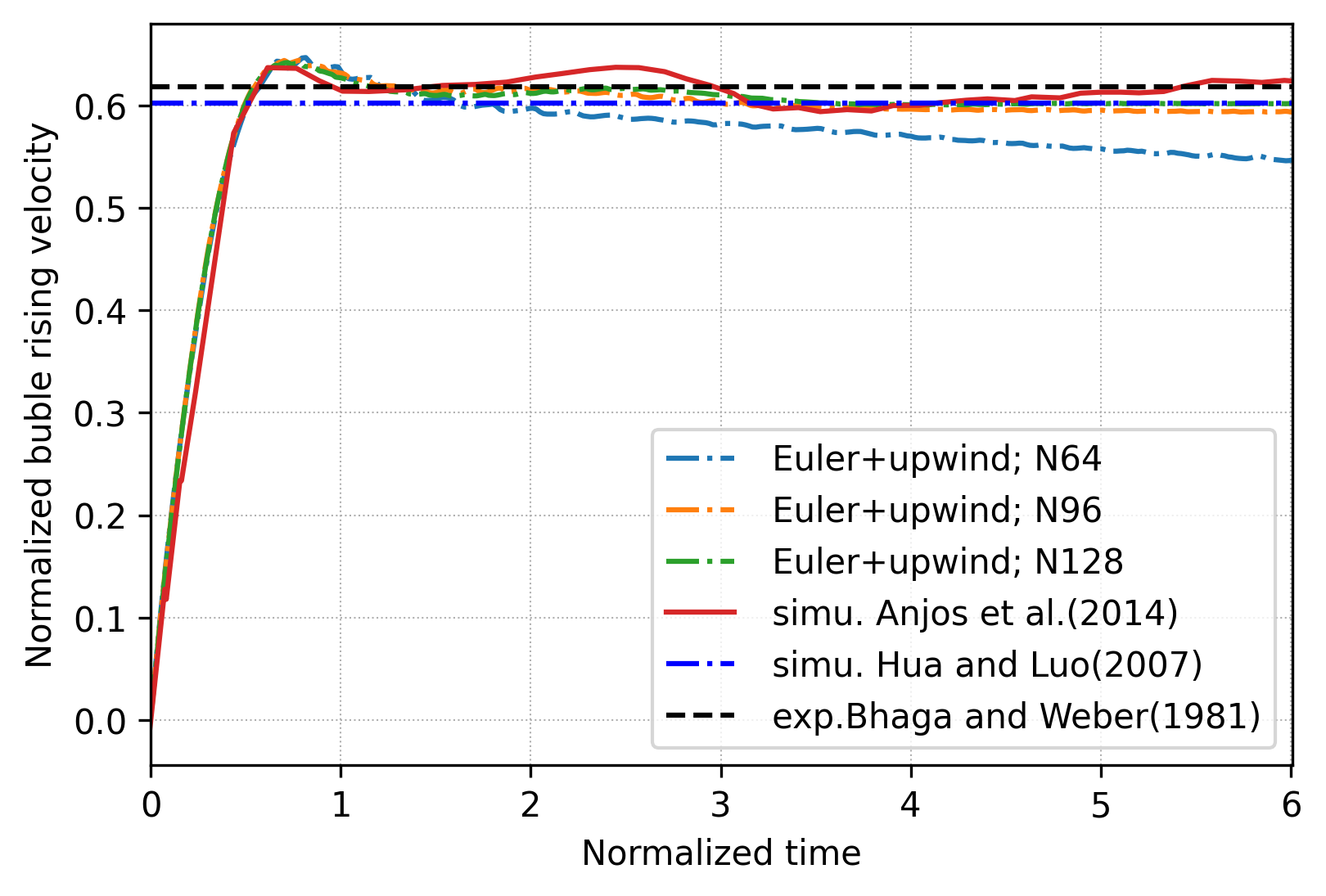}
      \caption{\textcolor{Reviewer2}{interIsoFoam: Mo$=1.31$}}
      \label{fig:risingBubble_isoAdv_1d31}
     \end{subfigure}
     \hfill
     \begin{subfigure}{.48\textwidth}
      \centering
      \includegraphics[width=\linewidth]{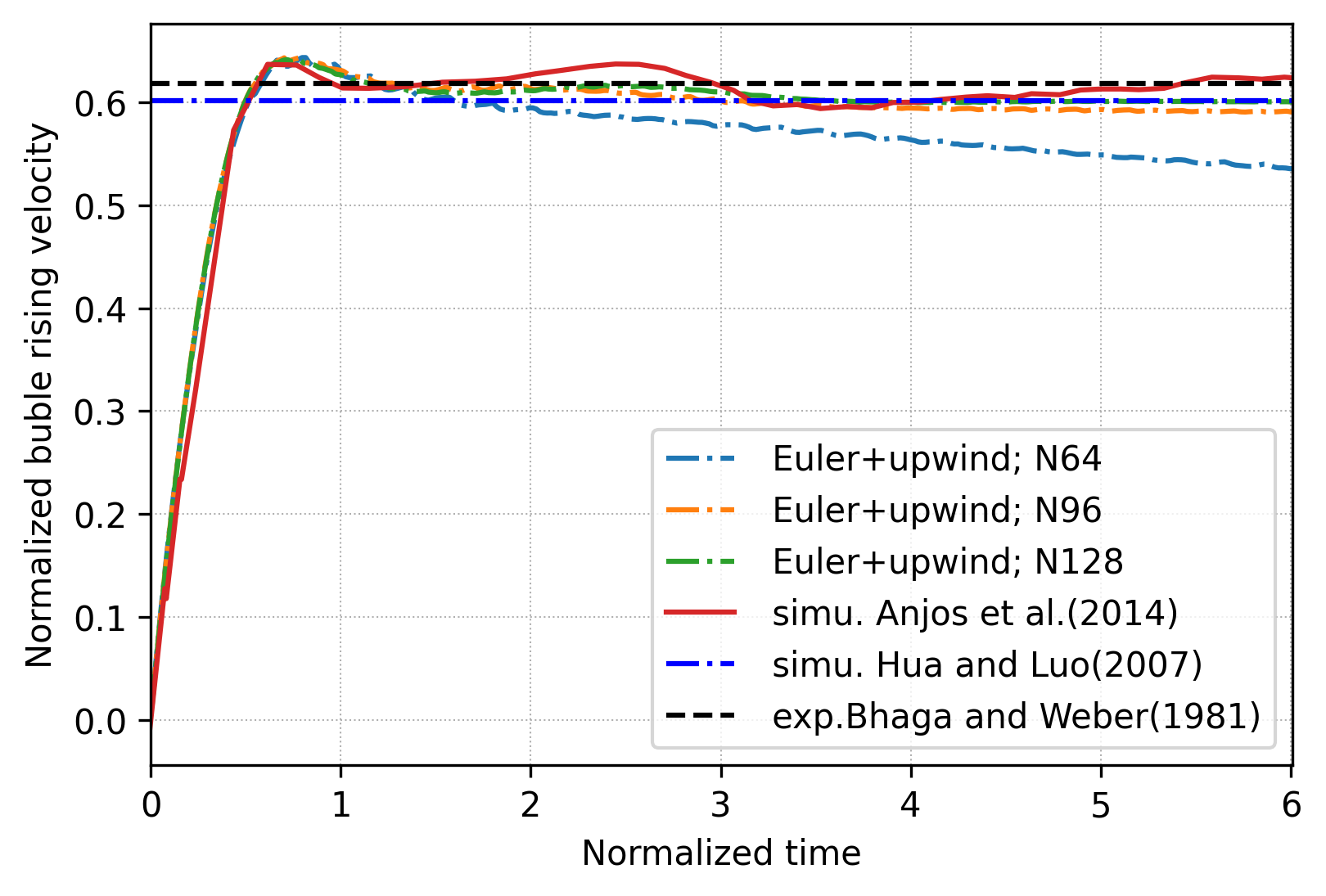}
      \caption{\textcolor{Reviewer2}{interIsoRhoFoam: Mo$=1.31$}}
      \label{fig:risingBubble_isoRho_1d31}
     \end{subfigure} 
    \caption{Temporal evolution of rising velocity using interIsoFoam and interIsoFoam: Euler $+$ upwind, $\rho^-/\rho^+\approx10^3$.}
    \label{fig:risingBubble_comp_risingVelocity}
\end{figure}

\begin{figure}[!htb]
     \begin{subfigure}{.48\textwidth}
      \centering
      \includegraphics[width=\linewidth]{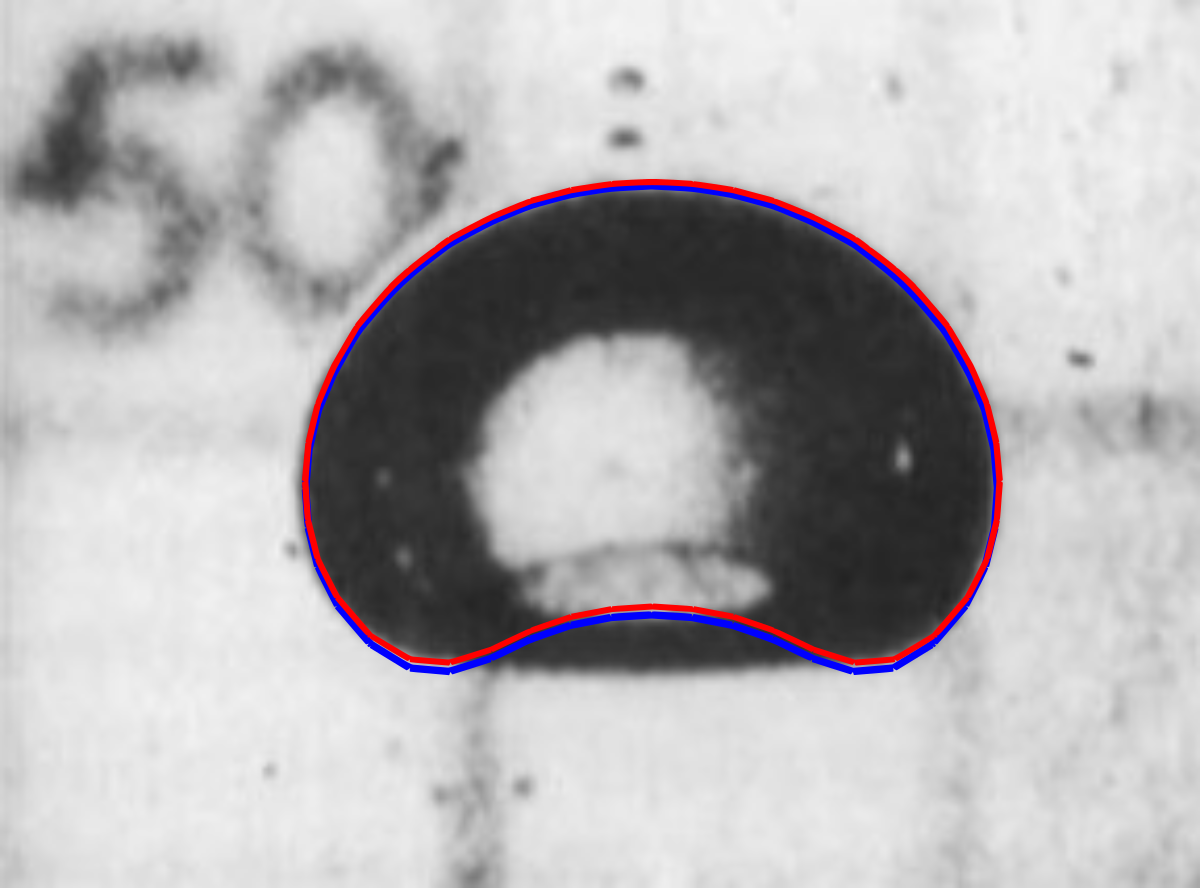}
     \caption{\textcolor{Reviewer1}{Mo$=848$}}
      \label{fig:risingBubble_shape_isoAdv_848}
     \end{subfigure}
     \hfill
     \begin{subfigure}{.45\textwidth}
      \centering
      \includegraphics[width=\linewidth]{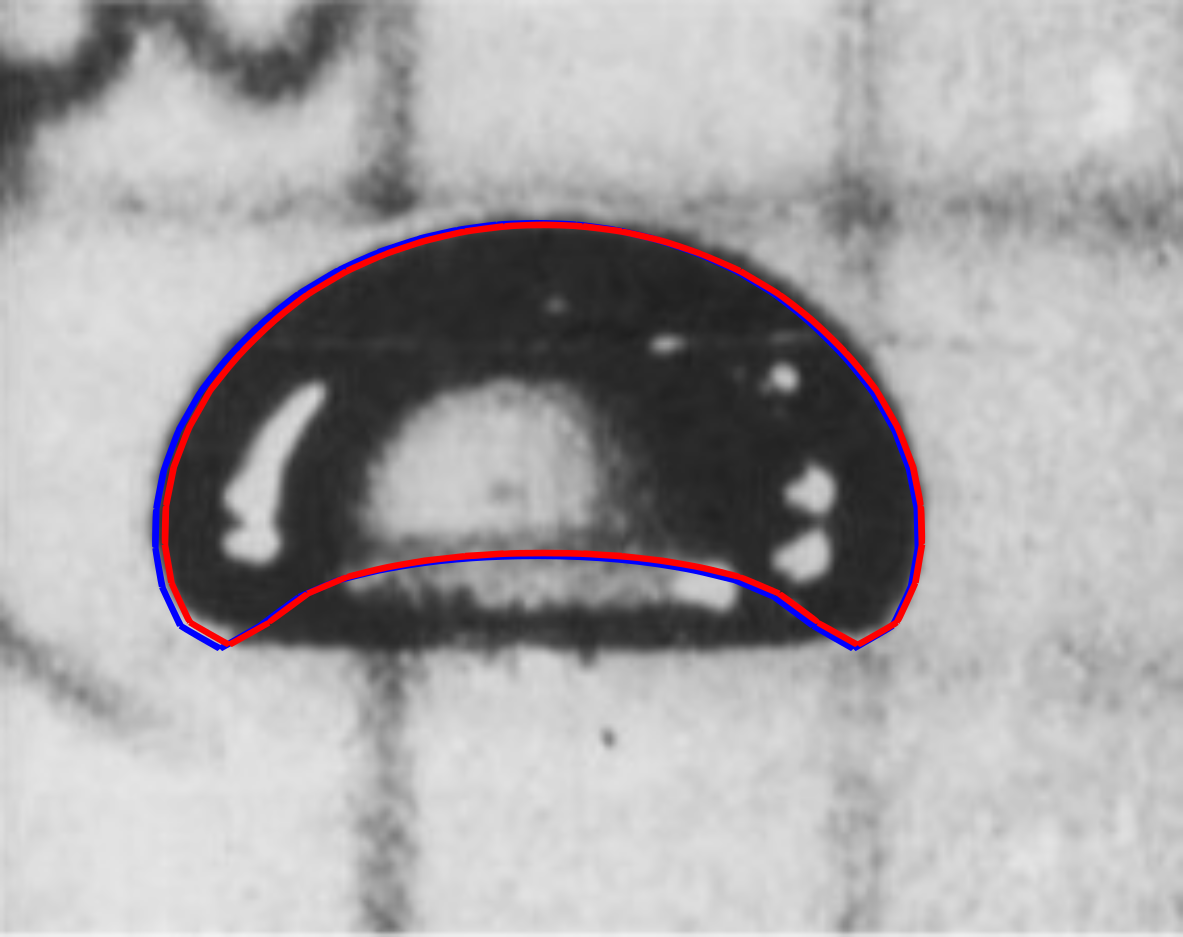}
      \caption{\textcolor{Reviewer1}{Mo$=41.4$}}
      \label{fig:risingBubble_shape_isoAdv_41d1}
     \end{subfigure}
      \begin{subfigure}{.45\textwidth}
      \centering
      \includegraphics[width=\linewidth]{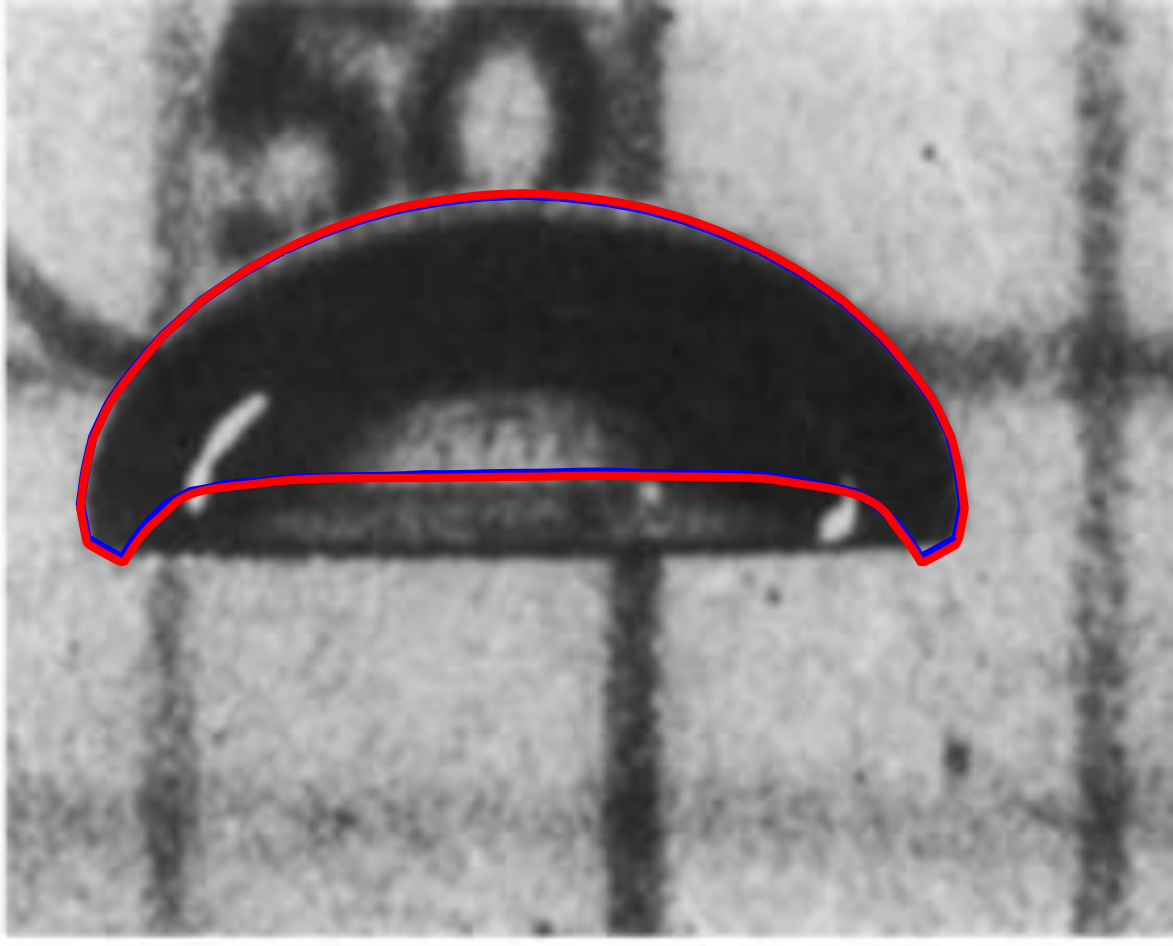}
     \caption{\textcolor{Reviewer1}{Mo$=1.31$}}
      \label{fig:risingBubble_shape_isoAdv_1d31}
     \end{subfigure}
    \caption{\textcolor{Reviewer1}{Comparisons of final shapes of rising bubbles using interIsoFoam and interIsoRhoFoam with the experimental visualization from \citet{Bhaga_weber_1981} (reprinted with permission): Euler $+$ upwind, red line from interIsoRhoFoam, blue line from interIsoFoam, $\rho^-/\rho^+\approx10^3$, $N=128$, $t=6$.}}
    \label{fig:risingBubble_comp_shape}
\end{figure}

As illustrated in \cref{fig:risingBubble_comp_risingVelocity}, the utilization of the Euler and upwind schemes ensures the preservation of equivalence between volume fraction and mass advection equation. Consequently, the results obtained from interIsoFoam and interIsoRhoFoam, when employing the identical Morton number, display a substantial level of similarity.  When considering Mo $=41.1$, as demonstrated in \cref{fig:risingBubble_isoAdv_41d1} and \cref{fig:risingBubble_isoRho_41d1}, the velocities acquired from both solvers demonstrate an initial increase until approximately $t=1$, followed by a subsequent decline leading to a stable state. Notably, the differences among the results become evident when dealing with the cases involving Mo $=41.1$. Specifically, in instances where a coarse mesh ($N=64$) is employed, the velocity decline is more pronounced, resulting in a smaller final stable velocity compared to the values presented in prior works \citep{Bhaga_weber_1981, Hua_2007, Anjos2014}. \textcolor{Reviewer2}{The larger final velocity disparity from the previous works, can be observed for Mo $=1.31$ with a mesh resolution $N=64$ in \cref{fig:risingBubble_isoAdv_1d31,fig:risingBubble_isoRho_1d31}, which is resulted from that the rising bubble with lower viscosities has a stronger deformation and has a thinner structure required to be captured. Using a coarse mesh loses this information and subsequently causes a larger error.} Conversely, higher resolutions yield stable velocities that agree well with the experimental data documented by \citet{Bhaga_weber_1981,Hua_2007}.  On the other hand, for cases with Mo $=848$, characterized by a larger viscosity, the impact of resolution is less conspicuous, as depicted in \cref{fig:risingBubble_isoAdv_848} and \cref{fig:risingBubble_isoRho_848}. The final velocities attained from different resolutions converge similarly to values that fall between the results obtained in the simulation conducted by \citet{Hua_2007} and the experimental study conducted by \citet{Bhaga_weber_1981}. 

Furthermore, the profiles obtained by slicing the surfaces of the droplets, passing through their centers, are compared with the experimental profiles from \citet{Bhaga_weber_1981}, as illustrated in \cref{fig:risingBubble_comp_shape}. For both solvers, the droplets with varying viscosities exhibit final shapes that closely align with the experimental visualizations, validating the hypothesis of the equivalence between Euler+upwind discretization of \cref{eq:momentum-transport} using the scaled mass flux $|V_f^\alpha|_s$ from \cref{eq:massfluxisoadvector}, and the solution of the auxiliary density equation using \cref{alg:interIsoRhoFoam}.

\subsection{Liquid jet in high speed gaseous cross-flow}\label{sec:jet-cross-flow}

\begin{figure}[H]
    \centering
    \def\svgwidth{0.6\textwidth}
    \footnotesize
    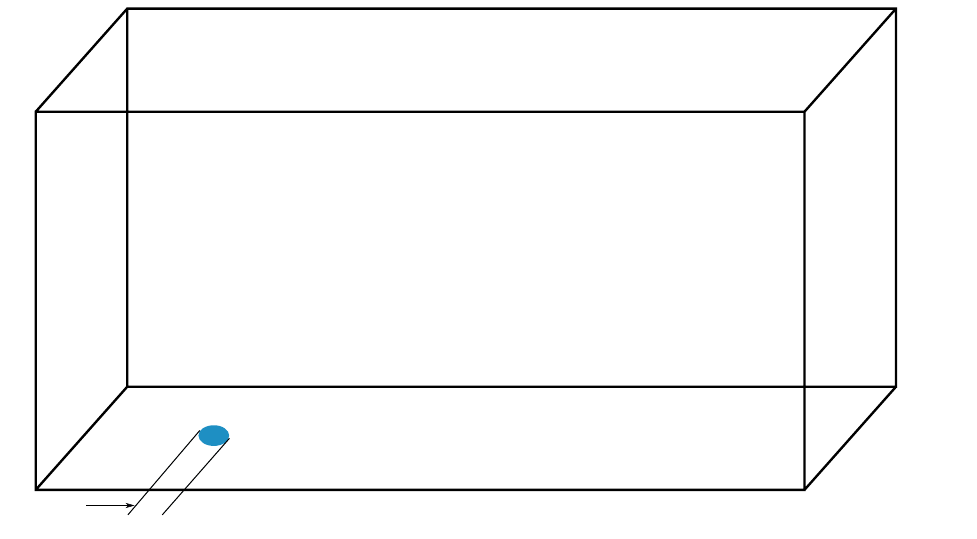%
    \caption{Liquid in a cross flow.}
    \label{fig:liquid-cross-flow}
\end{figure}

Different from the parallel velocities of the mixing layers in \cref{sec:mixing-layer-2D}, the liquid flows with a lower velocity is perpendicular to the velocity of the gaseous phase in this case, which is called the injection of a liquid jet in a gaseous cross-flow (LJCF) and is common in many engineering applications. The geometry and the physical properties are configured by referring to \citet{Zuzio2020}. The \textcolor{Tobi}{rectangular} computational domain $\Omega:[-0.01,0,-0.01]\times[0.03,0.02,0.01]\,\si{\m}$ has two inlets. The gas  flows in with a velocity $\v^+=[65,0,0]\,\si[per-mode=symbol]{\m\per\s}$ from the left boundary $x_{min}$. \citet{Thuillet2019} revealed the impact of the liquid inlet velocity profile on the jet trajectory. He simulated the jet with an uniform liquid inlet velocity profile, and with a velocity profile calculated through simulating the injector. The jet trajectory results from the case with calculated velocity profile showed better agreement with the experiment. We followed the calculated liquid injected velocity profile from \citep{Thuillet2019}, which is
\begin{equation*}
    \v^-_y = -21.434\left(\frac{r}{d}\right)^3 + 15.512\left(\frac{r}{d}\right)^2 + 8.6504
\end{equation*}
, where $\v^-_y$ is the $y-$component of the liquid inlet velocity $\v^-$, $r$ indicates the distance to the nozzle center, and $d$ is the diameter of the nozzle. The $x$, $z-$component of $\v^-$ are set to zero, whereas $d=\SI{0.002}{\m}$.
The nozzle's center locates at $[0,0,0]$ in the bottom boundary patch $y_{min}$. To save the computation resource, the uniform Cartesian mesh with a moderate resolution of $[N_x,N_y,N_z]=[128,64,64]$ has been adopted. \Cref{fig:liquid-cross-flow} depicts the flow domain. The physical properties are $\rho^-=\SI[per-mode=symbol]{1000}{\kilogram\per\cubic\metre}$, $\rho^+=\SI[per-mode=symbol]{1.225}{\kg\per\m^3}$, $\mu^-=1.0\times  \SI[per-mode=symbol]{e-3}{\kilogram\per\metre\per\second}$, $\mu^+=1.78\times  \SI[per-mode=symbol]{e-5}{\kilogram\per\metre\per\second}$, $\sigma = \textcolor{Reviewer2}{7.2\times\SI{e-2}{\N\per\meter}}$, $g=\SI[per-mode=symbol]{9.81}{\meter\per\square\second}$. 

\begin{figure}[!htb]
     \begin{subfigure}[t]{\textwidth}
      \centering
      \includegraphics[width=.49\linewidth]{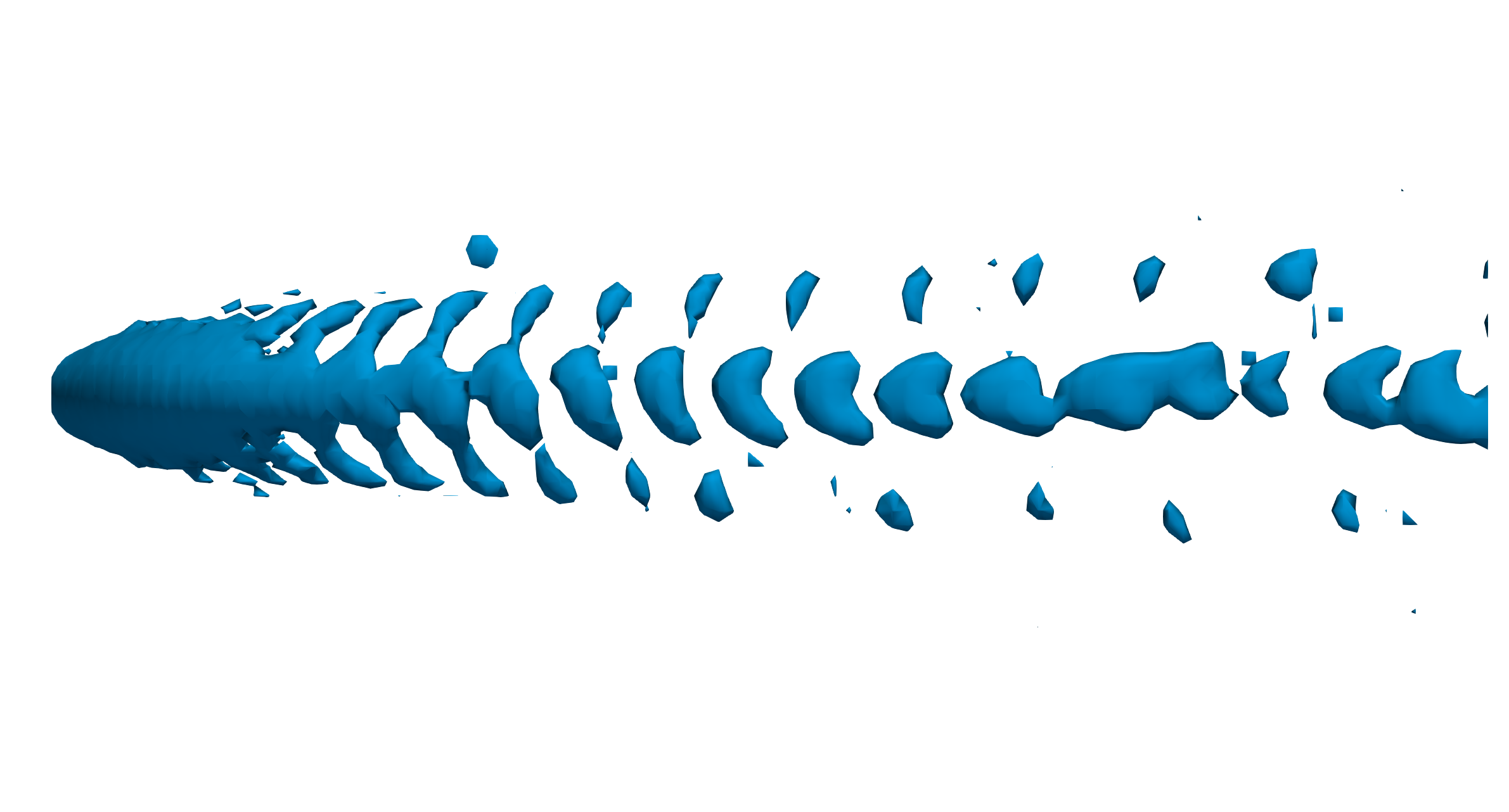}
      \hfill
      \includegraphics[width=.49\linewidth]{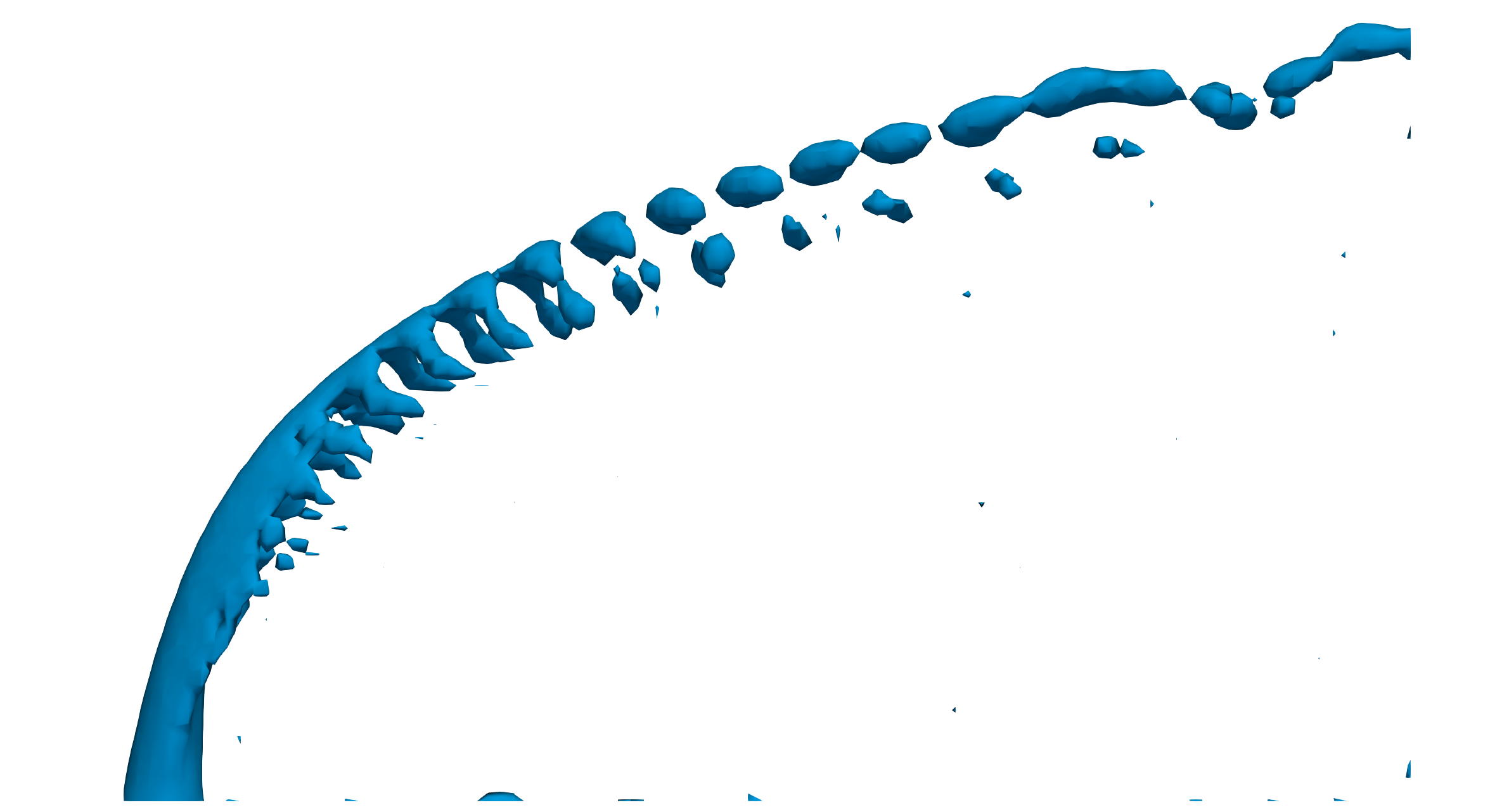}
      \caption{interIsoFoam: $t=\SI{7.1}{\ms}$}
      \label{fig:LJCF_interIso}
     \end{subfigure}
     \hfill
     \begin{subfigure}[t]{\textwidth}
      \centering
      \includegraphics[width=.49\linewidth]{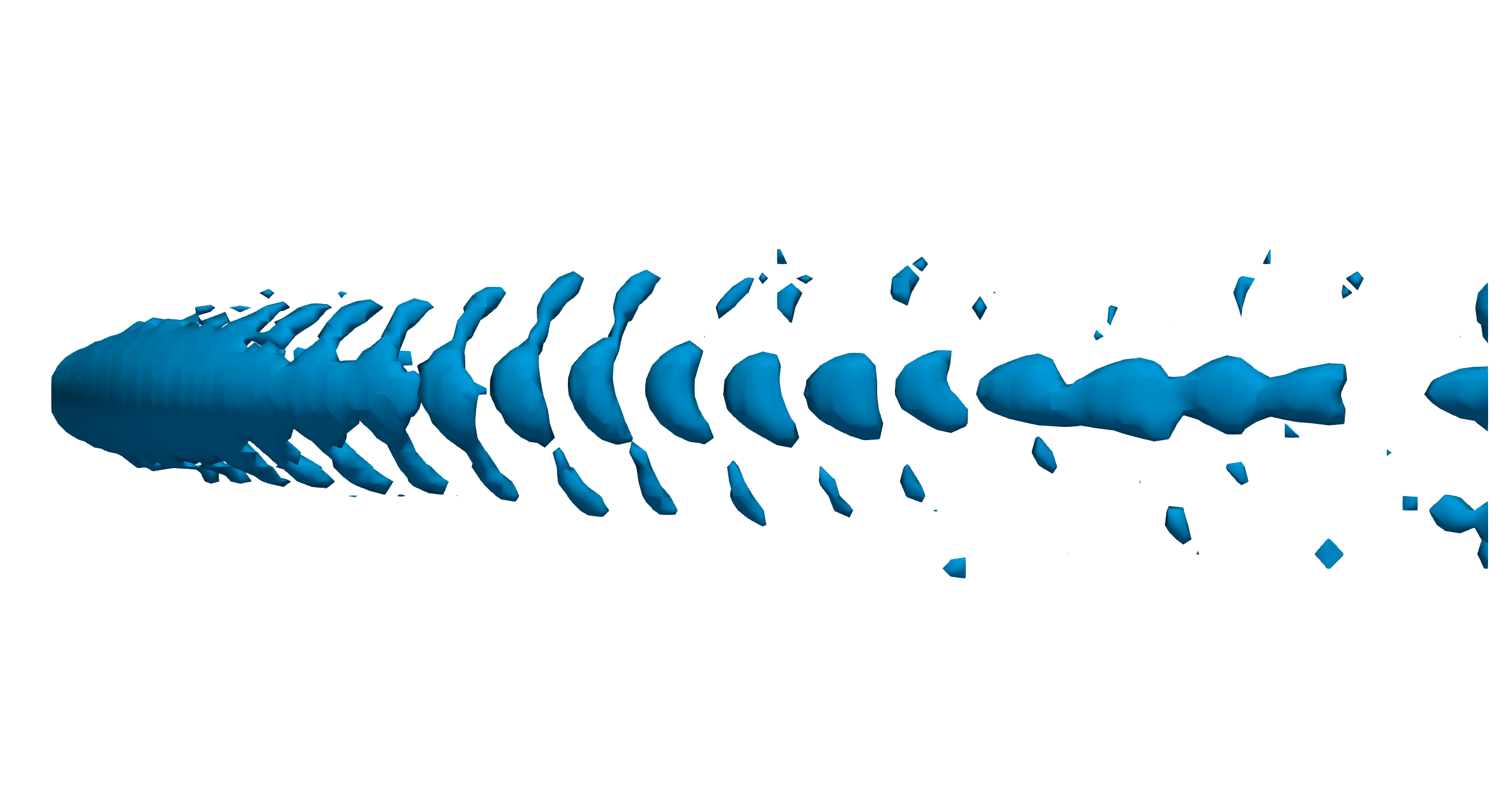}
      \hfill
      \includegraphics[width=.49\linewidth]{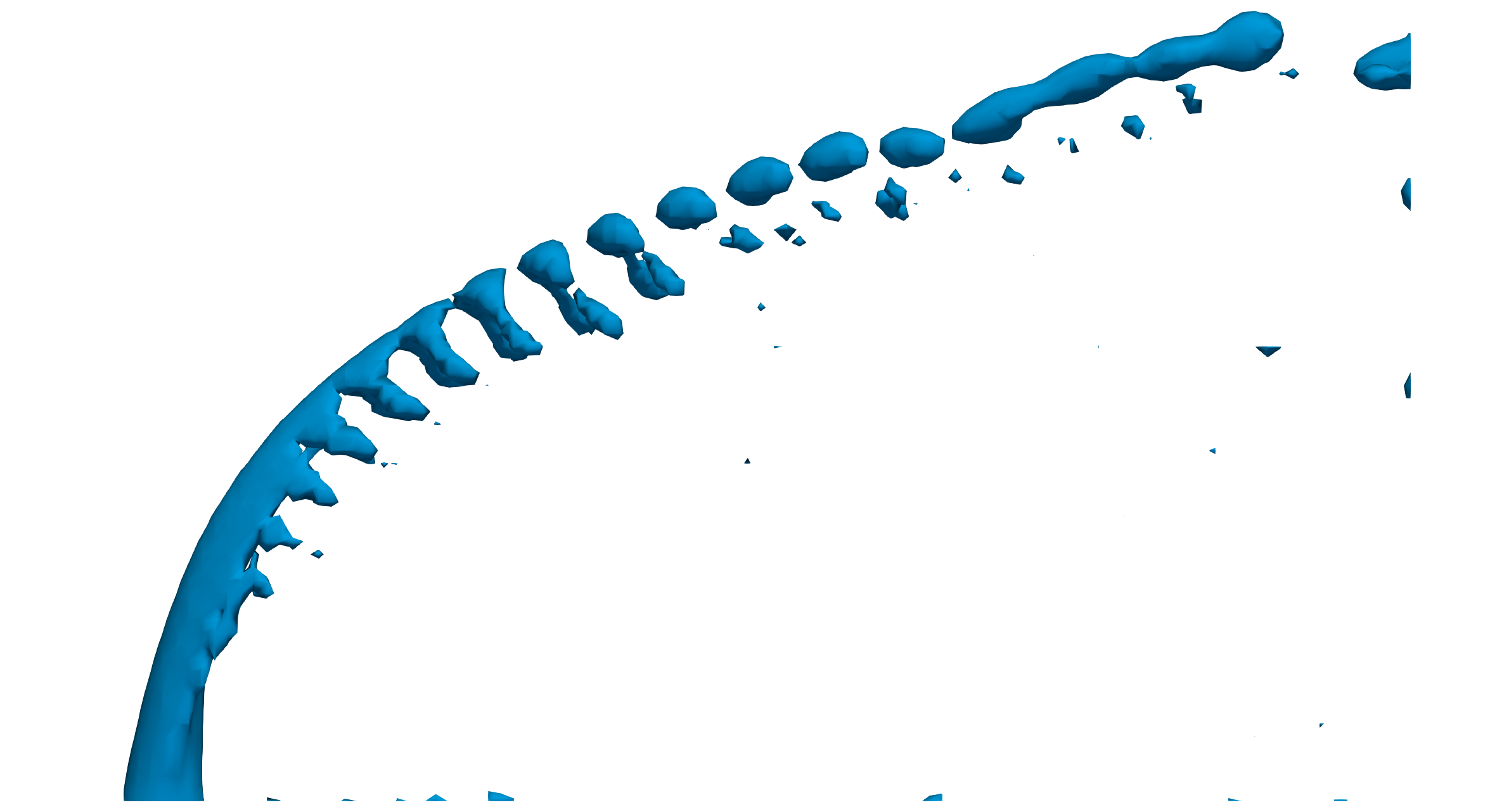}
      \caption{interIsoRhoFoam: $t=\SI{7.1}{\ms}$}
     \label{fig:LJCF_interIsoRho}
     \end{subfigure}   
     \hfill
     \begin{subfigure}[t]{\textwidth}
      \centering
      \includegraphics[width=.49\linewidth]{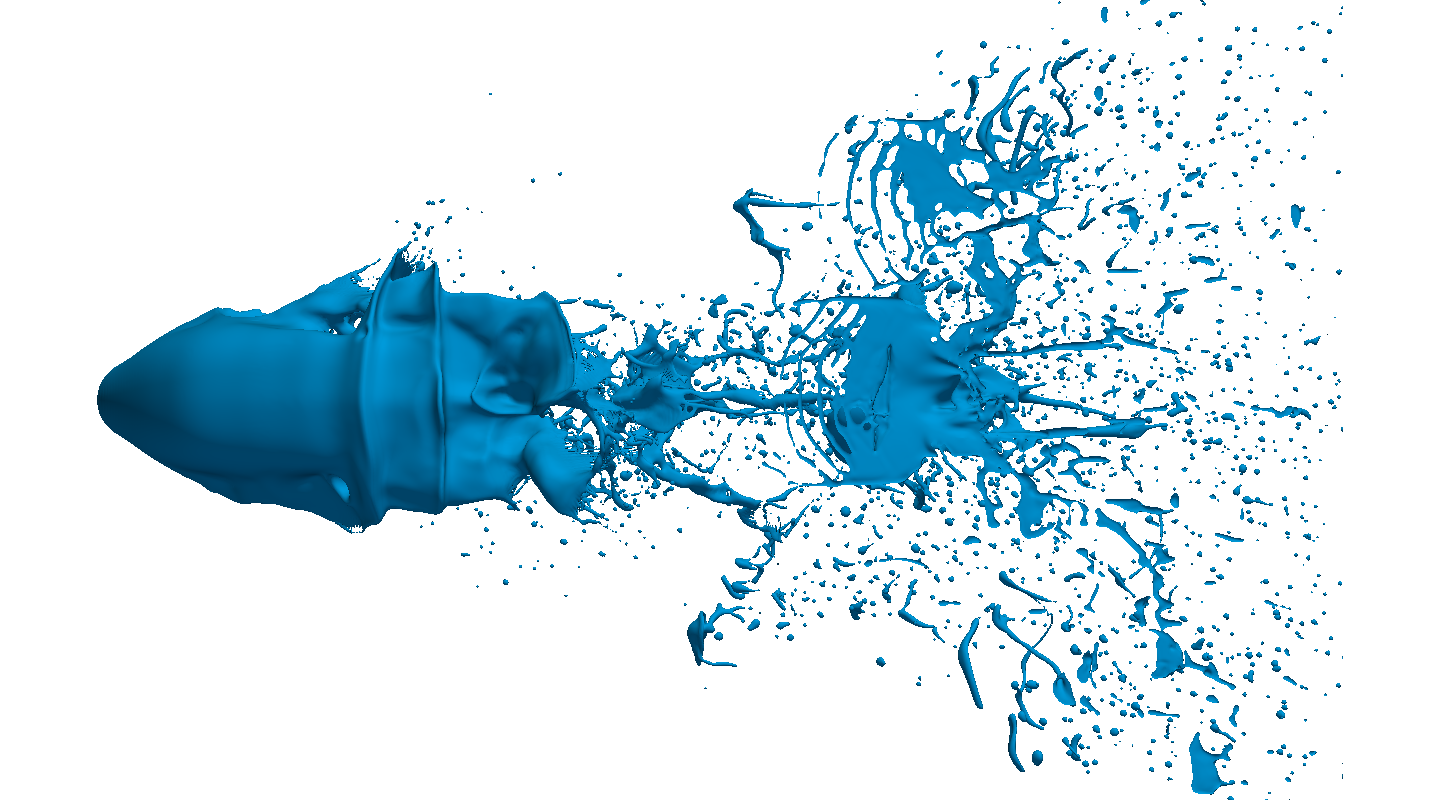}
      \hfill
      \includegraphics[width=.49\linewidth]{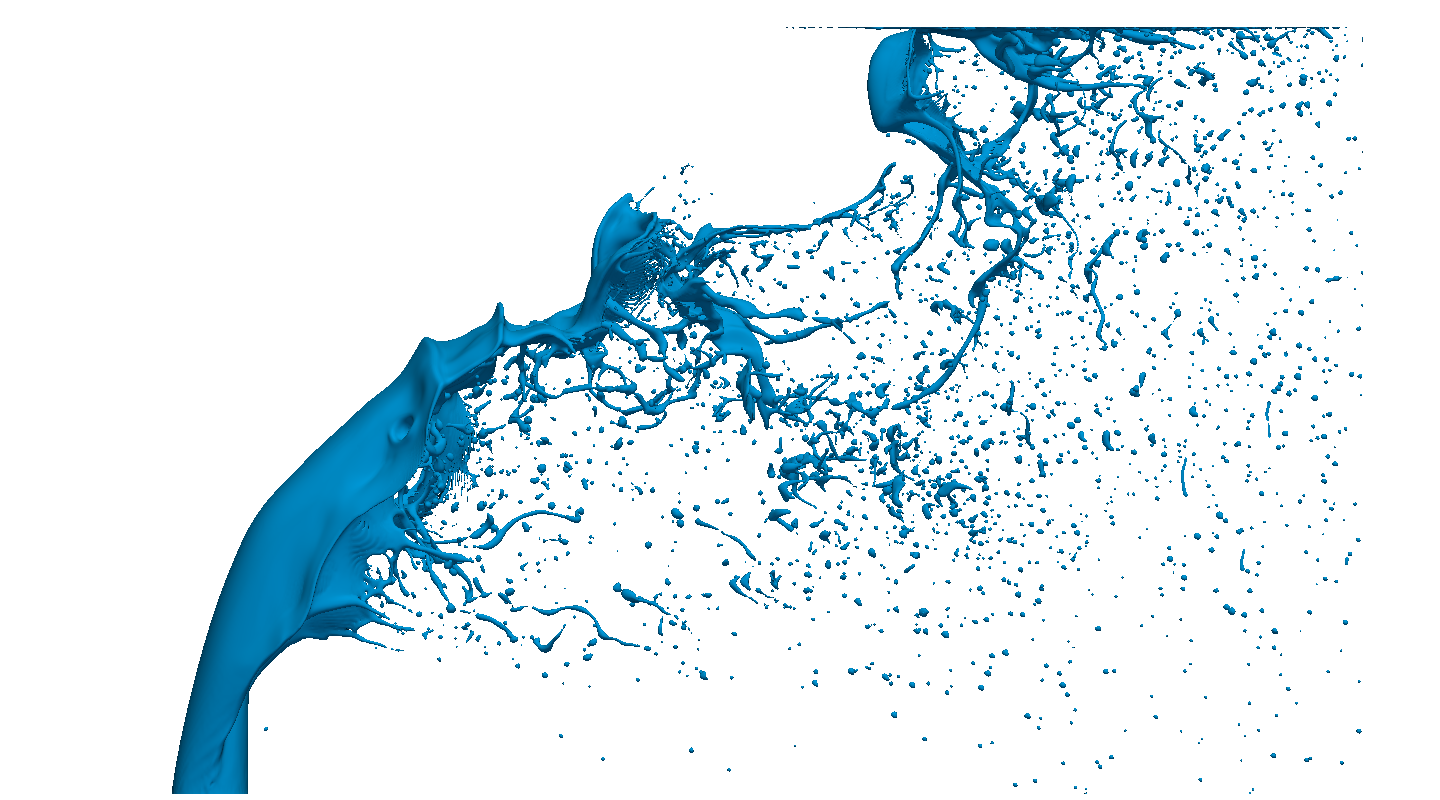}
      \caption{DYJEAT: $t=\SI{7.07}{\ms}$}
      \label{fig:LJCF_DYJEAT}
     \end{subfigure} 
    \caption{\textcolor{Davide}{The shape of the injected liquid with interIsoFoam and interIsoRhoFoam (Euler and Gauss upwind, density ratio: $816$, CFL number: $CFL=0.2$, resolution: $N_l=[128,64,64]$), and with DYJEAT (resolution: $N=[1024, 512, 512]$ \citep{Zuzio2020})}}
    \label{fig:LJCF_contour}
    
\end{figure}

\Cref{fig:LJCF_contour} shows the final state of the injected liquid at $t= \SI{7.1}{\ms}$ using interIsoFoam and interIsoRhoFoam with Euler and Gauss upwind regarding the iso-value of the reconstructed distance function $RDF=0$, as well as using DYJEAT \citep{Zuzio2020} with a high resolution of $[N_x,N_y,N_z]=[1024, 512, 512]$, whose shape is rendered by the iso-value $0.5$ of the volume fraction. \Cref{fig:LJCF_interIso} and \cref{fig:LJCF_interIsoRho} illustrate the ruptured liquid jet from the $y$-side view. The droplets' distributions of interIsoFoam and interIsoRhoFoam display many comparable characteristics. We can observe two strips of droplets and a strip of bag-like liquid structure. The outer liquid segregates into two $yz$-plane symmetric strips of droplets at an early stage, i.e., at a low penetration height, as shown in the right subfigure of \cref{fig:LJCF_interIso} and \cref{fig:LJCF_interIsoRho}. These droplets translate with the gas along the $x$-direction and spread spanwise in the $y$-direction. The remaining center liquid has a wavelike detachment and forms the bag-like structure in the middle of the strips. \Cref{fig:LJCF_DYJEAT} demonstrates the results from DYJEAT codes with a higher resolution. Similar liquid distributions can be observed: the liquid in the center zone of the nozzle propagates like a wave and breaks up into some large packets at a higher position, whereas the liquid in the periphery of the nozzle zone rips at a low penetration height. \textcolor{Reviewer1}{We also tested interIsoFoam with unstable schemes, and they fail. An example combination of Euler + cubic, is shown in \cref{fig:LJCF_Ecubic}, with the simulations on both coarser and finer mesh fail catastrophically.}

This validation case is a candidate for a benchmark case for validating two-phase flow numerical methods that consistently handle high density ratios, because the experimental form of the jet can be accurately reproduced on coarser mesh resolutions. Using a coarser mesh resolution interIsoFoam and interIsoRhoFoam of course do not capture the small structures such as liquid streaks, sacs and droplets, as shown in DYJEAT's results. However, the solvers accurately predict the jet curve, which can be used as a quantifiable argument for validity of a consistent method against experimental data. 

\begin{figure}[!htb]
     \begin{subfigure}[t]{.45\textwidth}
      \centering
      \includegraphics[width=\linewidth]{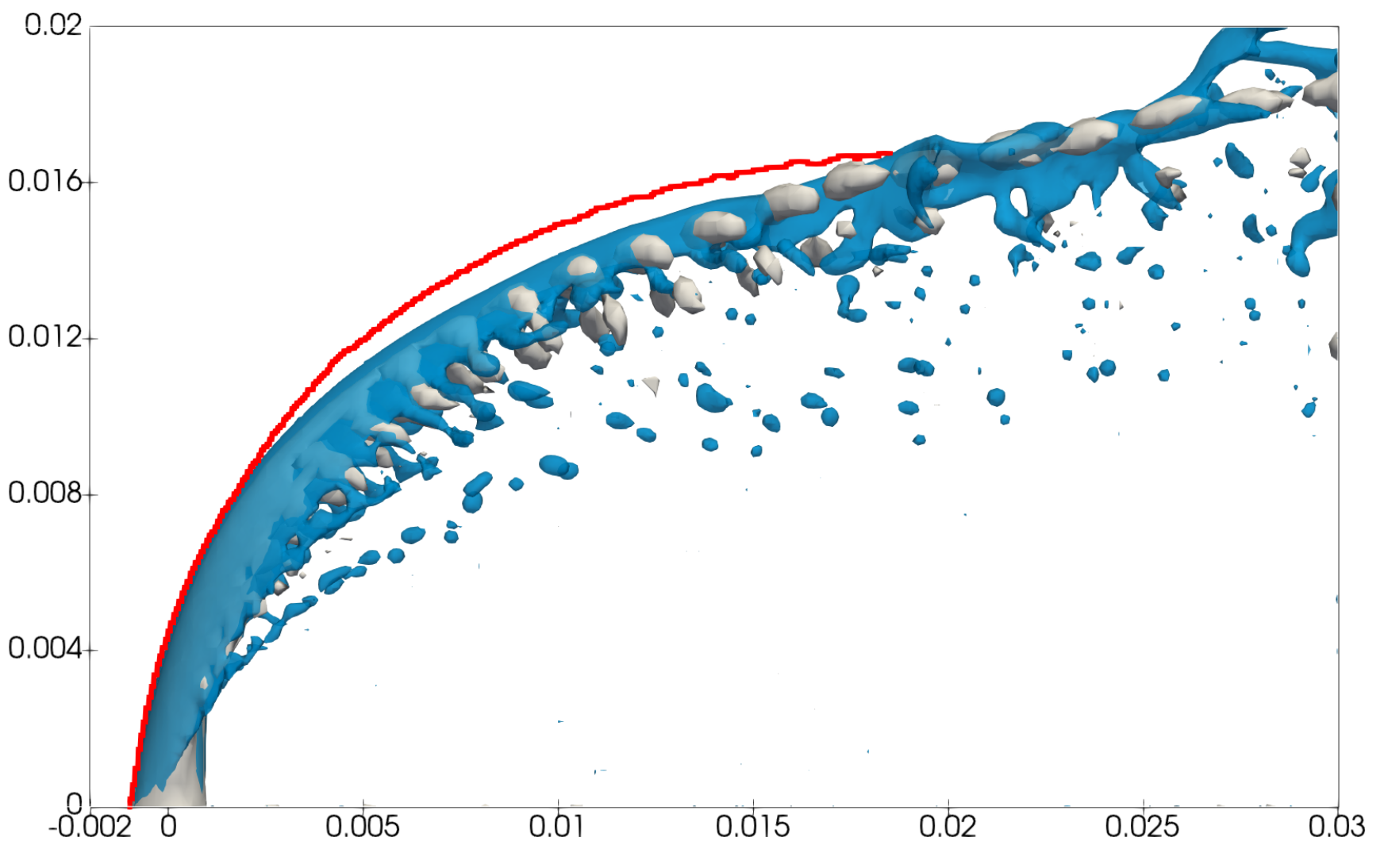}
      \caption{interIsoFoam}
      \label{fig:LJCF_interIso_comp}
     \end{subfigure}
     \begin{subfigure}[t]{.45\textwidth}
      \centering
      \includegraphics[width=\linewidth]{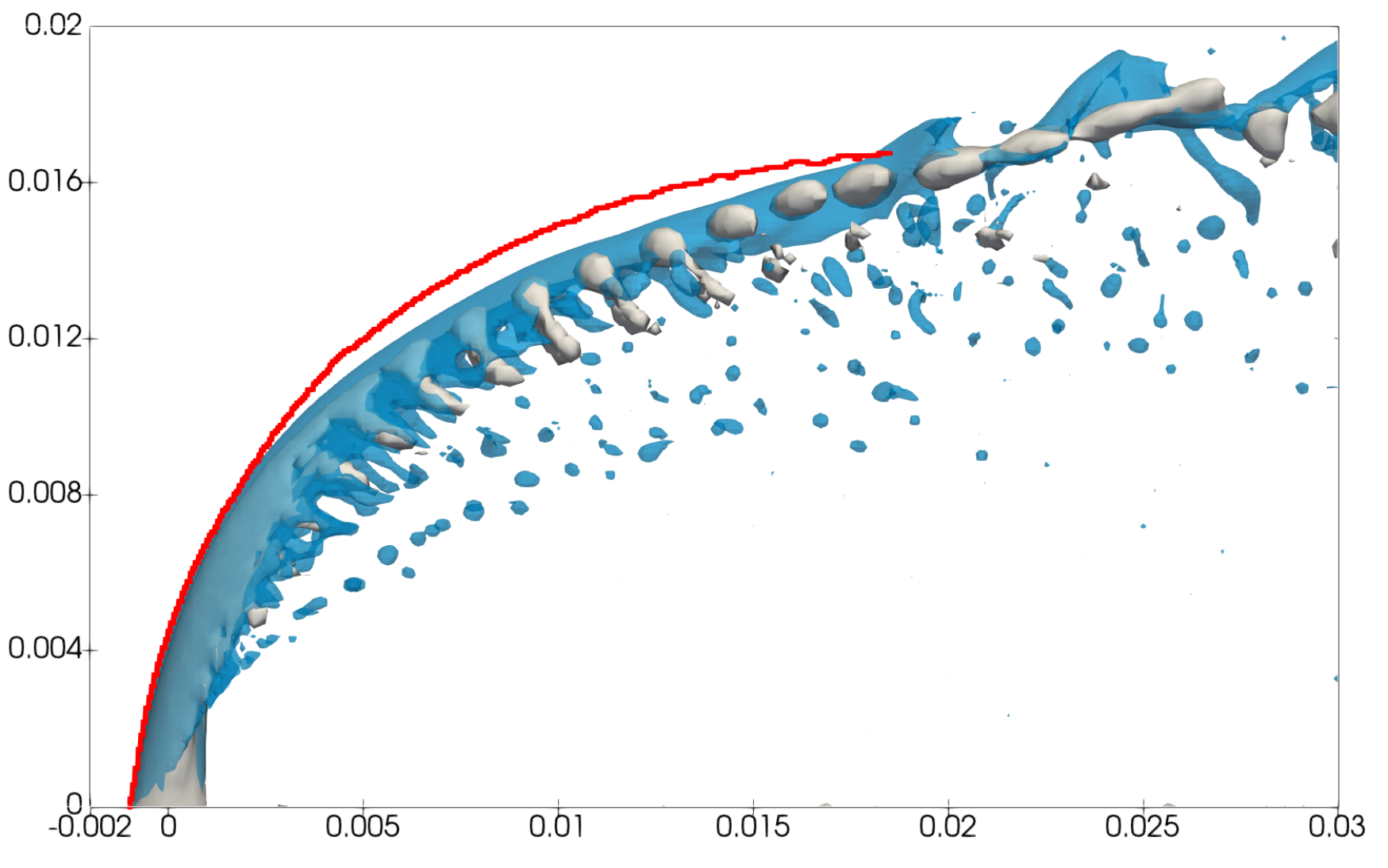}
      \caption{interIsoRhoFoam}
      \label{fig:LJCF_interIsoRho_comp}
     \end{subfigure}
    \begin{subfigure}[t]{.45\textwidth} 
      \centering
      \includegraphics[width=\linewidth]{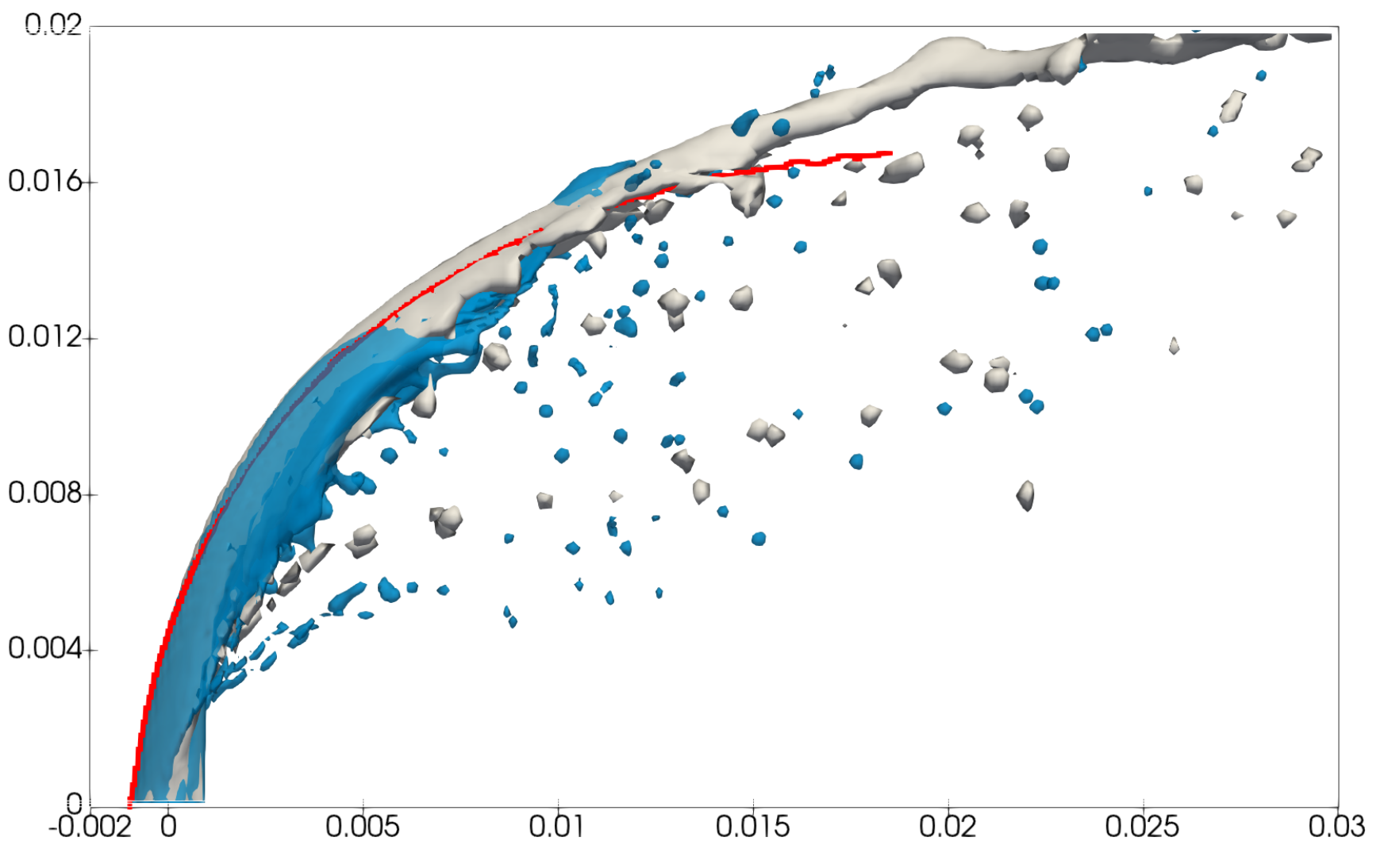}
      \caption{DYJEAT}
     \label{fig:LJCF_DYJEAT_comp}
     \end{subfigure}
    \caption{The instantaneous liquid jet shape at final time $t=7.1 ms$ with different resolutions (the blue translucent jet: $N_h=[256,128,128]$; the gray jet: $N_l=[128,64,64]$) and its comparison with the experimental results(the red line). This case makes it possible to evaluate performance on coarser meshes as resolving finer structures does not impact the jet trajectory.}
    \label{fig:LJCF_comp}
 \end{figure}

The \cref{fig:LJCF_comp} displays the final bent shape of liquid jet simulated by three solvers and their comparison with the experimental observation made by ONERA \citep{Desclaux2020,Bodoc2020}. The same case with two different resolutions are tested. We put the liquid jets' shape results in the same parallel view to compare them with each other and also with the experimental results marked by the red line. The blue translucent liquid jet represents the results from the case with a higher resolution $N_h=[254,128,128]$, while the gray liquid jet comes from the above low-resolution results. It can be seen from each \cref{fig:LJCF_interIso_comp,fig:LJCF_interIsoRho_comp,fig:LJCF_DYJEAT_comp} that more small droplets and complex structures can be captured when deploying the higher mesh resolution. Despite the different resolutions, there exists a very minor difference between the liquid jets with regard to the bent shape. The windward surfaces of the two liquid jets in each sub-figure almost attach to each other, which highlights that this case is insensitive to the mesh resolution.  As to the comparison with the experimental trajectories, both the jets' bent surface in \cref{fig:LJCF_interIso_comp,fig:LJCF_interIsoRho_comp} show good correspondence to the experimental shape, i.e. the red line at the low penetration height $<\SI{9}{\mm}$. The jets reattach to the red line in the upper-right zone. The maximal deviation between the simulated jets and the experiment shape is around $\SI{1}{mm}$, which is $5\%$ of the jet height. A more obvious discrepancy between the jets and the experiment is shown in \cref{fig:LJCF_DYJEAT_comp}. The liquid jet bent less than the experiment in the high-speed flow after the given time, which results in a wrong prediction of the impingement position.    

\section{Conclusions}
\label{sec:concl}

We analyze the equivalence between the mass conservation and volume fraction advection equation in the context of flux-based VOF methods. When an average mass flux is computed on the discrete level by scaling the fluxed phase-specific volume with the time step, we find that the equivalence is possible only when the first-order Euler temporal discretization is used for momentum conservation, no flux limiting is applied, and the flux-based VOF scheme uses first-order quadrature for the integration of the fluxed phase-specific volume. For any flux-based VOF method, the integration of the fluxed phase-specific volume $V_f^\alpha$ therefore plays a key role in the stability of two-phase flow simulations \textcolor{Tobi}{with} high density ratios, when the mass flux is estimated from it. Altering the mass flux $\rho_f F_f$ in the discrete momentum conservation equation by applying flux limiting schemes or blending schemes causes errors that are proportional to the density difference, and lead either to large errors in the interface shape and topology, or catastrophic failure. In other words, upwinding the two-phase momentum in a flux-based VOF method is consistent to the geometrical upwinding of the volume fraction by the flux-based VOF method.

We successfully apply the \rhoLENT method for high-density ratio flows to the isoAdvector-plicRDF Volume-of-Fluid method \citep{Scheufler2019}. The adaption of the \rhoLENT  method in the context of the geometrical Volume-of-Fluid method is straightforward, requiring only the geometrical calculation of the upwind area fraction $\alpha_f^{o}$ from the available geometric VOF interface (phase-indicator) approximation. 

We demonstrate the equivalence on the discrete level between the mass flux scaled from the phase-specific volume and the solution of an auxiliary density equation. We confirm this hypothesis by means of verification with challenging inviscid cases, as well as by validation against experiments.

\section{Acknowledgments}

The last author acknowledges funding by the German Research Foundation (DFG) – Project-ID 265191195 – SFB 1194. Calculations for this research were conducted on the Lichtenberg high performance computer of the TU Darmstadt. 

The computations with DYJEAT codes were granted access to the HPC resources of CALMIP supercomputing center under the allocation 2023 P18043.

\appendix
\section{Correct cyclic boundary condition for the plicRDF-isoAdvector method}
\label{Appdix:cyclic_BC}

\begin{figure}[!htb]
    \centering
    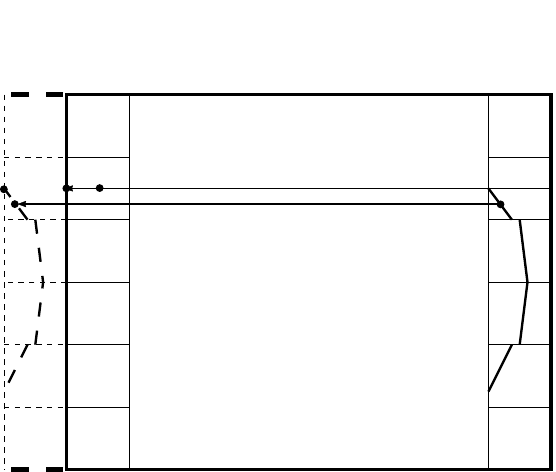
    \caption{The ghost cells layer to correct signed distance: the green line (\textcolor{mygreen}{\rule{5ex}{0.4mm}}) presents fixed RDF from cell centroid $\x_c^{nei}$ to interface in the cyclic neighbor cell, while the red line (\textcolor{myred}{\rule{5ex}{0.4mm}}) depicts the original unfixed RDF.}
    \label{fig:cyclicBC_ghostCellsLayer}
\end{figure}

A cyclic boundary condition (BC) treats two boundary patches as if they were physically connected, with their respective cell layers placed next to each other. For the geometrical VOF method such as the plicRDF-isoAdvector, the cyclic boundary condition impacts the interface reconstruction and the volume fraction advection. To achieve this, the cyclic BC performs \textcolor{Tobi}{calculations on} the so-called \emph{owner-patch}, and then reflects the result to the so-called \emph{neighbor-patch}, as shown schematically in \Cref{fig:cyclicBC_ghostCellsLayer} for a geometric VOF method in two dimensions.

For the advection discretization at cyclic BC, we noticed that the fluxed phase-specific volumes $V_f^{\alpha}$ field is not properly adjusted for the cyclic patches.

In the first step, the cyclic BC initializes $V_f^{\alpha}$ at every face of both cyclic patches using the upwind scheme, i.e.,
\begin{equation}
    V_{f}^{\alpha, init} = F_f\alpha_{U}\Delta t,
    \label{eq:initial_dvf}
\end{equation}
where $F_f:=\mathbf{v}_f \cdot \Sf$ is the volumetric flux at the centroid of the cell-face $S_f$, $\alpha_U$ is the  volume fraction of the upwind cell $U$ w.r.t the cell face $S_f$, and $\Delta t$ is the time step. 

In the second step, the cyclic BC computes the geometric $V_f^{\alpha}$ at the faces that belong to interface cells on both patches 
\begin{equation}
    V_f^{\alpha} =
    \int_{t^n}^{t^{n+1}} \int_{S_f} \chi \mathbf{v}  \cdot \mathbf{n} \, dS \, dt 
    =  \int_{t^n}^{t^{n+1}} \frac{F_f(t)}{|\mathbf{S}_f|}A_f(t)\, dt 
    \label{eq:orig_update_dvf}
\end{equation} 

The existing cyclic BC \citep{OF_V2306} does not consider cyclic boundary conditions. The discretization in cell layers adjacent to two cyclic boundary patches should handle the cell layers as if they are placed next to each other as shown in \Cref{fig:cyclicBC_ghostCellsLayer}. A phase-specific volume fluxed out of the domain on a face that belongs to the cyclic owner-patch, should be fluxed into the corresponding face in the cyclic-neighbor patch, as described in \Cref{alg:cyclicfix}.

\begin{center}
\begin{algorithm}[H]
    \centering
    \caption{The modified fluxed phase-specific volumes $V_f^{\alpha}$ update method in interIsoFoam.}
    \label{alg:modi_dvf_update}
    {\small
    \centering
    \begin{algorithmic}[1]
        \State Initialize $V_f^{\alpha, owner}$ using the upwind scheme. \Comment \Cref{eq:initial_dvf}
        \For{all boundary patches}
            \If{boundary patch is cyclic}
                \For{all cyclic-patch faces $f \in [1, |P_{cyclic}|]$}
                    \If{$F_f > 0$}
                        \State Geometrically compute the outflow $V_f^{\alpha}$.
                    \ElsIf{$F_f < 0$} \Comment The inflow $F_f < 0$ here is outflow $F_f > 0$ of the neighbor.
                        \State $V_f^{\alpha} = -V_f^{\alpha, neigbor}$
                    \EndIf
                \EndFor
            \EndIf        
        \EndFor
    \end{algorithmic}
    }
    \label{alg:cyclicfix}
\end{algorithm}
\end{center}

Aside from the consistent calculation of $V_f^{\alpha}$ in the advection, the cyclic BC allso affects the geometric interface reconstruction. The plicRDF reconstruction \citep{Scheufler2019} uses  the reconstructed distance function (RDF) and its gradient for improving discrete interface-normal vectors. 
When considering the cyclic boundary condition, the distance calculation should be treated carefully, especially in the case where two cells sharing a vertex are also attached to two cyclic patches. This issue is illustrated in \Cref{fig:cyclicBC_ghostCellsLayer}, where $\x_i^{own}$ denotes a center of a VOF interface polygon located in a corresponding interface cell $\Omega_i^{own}$ that belongs to the cell layer attached to the owner patch of the cyclic BC. 
In the plicRDF implementation in OpenFOAM-v2306 \citep{OF_V2306}, the cyclic BC falsely uses VOF interface polygon centers  $\x_i^{own}$ to compute signed distances at centers $\x_c^{nei}$ in the cell layer adjacent to the cyclic neighbor-patch. A correct implementation of the cyclic BC requires positions  $\x_i^{nei}$, as shown in \Cref{fig:cyclicBC_ghostCellsLayer}. The VOF interface centroids from the cell layer adjacent to the cyclic owner-patch $\x_i^{own}$ and the face centers of the cyclic owner-patch can be used to compute $\x_i^{nei}$ for every $\x_i^{own}$ and facilitate a correct cyclic (periodic) computation of the Reconstructed Distance Function (RDF) in the plicRDF-isoAdvector method. 

We define a transformation of the position of the interface centers using
\begin{equation}
        \x_i^{nei} := \x_i^{own} + (\x_f^{nei} - \x_f^{own}),
\end{equation}
where $(\x_f^{nei} - \x_f^{own})$ is the difference (displacement, or transformation) vector between the face centers of cyclic BC owner and neighbor patch face centers. 

Note that there is no need to tranform the interface normals $\n_{i}$, because they are orientation and not position vectors. 
The transformed PLIC polygon centroids together with the cyclic BC neighbor-patch interface normals  $\n^{nei}_{i}=\n^{own}_i$ build a cyclic ghost-data layer that is shown schematically by dashed cells in \Cref{fig:cyclicBC_ghostCellsLayer}.    

Signed distances in the cell layer adjacent to the cyclic owner-patch at cell centers $\x_i^{own}$, are thus computed from the transformed PLIC interface information from the cyclic patch-owner data.

If the cyclic-patch-adjacent cell $\Omega_c$ does contain its own PLIC interface with the PLIC centroid, then the signed distance to this PLIC interface is used, if it is closer to $\x_i$ of that cell.

\begin{figure}[!htb]
    \centering
    \includegraphics[width=0.35\textwidth]{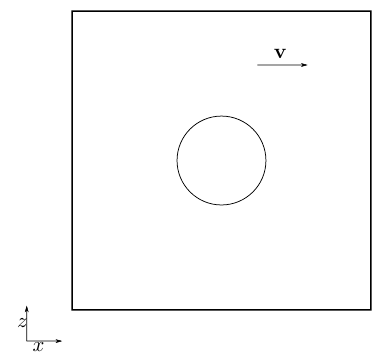}%
    \caption{A droplet moves with the ambient constant flow.}
    \label{fig:cyclicBC_description}
\end{figure}

We verify our discretization of the cyclic BC in the plicRDF-isoAdvector method using a constant flow case shown in \Cref{fig:cyclicBC_description}, where both the droplet and the ambient flow have the same initial velocity $\mathbf{v}=(v_x, 0)$. The left and right boundaries are set as cyclic (periodic). It is noteworthy that the momentum equation \Cref{eq:momentum-transport} is not solved in this test case, ensuring exactly constant velocity and pressure over time, aiming to only to test the periodic (cyclic) interface reconstruction and advection. We test two interface reconstruction methods, e.g. isoAlpha and plicRDF from \citep{Scheufler2019}.

\begin{figure}[!htb]
     \begin{subfigure}{.48\textwidth}
      \centering
      \includegraphics[width=\linewidth]{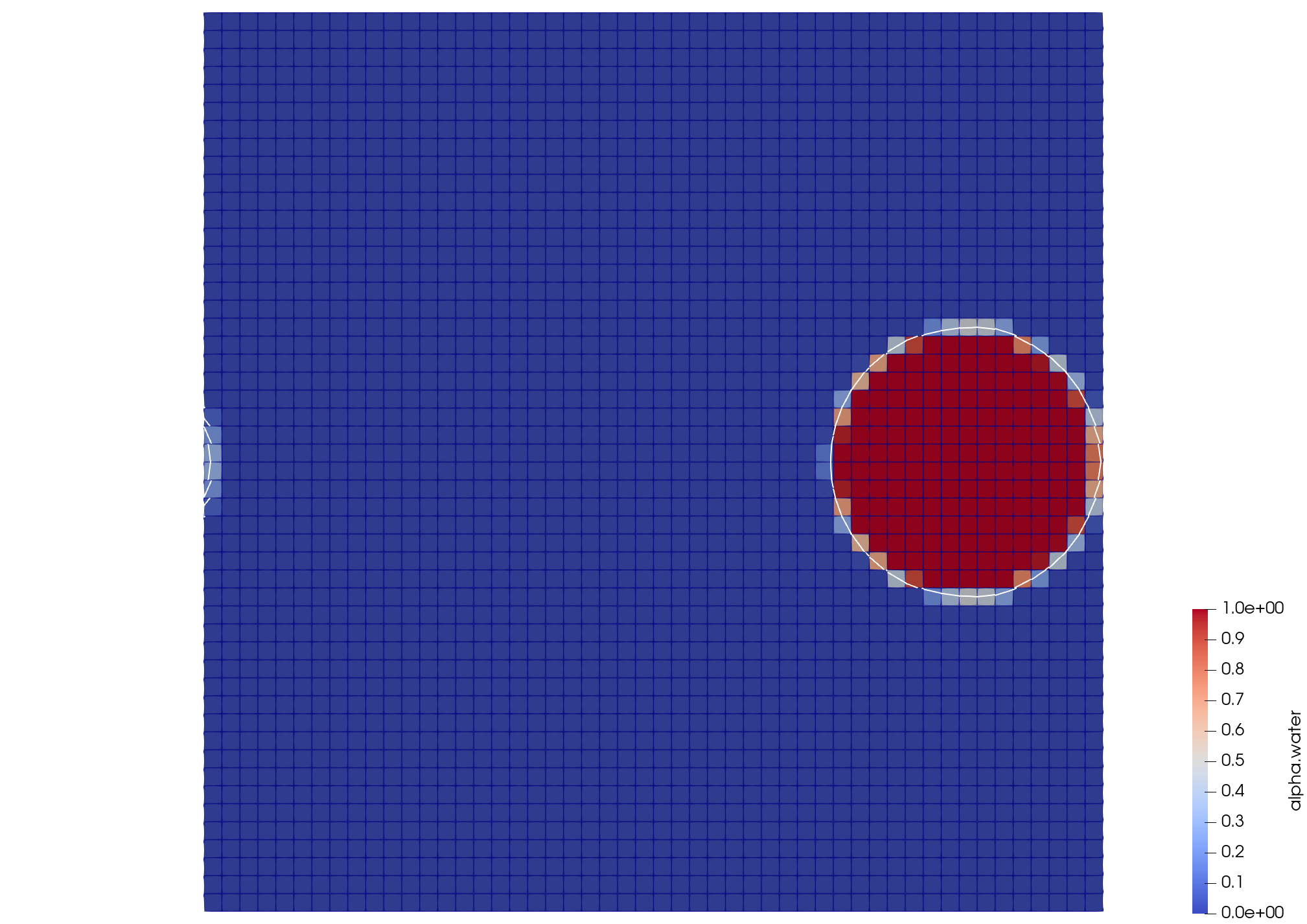}
      \caption{Unfixed}
      \label{fig:Cyclic_unfixed_isoAlpha}
     \end{subfigure}
     \hfill
     \begin{subfigure}{.48\textwidth}
      \centering
      \includegraphics[width=\linewidth]{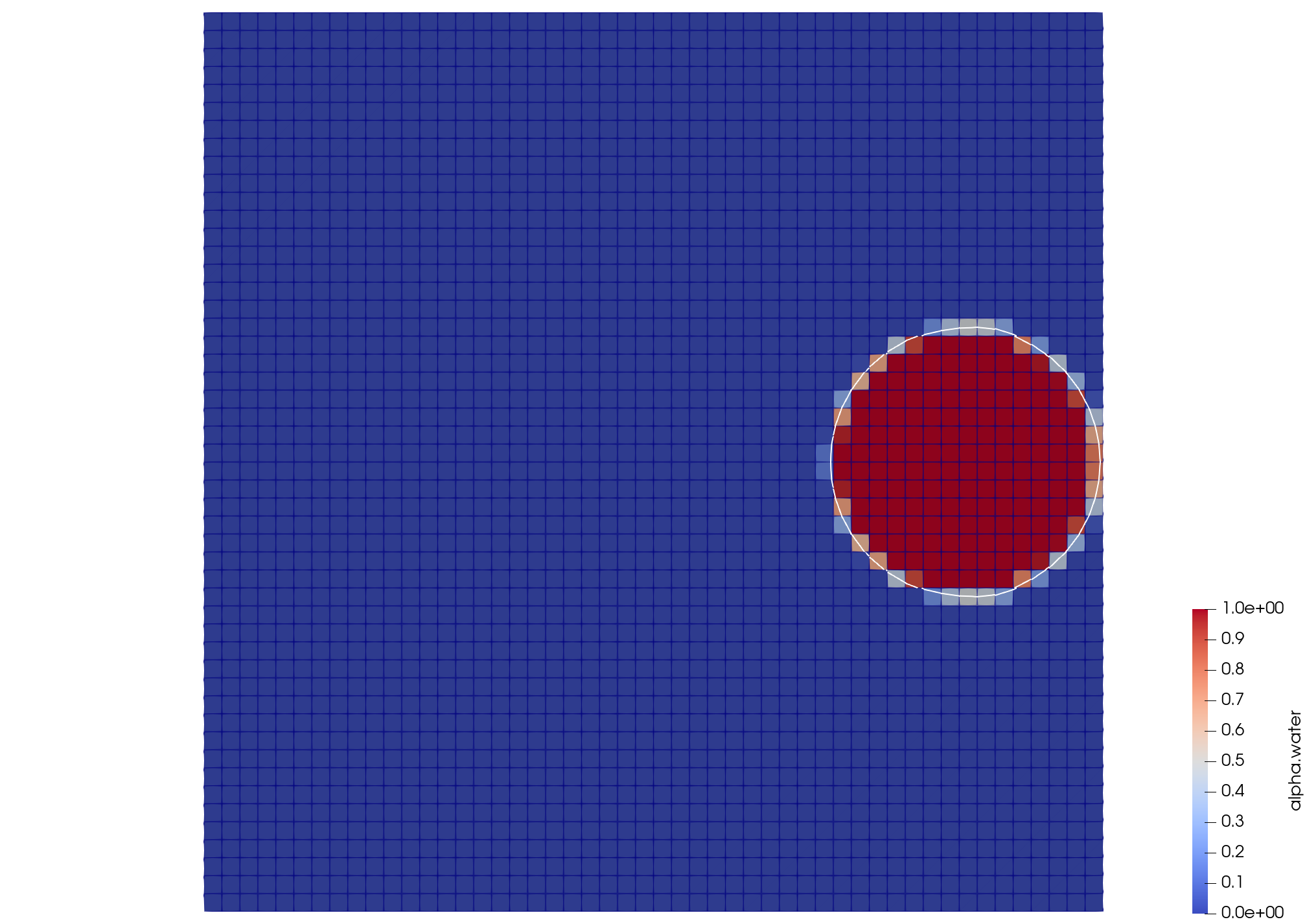}
      \caption{Fixed}
      \label{fig:Cyclic_fixed_isoAlpha}
     \end{subfigure}    
    \caption{The alpha field and interfaces reconstructed by iso-Alpha method reach to the right cyclic boundary; blue region: the ambient flow; red region: the liquid droplet; white line segments: the PLIC interfaces.}
    \label{fig:Cyclic_isoAlpha}
\end{figure}

\begin{figure}[!htb]
     \begin{subfigure}{.48\textwidth}
      \centering
      \includegraphics[width=\linewidth]{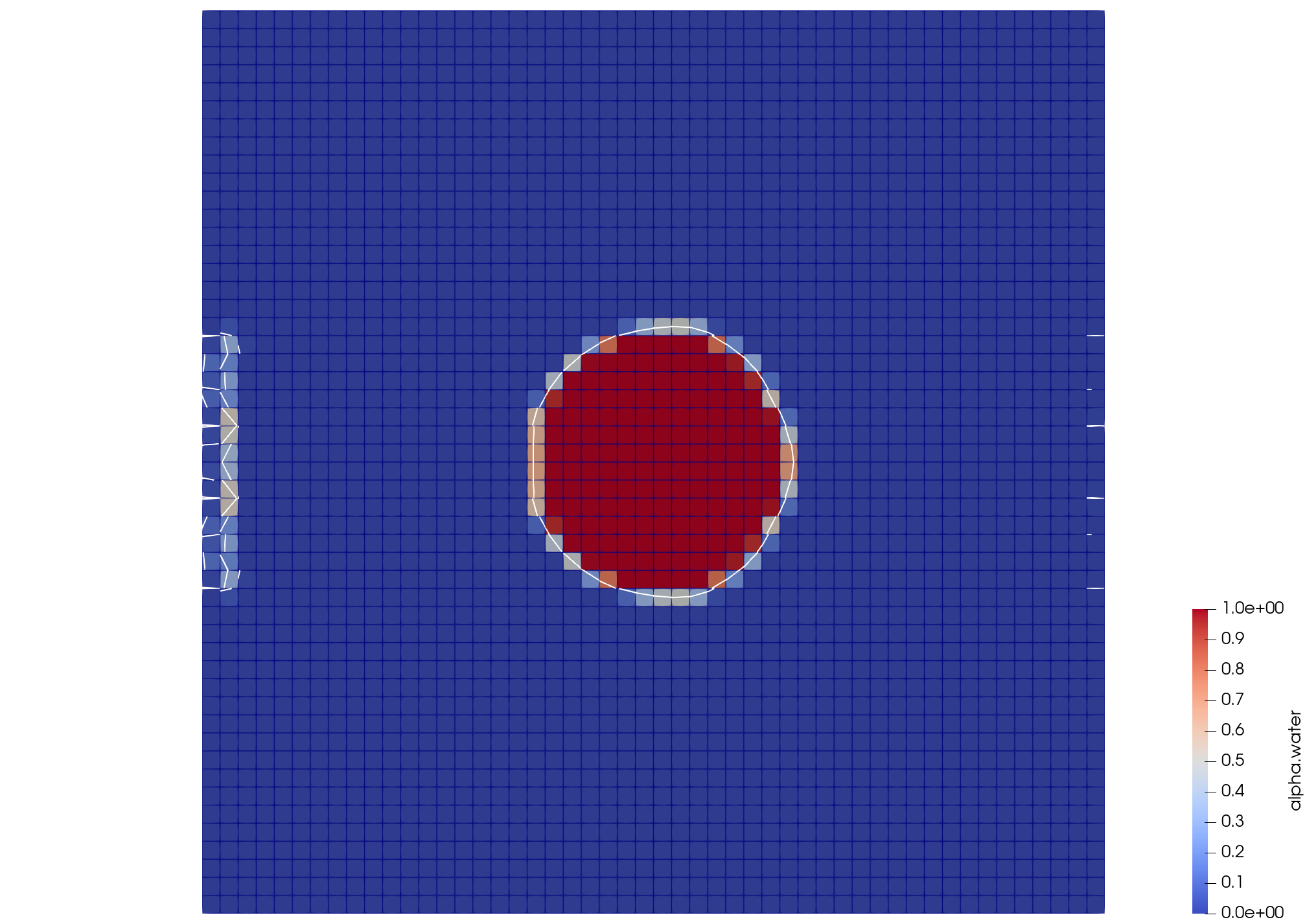}
      \caption{Unfixed}
      \label{fig:Cyclic_unfixed_plicRDF}
     \end{subfigure}
     \hfill
     \begin{subfigure}{.48\textwidth}
      \centering
      \includegraphics[width=\linewidth]{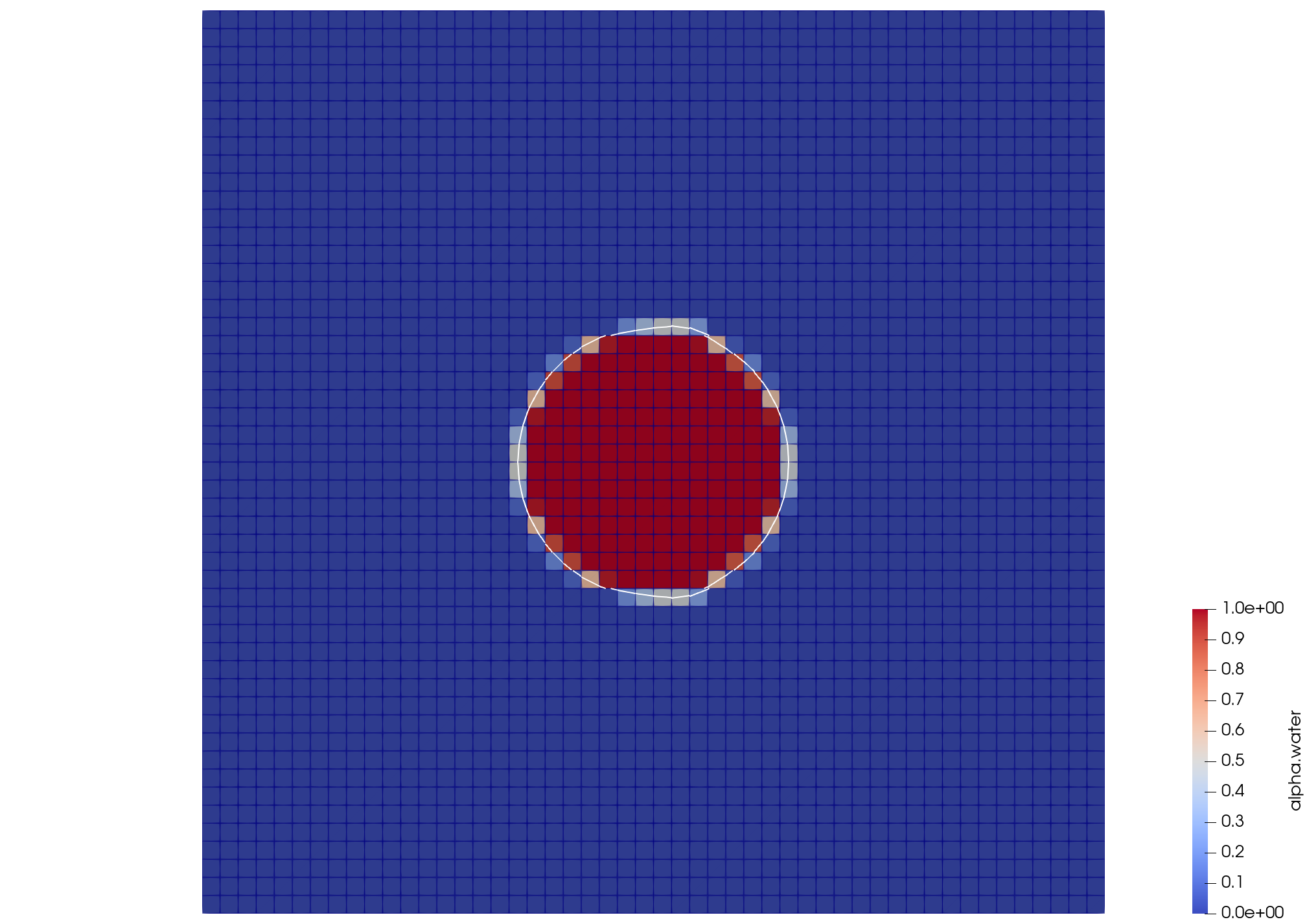}
      \caption{Fixed}
      \label{fig:Cyclic_fixed_plicRDF}
     \end{subfigure}    
    \caption{The alpha field and interfaces reconstructed by plic-RDF method cross the right cyclic boundary; blue region: the ambient flow; red region: the liquid droplet; white line segments: the PLIC interfaces.}
    \label{fig:Cyclic_plicRDF}
\end{figure}

\Cref{fig:Cyclic_isoAlpha} shows the droplet reaching the right cyclic boundary. With the erroneous calculation of the fluxed phase-specific volume $V_f^{\alpha}$ at the cyclic boundaries, as shown in \Cref{fig:Cyclic_unfixed_isoAlpha}, the interfaces appear in the patch neighbor cells incorrectly. The \Cref{fig:Cyclic_fixed_isoAlpha} show the accurate result of our modification with a single interface cell. The modification of $V_f^{\alpha}$ impacts both isoAlpha and plicRDF methods, while adapting the displacement is crucial only for the plicRDF reconstruction. As shown in \Cref{fig:Cyclic_unfixed_plicRDF}, without modifying the displacement vector, some liquid remains in the cell layer adjacent to the neighbor-patch after the droplet crosses the cyclic boundary, and reaches a location far from both cyclic boundaries. With applying our  modification from \Cref{fig:Cyclic_fixed_plicRDF}, the droplet retains its initial form.  The fixed cyclic boundary condition is available in \citep{rhoVoFCodes2023}.         

\section{\textcolor{Reviewer2}{Supplementary results}}
\label{Appdix:densityRatio1}
\begin{figure}[!htb]
     \begin{subfigure}{.96\textwidth}
      \centering
      \includegraphics[width=\linewidth]{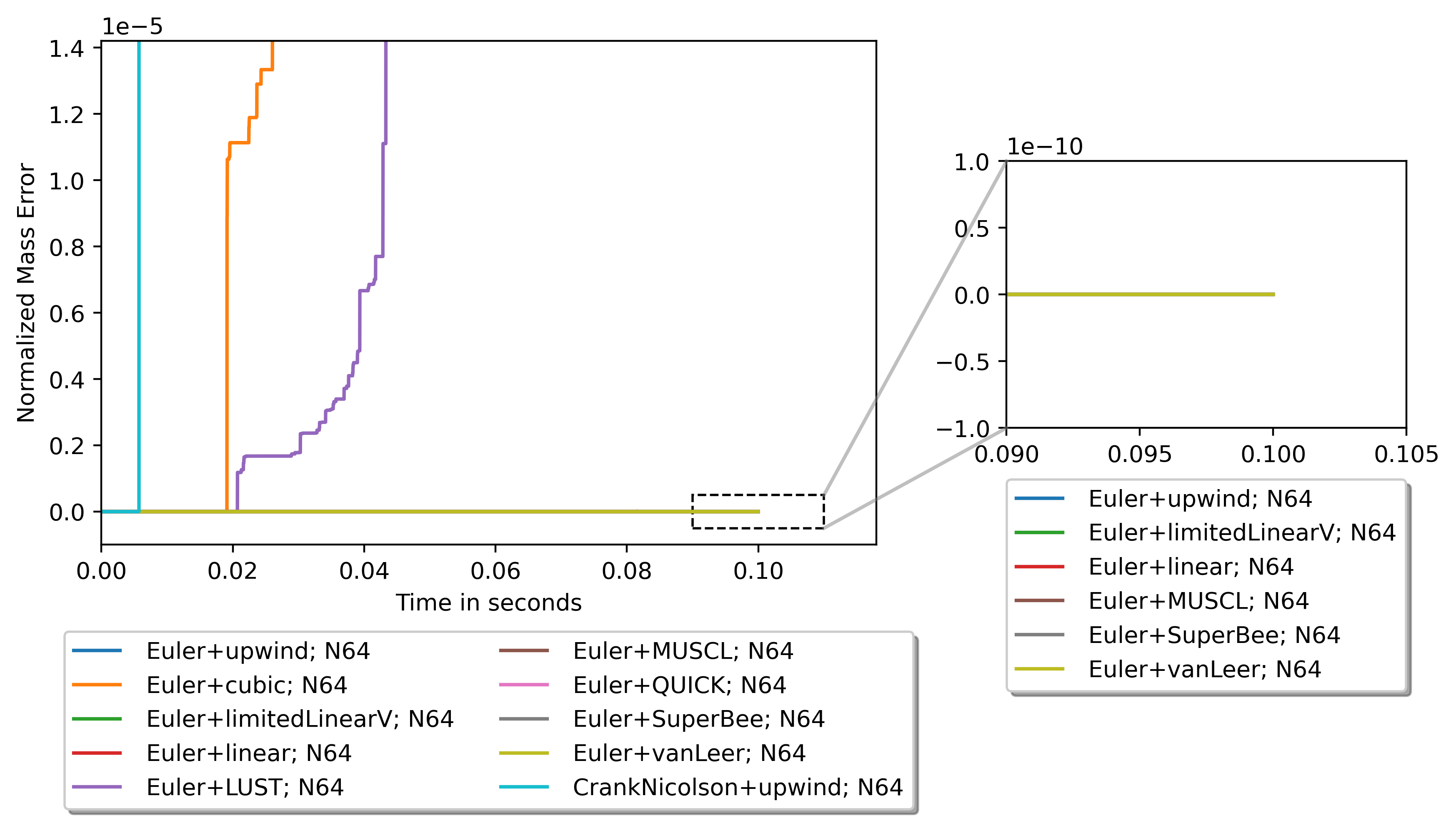}
      \caption{\textcolor{Reviewer21}{Mass error.}}
     \end{subfigure}
\end{figure}
\begin{figure}[!htb]\ContinuedFloat
     \begin{subfigure}{.96\textwidth}
      \centering
      \includegraphics[width=\linewidth]{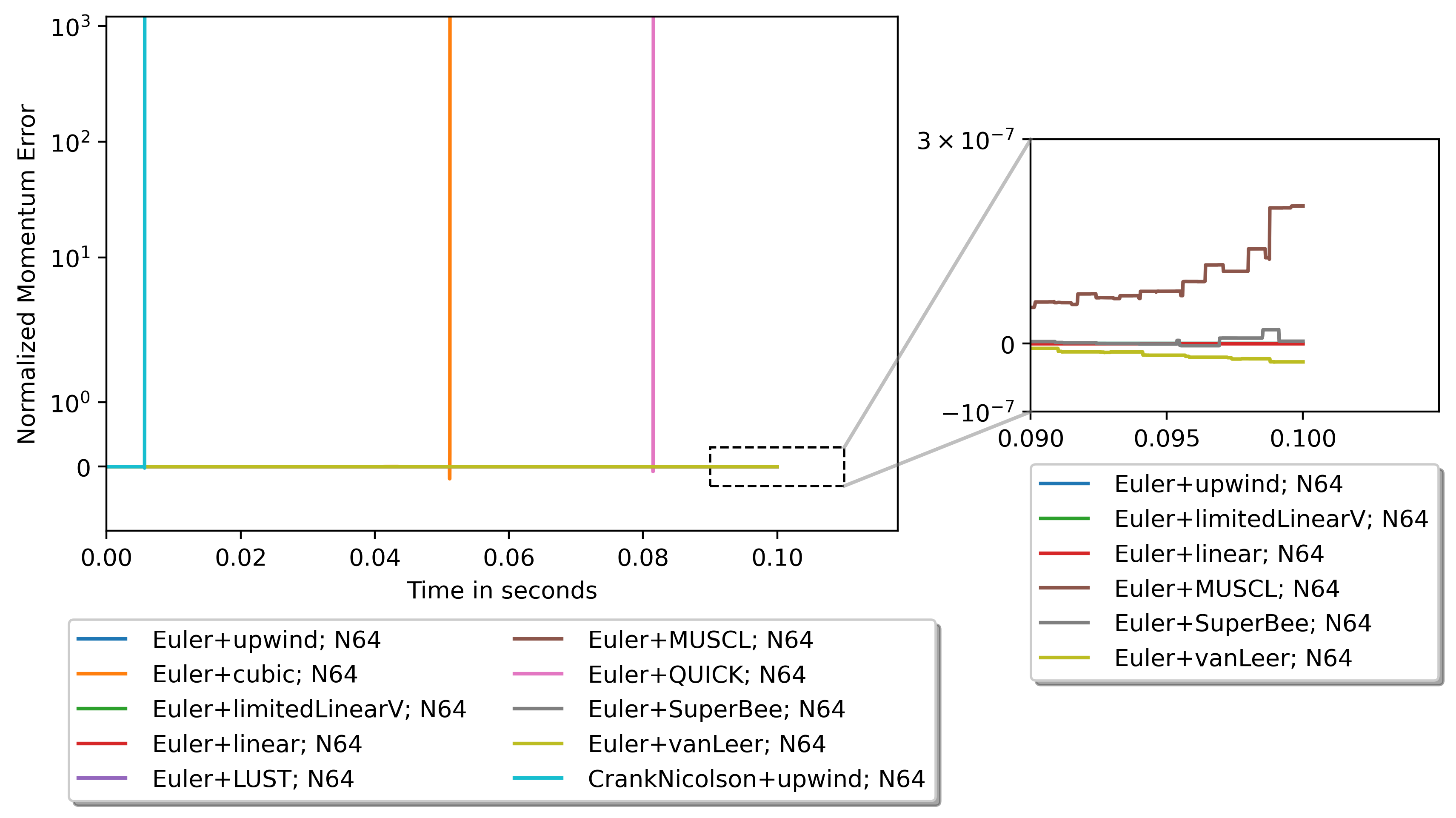}
      \caption{\textcolor{Reviewer21}{Momentum error.}}
     \end{subfigure} 
    \begin{subfigure}{.96\textwidth}
      \centering
      \includegraphics[width=\linewidth]{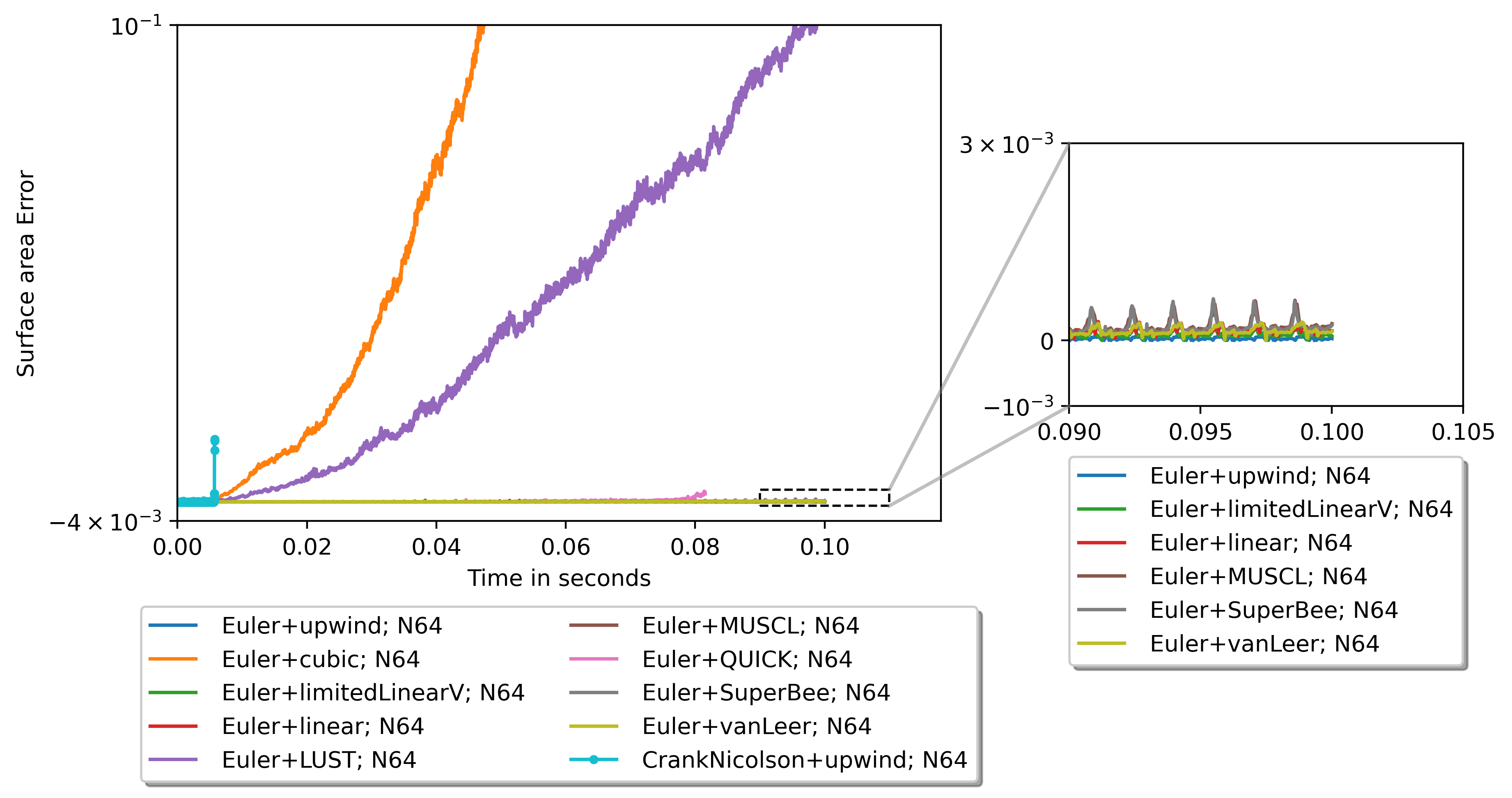}
      \caption{\textcolor{Reviewer21}{Sphericity error.}}
     \end{subfigure}  
    \caption{\textcolor{Reviewer21}{Temporal evolution of normalized mass, momentum conservation error and sphericity error with different schemes for the 2D case of Translating droplet in ambient flow: interIsoFoam, $N=64$, density ratio $=10^6$.}}
    \label{fig:Zuzoi_2D}
\end{figure}
\begin{figure}[!htb]
     \begin{subfigure}{.48\textwidth}
      \centering
      \includegraphics[width=\linewidth]{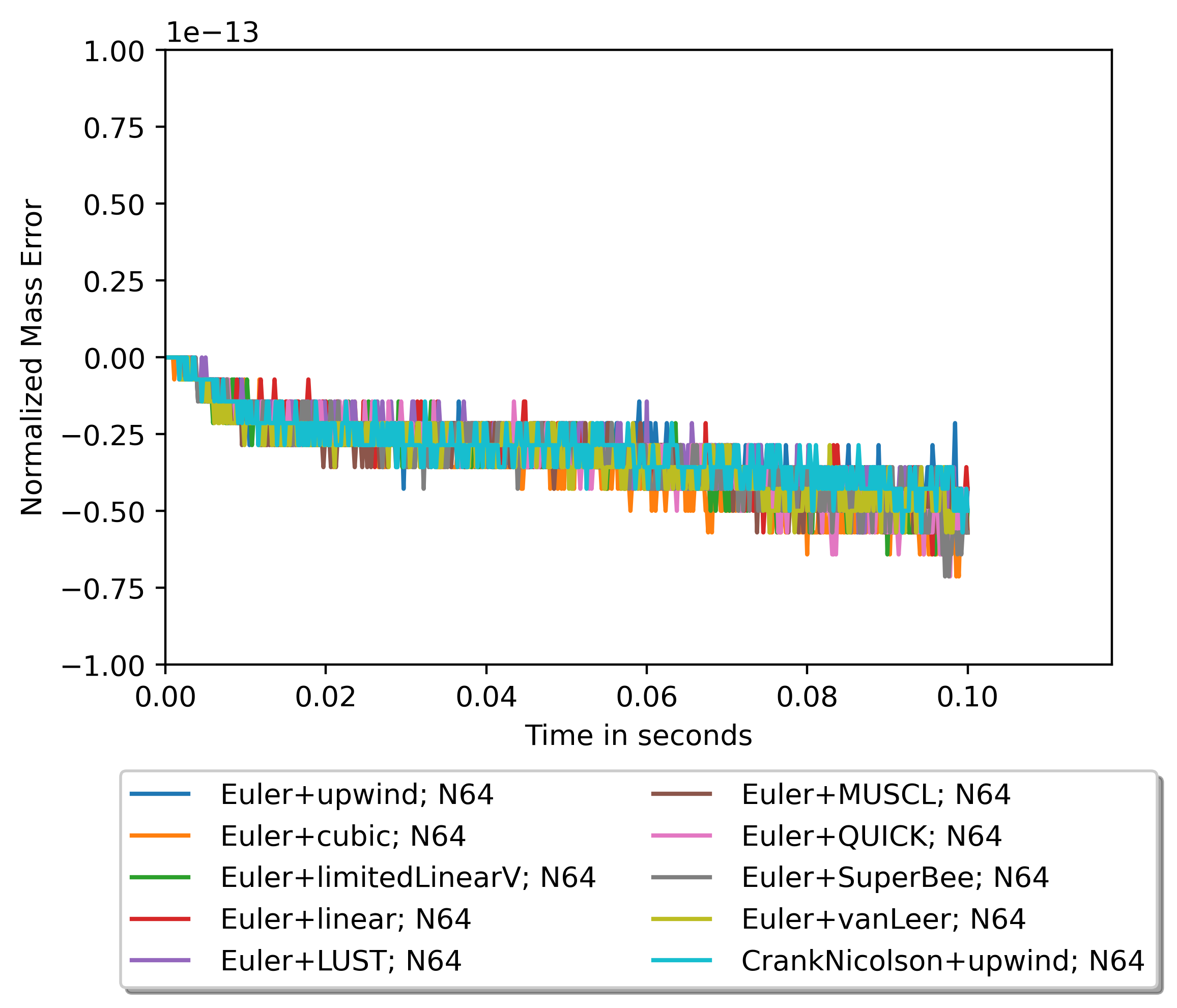}
      \caption{\textcolor{Reviewer21}{Mass error.}}
     \end{subfigure}
     \hfill
     \begin{subfigure}{.48\textwidth}
      \centering
      \includegraphics[width=\linewidth]{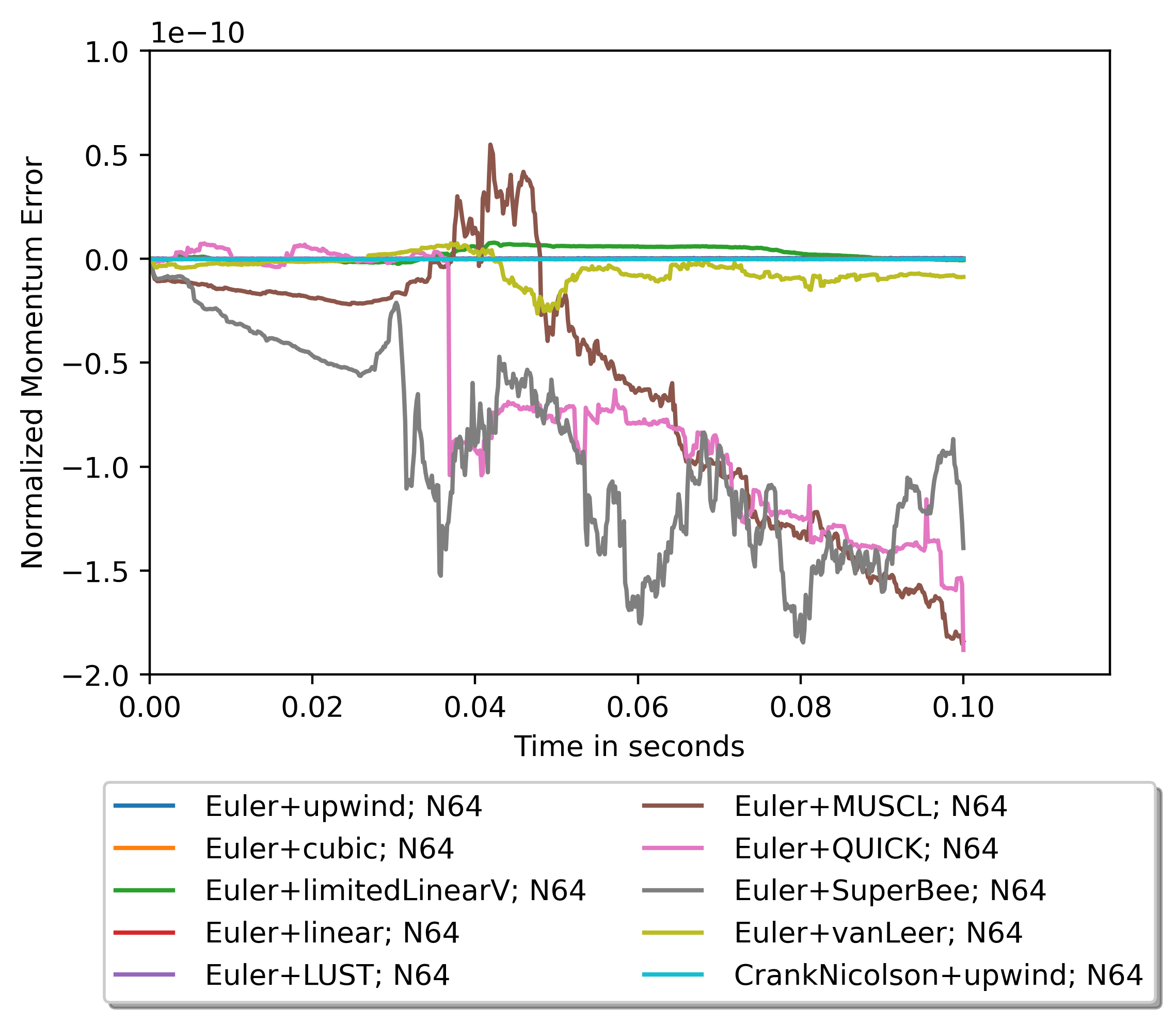}
      \caption{\textcolor{Reviewer21}{Momentum error.}}
     \end{subfigure}    
    \caption{\textcolor{Reviewer21}{Temporal evolution of normalized mass and momentum conservation error with different schemes for the case of Translating droplet in ambient flow: interIsoFoam, $N=64$, density ratio $=1$.}}
    \label{fig:densityRatio1}
\end{figure}
\begin{figure}
    \centering
    \includegraphics[width=.58\textwidth]{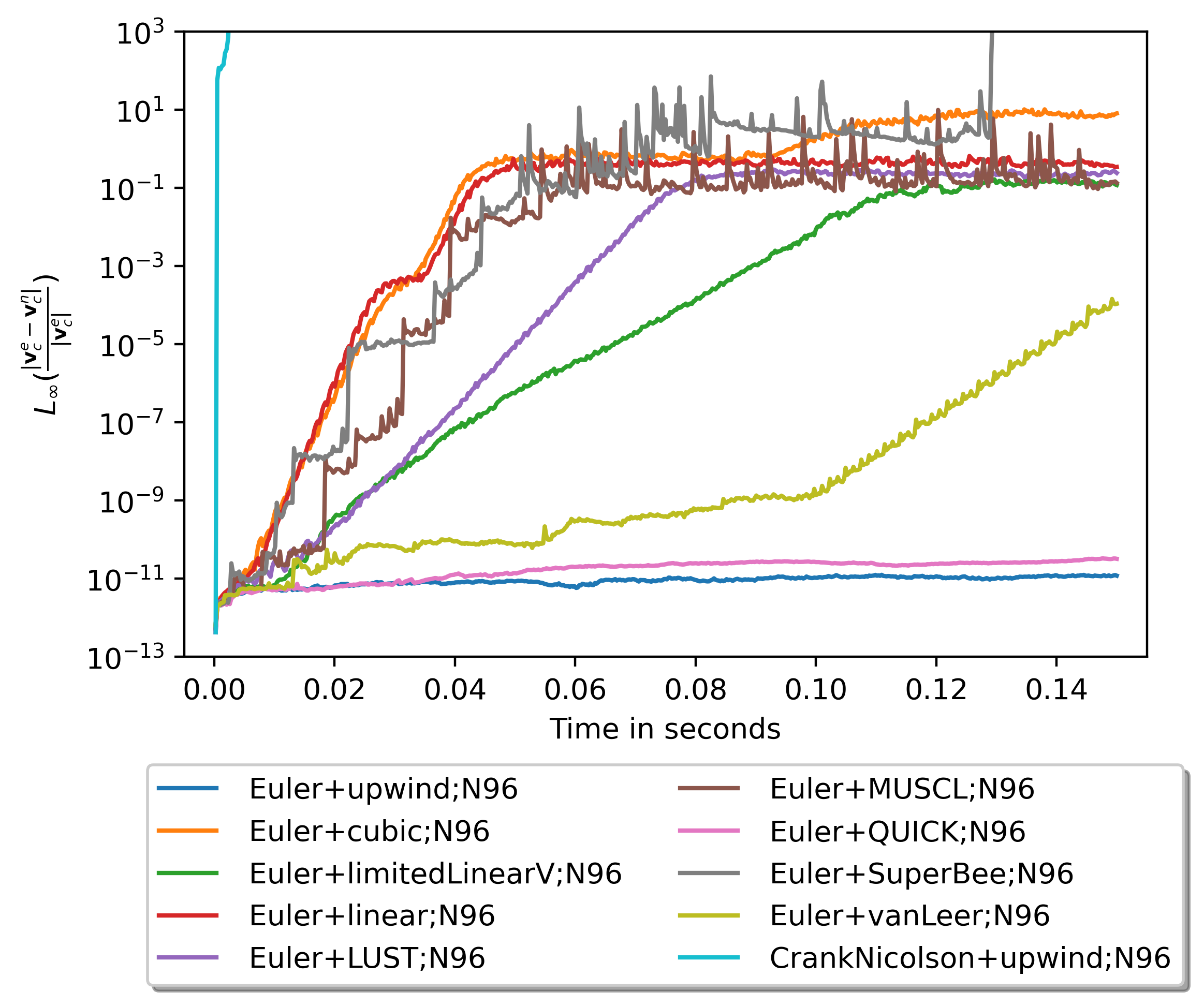}
    \caption{\textcolor{Reviewer21}{Temporal evolution of the velocity error norm $L_{\infty}(\v)$ with 2D pure advection - combining $10$ schemes,  density ratio is $10^4$, mesh resolution is $N=96$.}}
    \label{fig:TranslatingDroplet3D_schemes_comp_2D}
\end{figure}
\begin{figure}
    \centering
    \includegraphics[width=.58\textwidth]{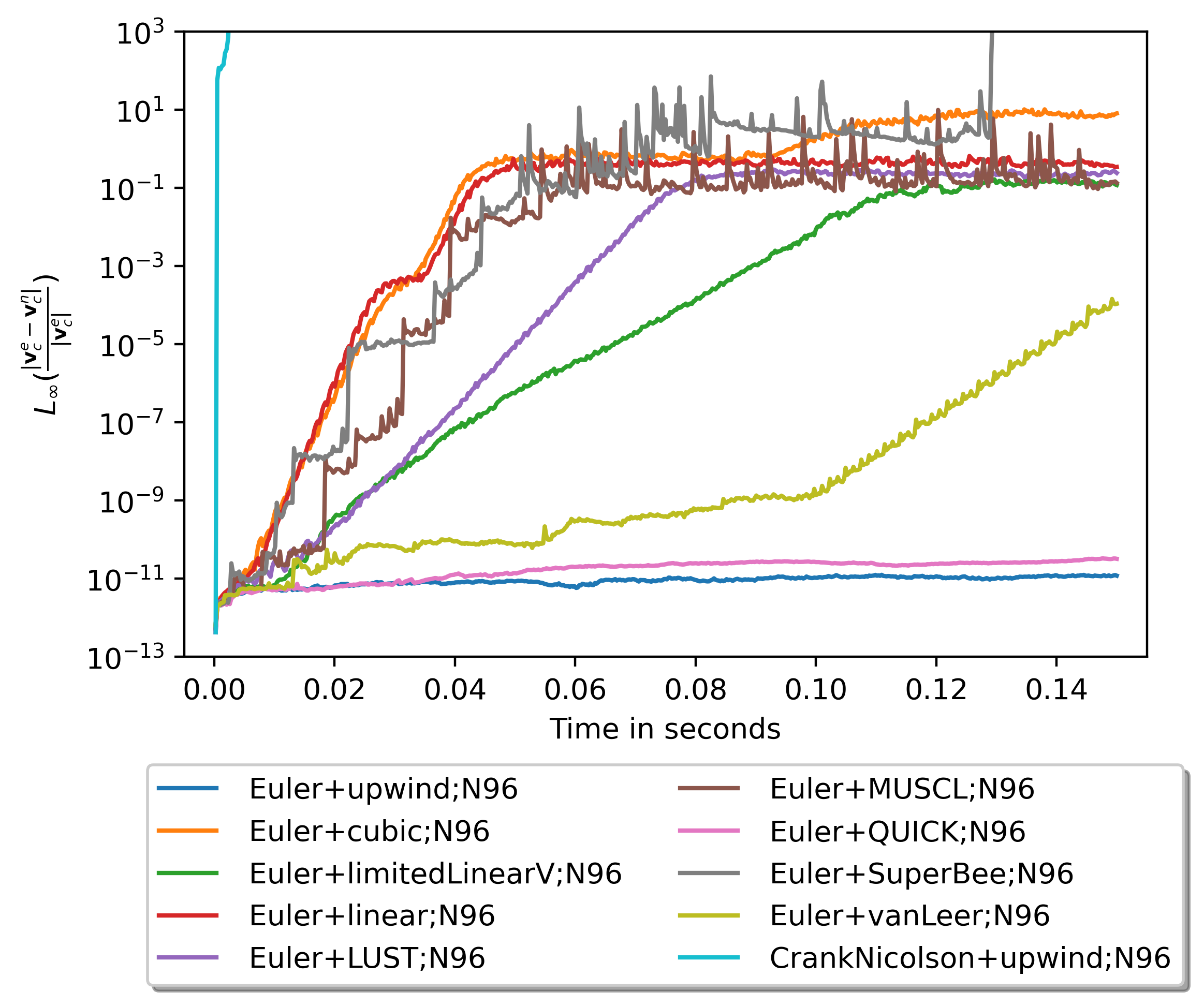}
    \caption{\textcolor{Reviewer2}{Temporal evolution of the velocity error norm $L_{\infty}(\v)$ with pure advection - combining $10$ schemes,  density ratio is $1$, mesh resolution is $N=96$. All schemes are stable.}}
    \label{fig:TranslatingDroplet3D_schemes_comp_densityRatio1}
\end{figure}

\begin{figure}[!htb]
     \begin{subfigure}{.48\textwidth}
      \centering
      \includegraphics[width=\linewidth]{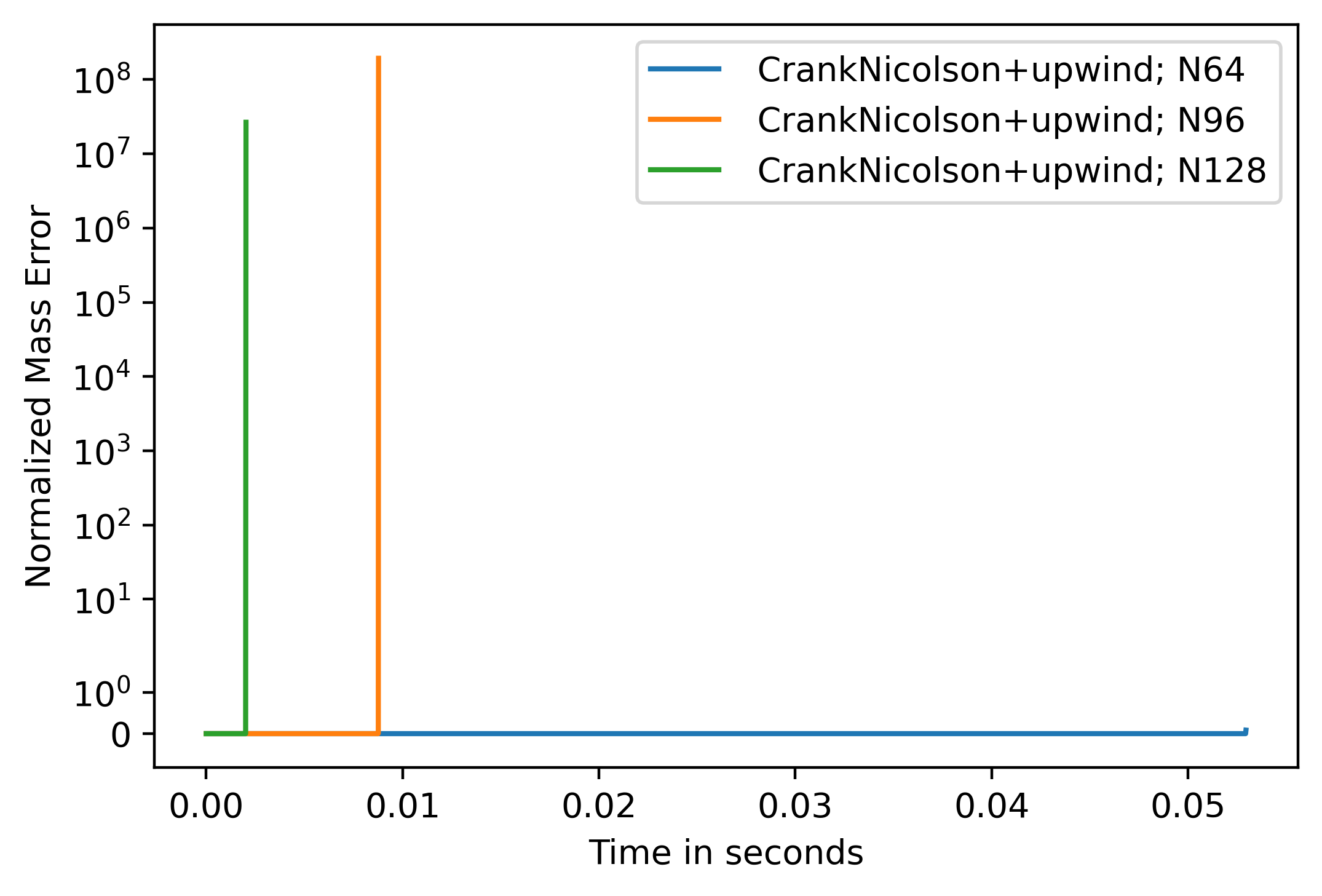}
      \caption{\textcolor{Reviewer2}{Mass error.}}
     \end{subfigure}
     \hfill
     \begin{subfigure}{.48\textwidth}
      \centering
      \includegraphics[width=\linewidth]{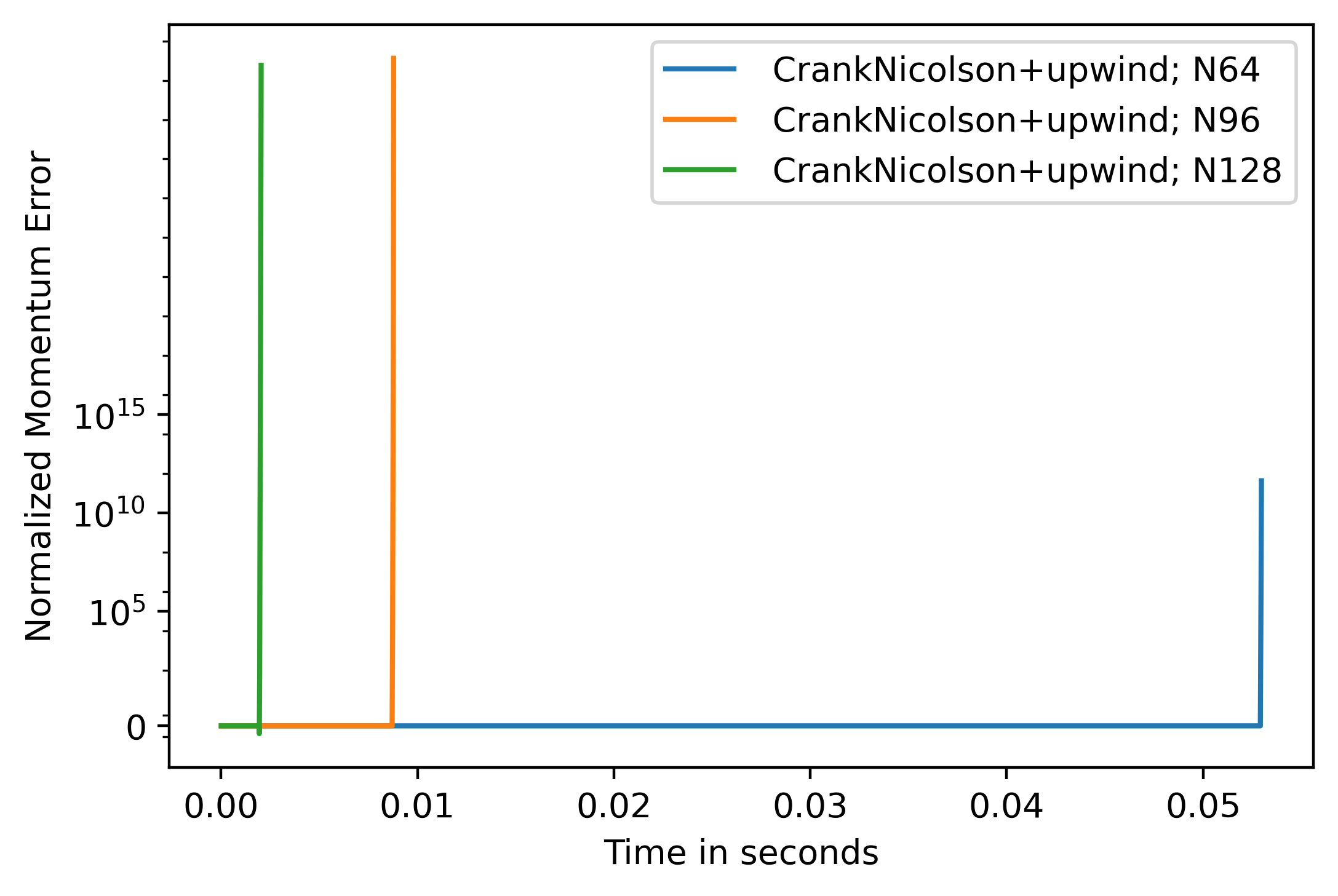}
      \caption{\textcolor{Reviewer2}{Momentum error.}}
     \end{subfigure}    
    \caption{\textcolor{Reviewer2}{Temporal evolution of normalized mass and momentum conservation error using CrankNicolson + upwind with different resolutions for the case of Translating droplet in ambient flow: interIsoFoam, $N=64,96,128$, density ratio $=\ 10^6$.}}
    \label{fig:CNSuddenJump}
\end{figure}

\begin{figure}[!htb]
     \begin{subfigure}[t]{.48\textwidth}
      \centering
      \includegraphics[width=\linewidth]{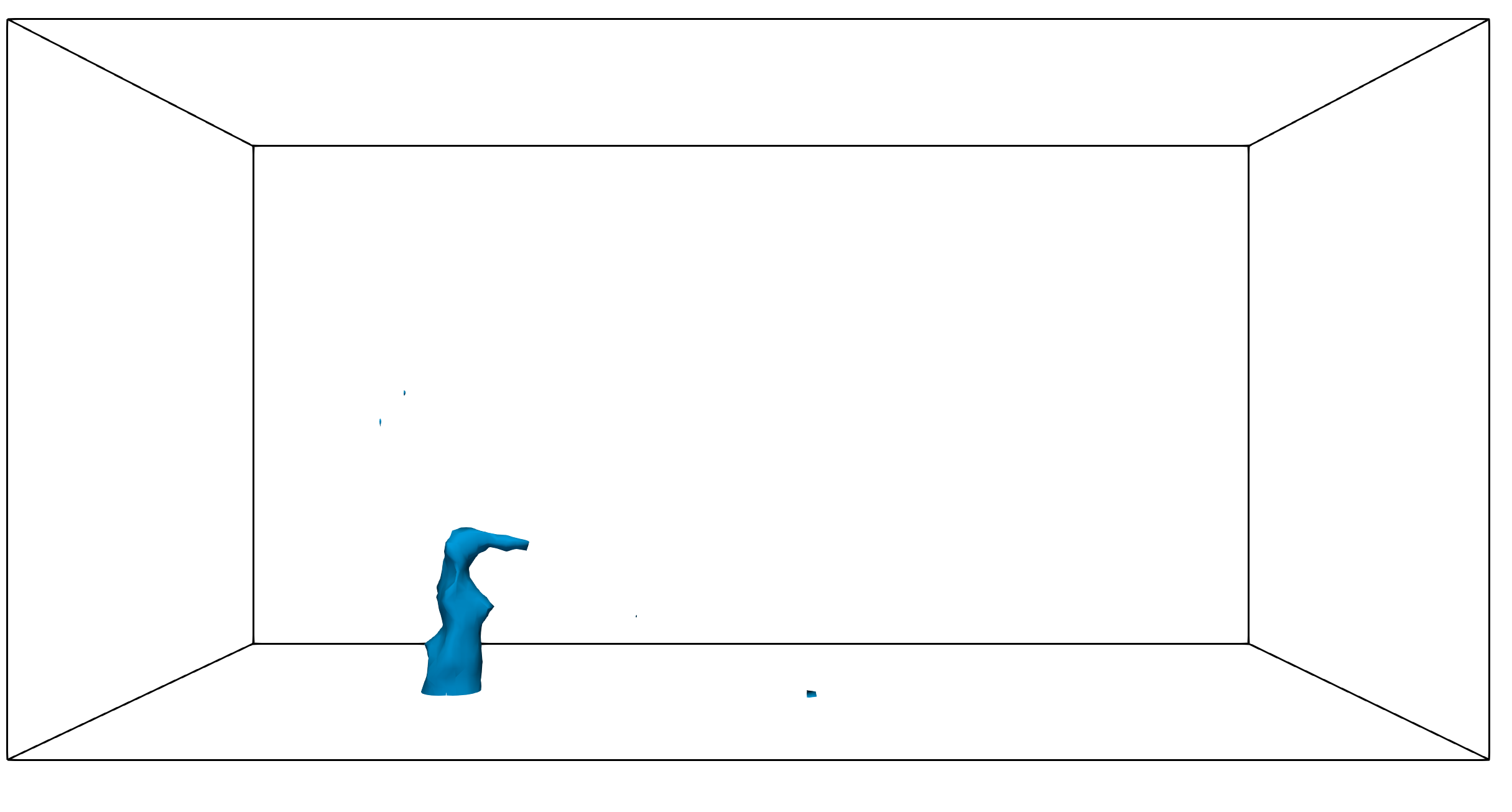}
      \caption{\textcolor{Reviewer1}{Resolution: $N_l=[128,64,64]$; final time: $t=\SI{0.7}{\ms}$.}}
     \end{subfigure}
     \hfill
     \begin{subfigure}[t]{.48\textwidth}
      \centering
      \includegraphics[width=\linewidth]{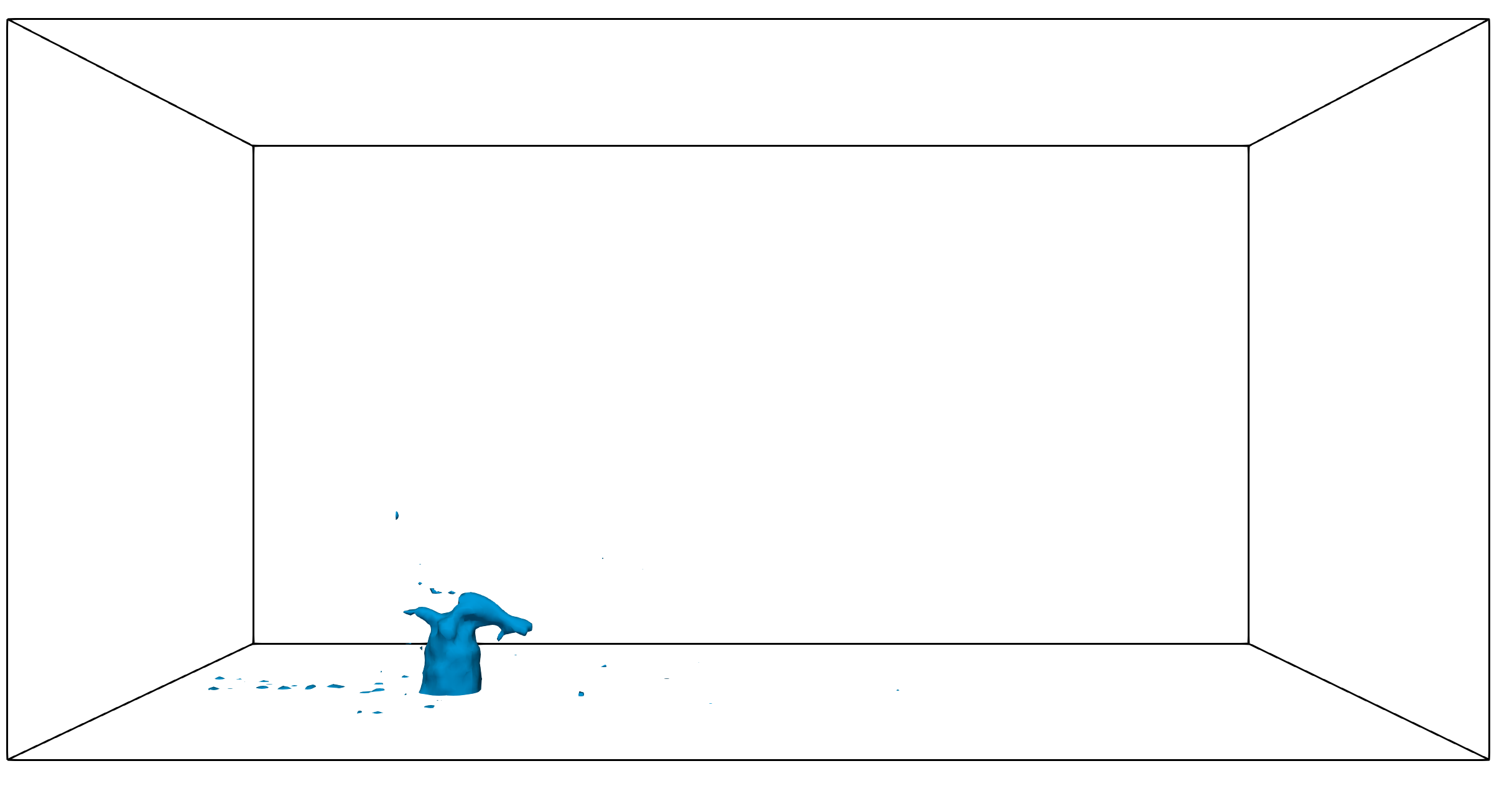}
      \caption{\textcolor{Reviewer1}{Resolution: $N_h=[256,128,128]$; final time: $t=\SI{0.4}{\ms}$.}}
     \end{subfigure}    
    \caption{\textcolor{Reviewer1}{The shape of the exploded injected liquid with interIsoFoam (Euler and cubic, density ratio: $816$, CFL number: $CFL=0.2$.}}
    \label{fig:LJCF_Ecubic}
\end{figure}



\bibliographystyle{elsarticle-num-names} 
\bibliography{literature}


\clearpage

\end{document}